\documentclass[useAMS,usenatbib]{mn2e}
\usepackage{graphicx}
\usepackage{amsmath}

\newcommand{\ha}{H$\alpha$} 
\newcommand{\hbeta}{H$\beta$}
\newcommand{\hc}{H$\gamma$}
\newcommand{\hdelta}{H$\delta$}

\newcommand{\neon}{[Ne~{\sc iii}]}
\newcommand{\neoniv}{[Ne~{\sc iv}]}
\newcommand{\neonHI}{[Ne~{\sc iii}] + H{\sc i}}
\newcommand{\argon}{[Ar~{\sc iv}]}

\newcommand{\argoniii}{[Ar~{\sc iii}]}
\newcommand{\argonV}{[Ar~{\sc v}]}
\newcommand{\helium}{He~{\sc i}}
\newcommand{\heliumb}{He~{\sc ii}}

\newcommand{\nitrogen}{[N~{\sc ii}]}
\newcommand{\nitrogena}{[N~{\sc i}]}
\newcommand{\ironiii}{[Fe~{\sc iii}]}
\newcommand{\oxygeniii}{[O~{\sc iii}]}
\newcommand{\oxygeni}{[O~{\sc i}]}
\newcommand{\oxygenii}{[O~{\sc ii}]}

\newcommand{\sulfur}{[S~{\sc iii}]}
\newcommand{\sulfurt}{[S~{\sc ii}]}
\newcommand{\chloro}{[Cl~{\sc iii}]}

\newcommand{\kripto}{[K~{\sc iv}]}

\newcommand{\degree}{$^{\circ}$}
\def\vhel{\ifmmode{V_{{\rm HEL}}}\else{$V_{{\rm HEL}}$}\fi}
\def\vsys{\ifmmode{V_{\rm sys}}\else{$V_{\rm sys}$}\fi}
\def\kms{\ifmmode{~{\rm km\,s}^{-1}}\else{~km~s$^{-1}$}\fi}
\def\vlsr{\ifmmode{v_{\rm lsr}}\else{$v_{\rm lsr}$}\fi}

\title[Low-ionization structures in planetary nebulae]
{Low-ionization structures in planetary nebulae -- I: physical, kinematic and excitation properties} 

\author[Akras \& Gon\c calves] 
{Stavros Akras $\thanks{e-mail:akras@astro.ufrj.br}$,
Denise R. Gon\c calves\\
Observat\'orio do Valongo, Universidade Federal do Rio de Janeiro, Ladeira Pedro Antonio 
43, 20080-090, Rio de Janeiro, Brazil}

\begin{document}  

\date{Received **insert**; Accepted **insert**}

\pagerange{\pageref{firstpage}-\pageref{lastpage}}

\maketitle
\label{firstpage}

\begin{abstract}
Though the small-scale, low-ionization knots, filaments and jets (LISs) of planetary nebulae (PNe) are known for $\sim$30~yr, 
some of their observational properties are not well established. In consequence our ability to include them in the wider context of the 
formation and evolution of PNe is directly affected. Why most structures have lower densities than the PN shells hosting them? 
Is their intense emission in low-ionization lines the key to their main excitation mechanism? 
Therefore, if considered altogether, can LISs line ratios, chemical abundances and kinematics enlighten the interplay between the different 
excitation and formation processes? Here we present a spectroscopic analysis of five PNe that possess LISs confirming that all nebular components 
have comparable electron temperatures, whereas the electron density is systematically lower in LISs than in the surrounding nebula. 
Chemical abundances of LISs versus other PN components do not show significant differences as well. By using diagnostic diagrams 
from shock models, we demonstrate that LISs' main excitation is due to shocks, whereas the other components are mainly photo-ionized. 
We also propose new diagnostic diagrams involving a few emission lines (\nitrogen, \oxygeniii, 
\sulfurt) and  $\rm{log}$(f$_{shocks}$/f$_{\star}$), where f$_{shocks}$ and f$_{\star}$ are the 
ionization photon fluxes due to the shocks and the central star ionizing continuum, respectively. 
A robust relation differentiating the structures is found, with the shock-excited clearly having $\rm{log}$(f$_{shocks}$/f$_{\star}$)$>$-1; 
while the photo-ionized show $\rm{log}$(f$_{shocks}$/f$_{\star}$)$<$-2. A transition zone, with -2$<\rm{log}$(f$_{shocks}$/f$_{\star}$)$<$-1 
where both mechanisms are equally important, is also defined.

\noindent
\end{abstract}


\section{Introduction}
Over the last 30 years, imaging surveys of planetary nebulae (PNe) have brought to light shapes and 
structures with high complexity (Manchado et al. 1996; G\'orny et al. 1999; Boumis et al. 2003, 2006; 
Parker et al. 2006; Sahai 2011; Sabin et al. 2014). Several works have been carried out thus far attempting to group 
PNe based on their morphologies (Balick 1987; Schwarz et al. 1993; Manchado et al. 1996; Sahai et al. 2011). 
Any morphological classification of PNe relies on the shape of the large-scale structures (nebular 
components) such as rims, attached shells and halos, and divide them into five main classes: point-symmetric, 
elliptical (that includes round), bipolar, multi-polar and irregular. These large-scale structures are better identified in 
hydrogen recombination lines as well as in bright forbidden O$^{++}$ emission lines. 

The formation of spherically symmetric PNe can adequately be explained under the interacting stellar winds 
model proposed by Kwok et al. (1978). Although, the deviation from spherical symmetry is still an open 
question (see Balick \& Frank 2002; Shaw 2012). A considerable effort has been made to explain the formation 
of complex PNe morphologies. Most models consider the presence of an equatorial density enhancement in order to 
collimate the fast stellar wind and result in the formation of the axis-symmetric PNe (e.g. Mellema et al. 1991; Mellema 
1995). The former equatorial density enhancement is usually attributed to the mass-exchange interaction between 
the components in a close binary system (Soker \& Livio 1994; Nordhaus \& Blackman 2006). Single rotating stars, 
with or without magnetic fields, have also been investigated as potential origin of the equatorial density enhancement 
and therefore, the formation of aspherical PNe (Garc\'ia-Segura et al. 1999). Although, very recent magneto--hydrodynamic 
models have shown that complex PNe, such as bipolar, cannot be descendant of single stars (Garc\'ia--Segura et al. 2014). 

Another enigmatic and poorly understood component in PNe are the structures on smaller scales than the 
main nebular components (Balick et al. 1998; Corradi et al. 1996; Gon\c calves et al. 2001). These structures are seen 
mostly in the light of low-ionization lines, such as \nitrogen, \sulfurt, \oxygenii\ and \oxygeni. Corradi et al. (1996) 
unveiled their presence in various PNe, independently the morphological type, by using the image \nitrogen/\oxygeniii\ line ratio 
 technique, which emphasizes the excitation degree through the nebulae.

These small-scale, low-ionization structures (hereafter LISs; Gon\c calves et al. 2001) exhibit a variety of morphological types 
such as knots, jets, jet-like, filaments and head-tails (Balick et al. 1998; Corradi et al. 1996; Gon\c calves et al. 2001), 
whilst they cover a wide range of expansion velocities from few tens to hundreds of \kms.
Based on their expansion velocities, they are labelled as fast, low-ionization emission regions (FLIERs; Balick et al. 1993),  
bipolar, rotation episodic jets (BRETs; L\'opez et al. 1995) or slow moving low ionization emitting regions 
(SLOWERs; Perinotto 2000). Gon\c calves et al. (2001), studying a large number of PNe with LISs, reached the conclusion that LISs 
appear in all morphological classes of PNe, suggesting that they may not be associated with the mechanisms 
responsible for the deviation of PNe structures from the spherical symmetry. 

Several models have been proposed to explain the formation of LISs, but none of them can provide a general mechanism for 
all the different types. Jet formation in PNe is usually associated with the interaction of a binary system and 
an accretion disk around a companion (e.g. Soker \& Livio 1994; Blackman et al. 2001). As for the knots, any clue for their 
formation also needs to consider if they appear in pairs or are randomly distributed (Gon\c calves et al. 2001). 
Symmetric pairs of low-ionization knots seem to appear due to dynamical or/and radiation instabilities (Garc\'ia--Segura 
et al. 1999). For the isolated knots, {\it in situ} instabilities Soker \& Reveg (1998), and the stagnation knot's model 
(Steffen et al. 2001) seem to be plausible formation mechanisms that explain some of their characteristics, as much as 
the shock models from Dopita (1997). For more details on formation mechanisms of LISs see the comprehensive review by 
Gon\c calves et al. (2004). A common characteristic of the models for the formation of LISs is the determination 
of their age with respect to the main body of the nebula. It has been observationally found that  
some jets predate the main nebular structure, whereas others appear to be younger, like that of NGC~6337 (Tocknell 
et al. 2014; Jones et al. 2014) and the jets and knots of the IC~4846, Wray~17-1 and K~1-2 studied in this paper. 
Altogether the latter two papers quoted, and previous results for the age of LIS in PNe (Gon\c calves 2004) reveal 
a likely connection of LISs with the evolution of the PNe central stars.

Miszalski et al. (2009a) had also proposed a possible link between LISs and the central stars of PNe. 
These authors argue that LISs seem to be very common in PNe with hydrogen-deficient stars,  
e.g [WR]. In this scenario, LISs are likely associated with the turbulence of the strong stellar winds 
from these, though this connection is still lacking further confirmation. In addition, 
the same authors claim that there may also exist a relation between the presence of LISs and the binarity of the 
central stars, with 40\% of post common envelope (CE) PN possessing LISs.

The physical and chemical properties e.g. electron temperature ($\it{T}_{\rm{e}}$), electron density ($\it{N}_{\rm{e}}$), 
ionic and elemental abundances of LISs have also been studied by different groups over the past e decades  
(Balick et al. 1993, 1998; Hajian et al. 1997; Gon\c calves et al. 2003, 2004, 2009; Leal--Ferreira et. al. 
2011; Monteiro et al. 2013). No significant difference between the $\it{T}_{\rm{e}}$ of LISs and that of 
the higher excitation nebular component (e.g. core, rims, shells) has been found, 
contrary to the $\it{N}_{\rm{e}}$, which is found to be systematically lower in LISs when compared to the surrounding medium. 
This result is in conflict with the theoretical predictions from the formation models of LISs, 
which predict LISs to have several times higher density than the surrounding ionized medium. Gon\c calves et al. (2009) claim 
that the mass of LISs must be mostly neutral in order to overcome the discrepancy between models and observations, 
since all the formation models of LISs calculate the total density of gas (dust, atomic and molecular) 
and not the electron density, which corresponds only to the ionized fraction of the gas.

The strong \nitrogen\ emission line found in LISs compared to the surrounding ionized nebula 
was firstly attributed to a significant local overabundance of nitrogen (Balick et al. 1994). 
Mellema et al. (1998) and, more recently, Gon\c calves et al. (2006) reached the conclusion that 
an overabundance of N is not a necessary condition for getting such a strong line-emission. 
Based on photo-ionization models, Gon\c calves et al. (2006) found that the N$^+$/N = O$^+$/O 
assumption of the ionization correction factor (ICF, Kingsburgh \& Barlow 1994) should be wrong, 
and resulting in seemingly overabundance of N in LISs. 

Accordingly to the theoretical work by Aleman \& Gruenwald (2011), the peak intensity of 
the low-ionization optical lines (\oxygenii, \nitrogen, \sulfurt, \nitrogena\ and \oxygeni), as well as the 
H$_2$ co-rotational lines, occurs in the same narrow transition zone between the ionized and neutral 
(photo-dissociation) regions. Given that molecular hydrogen can be excited either by shocks or 
by absorbing UV-photons emitted from the central star, the enhancement of low-ionization lines in LISs may be 
attributed to these mechanisms. Raga et al. (2008) show that shock models with low and moderate photo-ionization rates 
are able to reproduce emission line ratios typical of shock-excitation regions, whereas those models with 
high photo-ionization rates result in emission line ratios similar to photo-ionized nebulae. Therefore, besides 
the expansion velocities, the evolutionary stage (e.g. the age) and the photo-ionization rate (or the strength) 
of the central star are crucial parameters in order to distinguish between predominantly shock- and photo-excited structures.

In this series of papers, new spectroscopic data from a sample of 10 PNe that possess LISs are combined with 
kinematic data in order to study, at the same time, the physico-chemical and morpho-kinematic properties of LISs and 
other nebular components. The first 5 PNe, with LISs embedded in the main nebular shells,  
are presented here. The remaining PNe will appear in a forthcoming publication. The observations and data 
analysis are described in Section 2. The emission line fluxes, the physico-chemical properties, and the 
morpho-kinematic characteristics of the PNe themselves, as well as of their LISs, are presented in Section 3 
and 4, respectively. The results of this work are discussed in Section 5, and 
we wrap up with the conclusions in Section 6.

\section{Observations}

\begin{figure}
\centering
\includegraphics[scale=0.425]{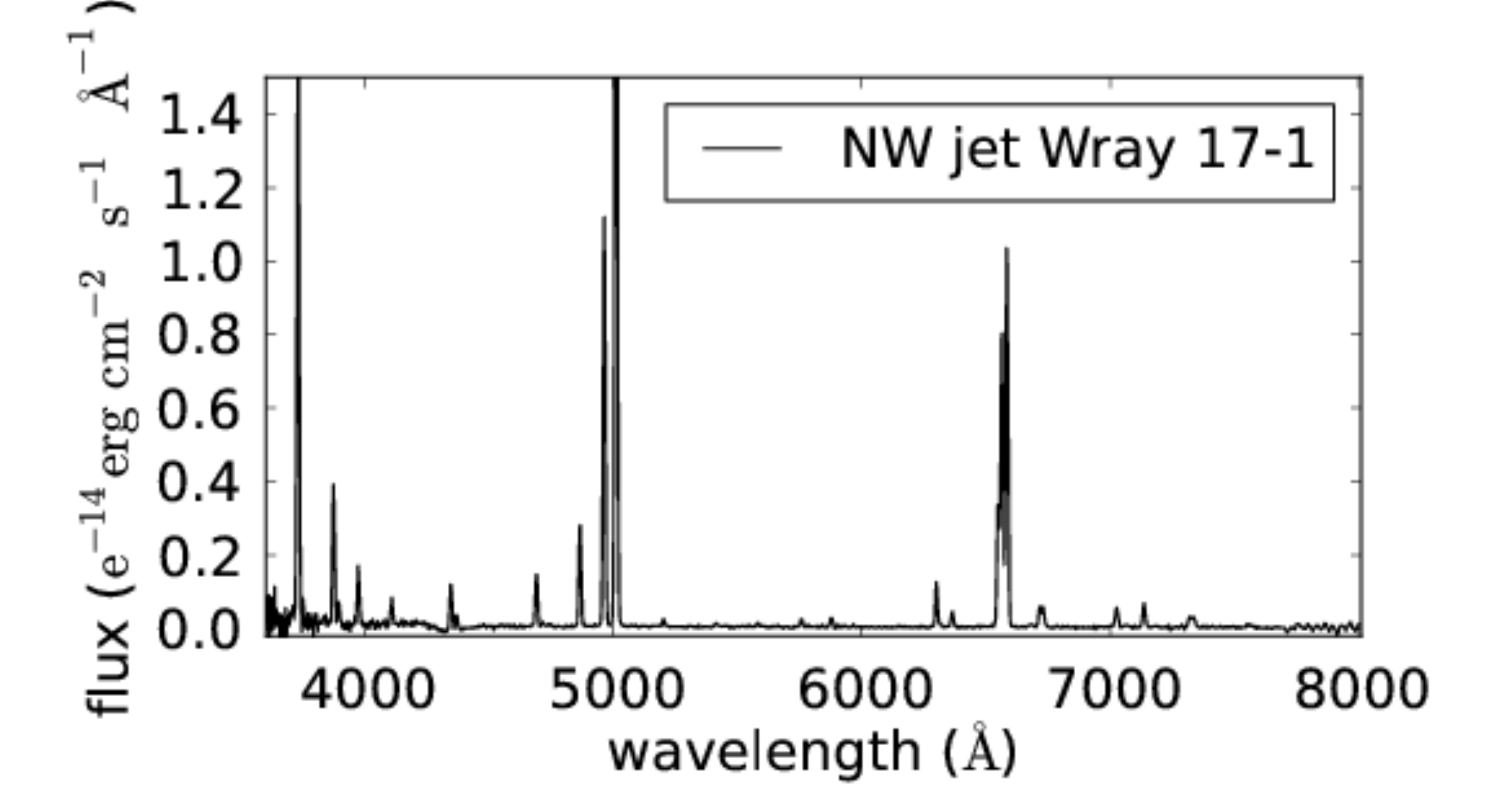}
\includegraphics[scale=0.425]{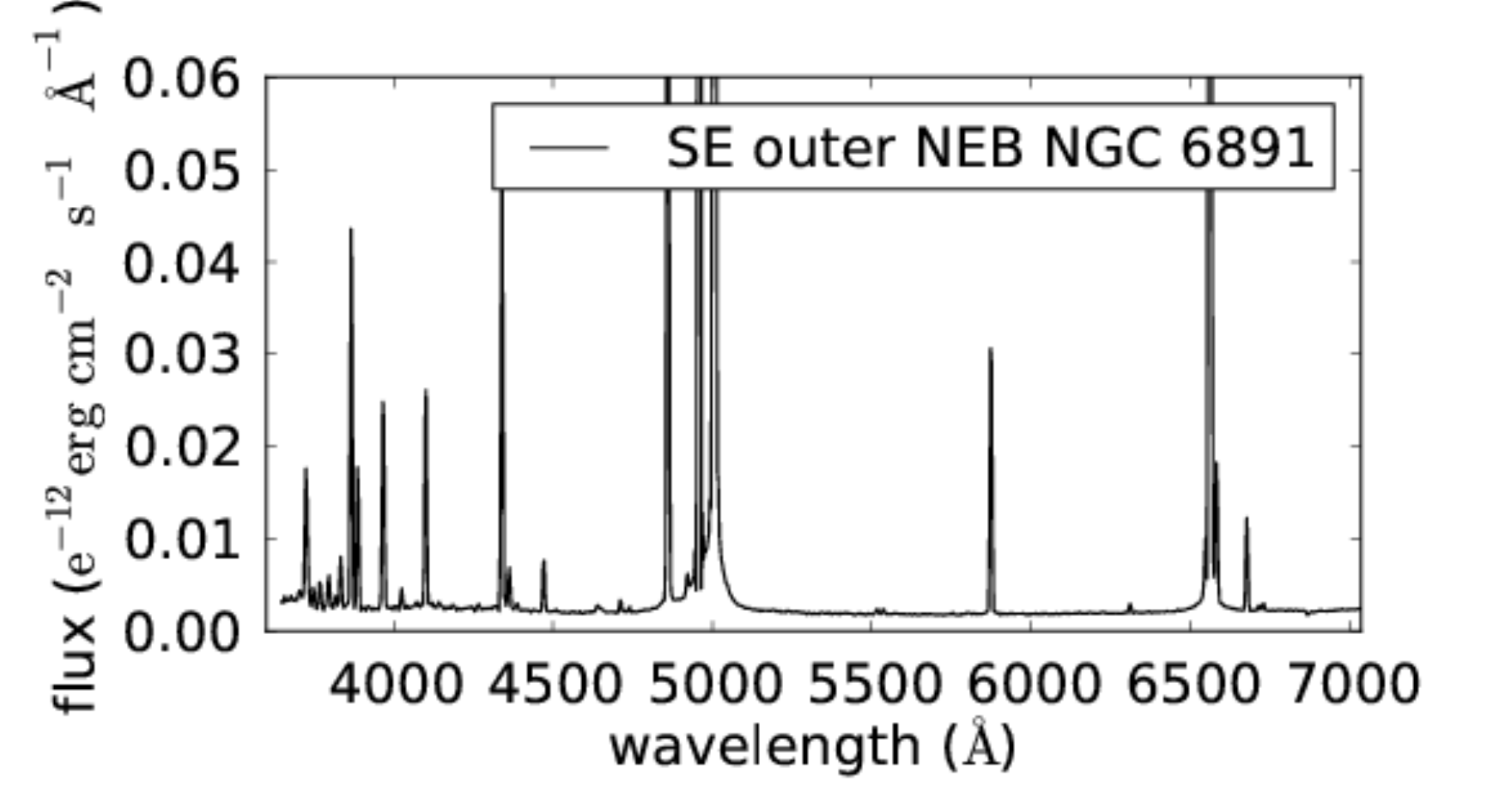}
\caption[]{Observed spectra of the NW jet in Wray~17-1 (upper panel), covering the wavelength range of 3600--8000~\AA, obtained 
with DFOSC@Danish, and SE outer NEB in NGC~6891 (lower panel) with a wavelength range of 3650--7050~\AA, and observed with the 
IDS attached to the Isaac Newton telescope.}
\end{figure}

\subsection{Intermediate-dispersion Spectroscopy}
Intermediate-dispersion spectra of three PNe (NGC~6891 and NGC~6572, IC~4846) were obtained with the
2.5~m Isaac Newton Telescope (INT) at the Observatorio del Roque de los Muchachos, Spain, 
on 2001 August 30, 31 and September 4, respectively. The Intermediate 
Dispersion Spectrograph (IDS) was used with the 235~mm camera and the R300V grating resulting in a scale of 
3.3 \AA\ pixel$^{-1}$ and covering the wavelengths range of 3650-7000 \AA. The spatial scale 
of the instrument was 0.70~arcsec~pixel$^{-1}$, with the TEK5 CCD. The slit width was 1.5~arcsec and 
the slit length 4~arcmin. 

The spectra of K~1-2 and Wray 17-1 were obtained with the 1.54 m Danish telescope at the European
Southern Observatory (ESO), La Silla (Chile), on 1997 April 10 and 11, respectively. 
The Danish Faint Object Spectrograph and Camera (DFOSC) was used with the 2000 $\times$ 2000 CD (15$\mu$m 
pixel) and the Grism \#4 (300 lines mm$^{-1}$), resulting in 2.2 \AA\ pixel$^{-1}$, from 3600 to 8000 \AA. 
The spatial scale of the instrument was 0.40~arcsec~pixel$^{-1}$. The slit width and length were 1.0~arcsec 
and $>$13.7~arcmin, respectively. 

Spectro-photometric standard stars were also observed with the INT and Danish telescopes, 
in order to calibrate the spectra, while the fluxes were corrected for the atmospheric extinction 
and interstellar reddening. During the night, bias frames, twilight and tungsten flat-field exposures, 
and wavelength calibrations were also obtained. Data reduction was performed using the standard \textsc{iraf} 
instructions for long-slit spectra. Individual images were bias subtracted and flat-field corrected using a 
series of twilight flat frames. Two examples of the spectra under analysis, one for each telescope configuration, 
are presented below in Fig.~1.

\subsection{High-dispersion Spectroscopy}
High-dispersion, long-slit spectroscopy in the \ha+\nitrogen\ and/or 
\oxygeniii\ emission lines, for 4 out of the 5 PNe in our sample, were acquired from the SPM Kinematic Catalogue 
of Galactic Planetary Nebulae (L\'opez et al. 2012; hereafter KCGPN). The spectra of three PNe (NGC~6572, NGC~6891 
and K~1-2) were obtained at the 2.1-m, f/7.5 telescope in San Pedro M\'artir National Observatory (Mexico), whereas for 
Wray~17-1, the data were obtained from the 3.9-m, f/8 Anglo--Australian telescope in the Siding Spring Observatory (Australia), 
and in both cases by using the Manchester Echelle Spectrometer (MES--SPM; Meaburn et al. 2003). Two slit widths of 
70 and 150$\mu$m were used, corresponding to a velocity resolution of 9.2 and 11.5~\kms, respectively. 
Additional kinematic information were also drawn for the literature.

\section{Results: physico-chemical properties}

In the even numbered tables, from Table~2 to 16, we list the emission line fluxes, corrected for 
atmospheric extinction and interstellar reddening, for several nebular components, e.g. LISs (knots, 
jets, filaments), inner and outer main nebular regions (NEB) and for the entire PN, whenever possible. 
All fluxes are normalized to $F$(\hbeta)=100. The interstellar extinction $c$(\hbeta) was derived from the 
Balmer \ha/\hbeta\ ratio (eq.~1), using the interstellar extinction law by Fitzpatrick (1999) and R$_{\rm{v}}$=3.1,
 
\begin{equation}
c(\rm{H}\beta)=\frac{1}{0.348}\rm{log}\frac{\it{F}(\rm{H}\alpha)/\it{F}(\rm{H}\beta)}{2.85}
\end{equation}

\noindent
where 0.348 is the relative logarithmic extinction coefficient for \hbeta/\ha. 

$\it{T}_{\rm{e}}$ and $\it{N}_{\rm{e}}$ for each nebular component were derived by using the {\sc temden} task in \textsc{iraf} 
(Shaw \& Dufour 1995) and are presented in the lowermost part of the flux tables. 
Given that the abundances derived from collisionally excited lines are strongly dependent 
on $\it{T}_{\rm{e}}$, two combinations of $\it{T}_{\rm{e}}$ and $\it{N}_{\rm{e}}$ have been used depending on the 
degree of ionization of each ion, whenever possible. The combination of $\it{T}_{\rm{e}}$\nitrogen\ and 
$\it{N}_{\rm{e}}$\sulfurt\ covers the low-excitation regions (N$^{+}$, O$^+$ and S$^+$), whereas the combination 
of $\it{T}_{\rm{e}}$\oxygeniii\ and $\it{N}_{\rm{e}}$\sulfurt\ covers the medium-excitation regions of these PNe 
(e.g. O$^{++}$, S$^{++}$, Ne$^{3+}$). Significant differences between the two diagnostic temperatures are found 
in IC~4846, K~1-2 and NGC~6572, which make the usage of the two diagnostic temperatures necessary for properly 
deriving the ionic and chemical abundances.

In addition, for a proper estimation of $\it{N}_{\rm{e}}$ from the \argon\ diagnostic lines, we calculated and subtracted 
the contribution of the \helium\ $\lambda 4712$ recombination line and those of the \neoniv\ $\lambda\lambda 4724$, 
4726 emission lines to the \argon\ $\lambda 4711$ one. In particular, the contribution of the \helium\ $\lambda 4712$ line 
was calculated using the theoretical work by Benjamin et al. (1999), whereas the contribution of the \neoniv\ $\lambda\lambda 
4724$, 4726 emission lines were considered negligible in all cases, given that the central stars of the PNe in this work are not 
hot enough to ionize Ne$^{3+}$ (P.I.=97.11 eV), except the central star of Wray~17-1 ($T_{\rm eff}$=140000~K; Rauch and Werner 1997). 
Although, the $\it{N}_{\rm{e}}$ of this nebula is extremely low, close to the lower limit, and small changes in the \argon\ 4711/4741 line ratio  
due to the contribution of the \neoniv\ lines will not have a significant 
impact in the $\it{N}_{\rm{e}}$. In short, a comparison of $\it{N}_{\rm{e}}$ derived from the different diagnostics lines does not show significant 
differences that could be associated with the \neoniv\ lines. We have to mention here the extreme case of NGC~6891, for which 
we find significantly higher $\it{N}_{\rm{e}}$\argon, compared to $\it{N}_{\rm{e}}$\sulfurt\ and $\it{N}_{\rm{e}}$\chloro, by factors 
of 6 and 3.5, respectively. This probably indicates strong density stratification, with denser inner regions with respect to the
outer ones.

\begin{figure}
\centering
\includegraphics[scale=0.29]{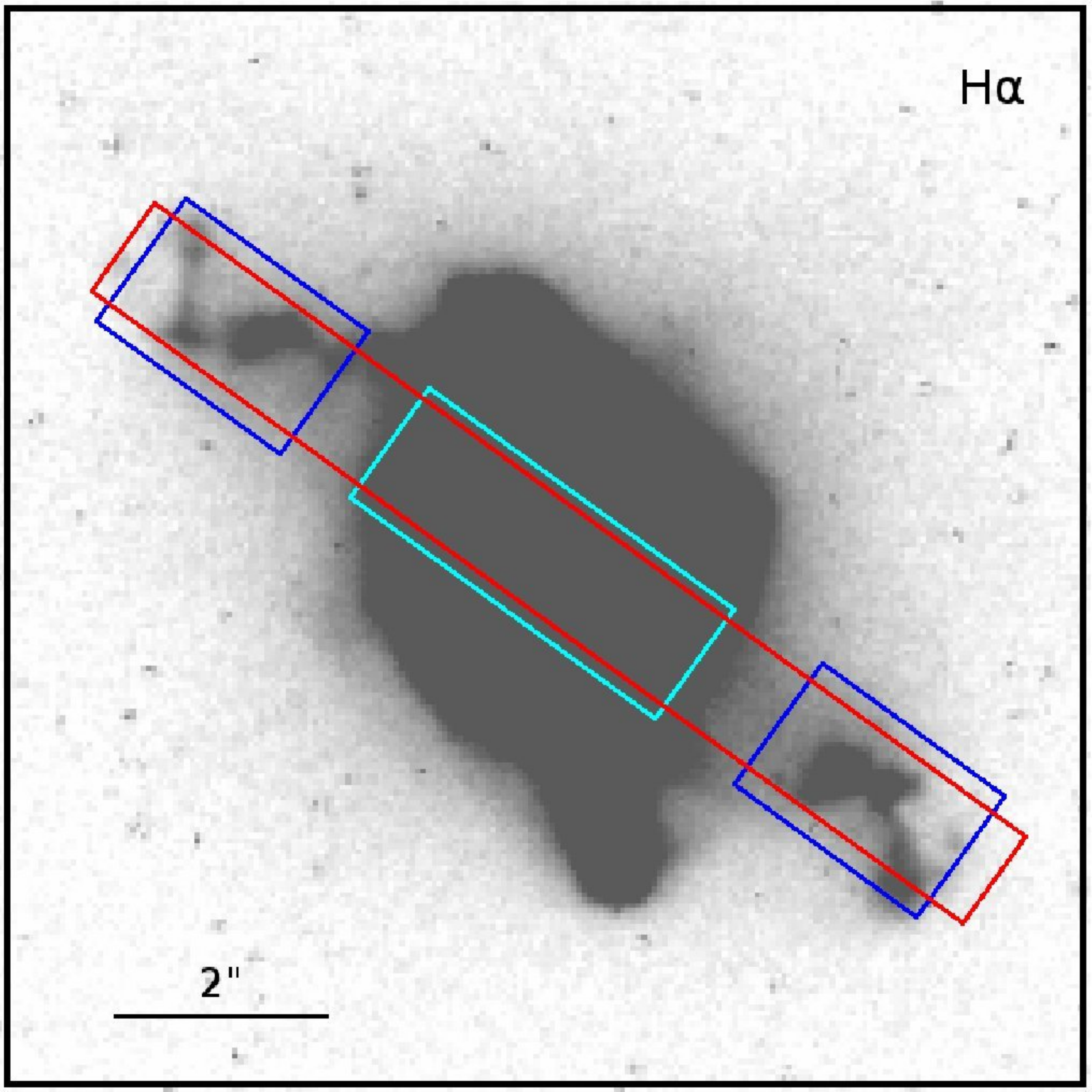}
\includegraphics[scale=0.43]{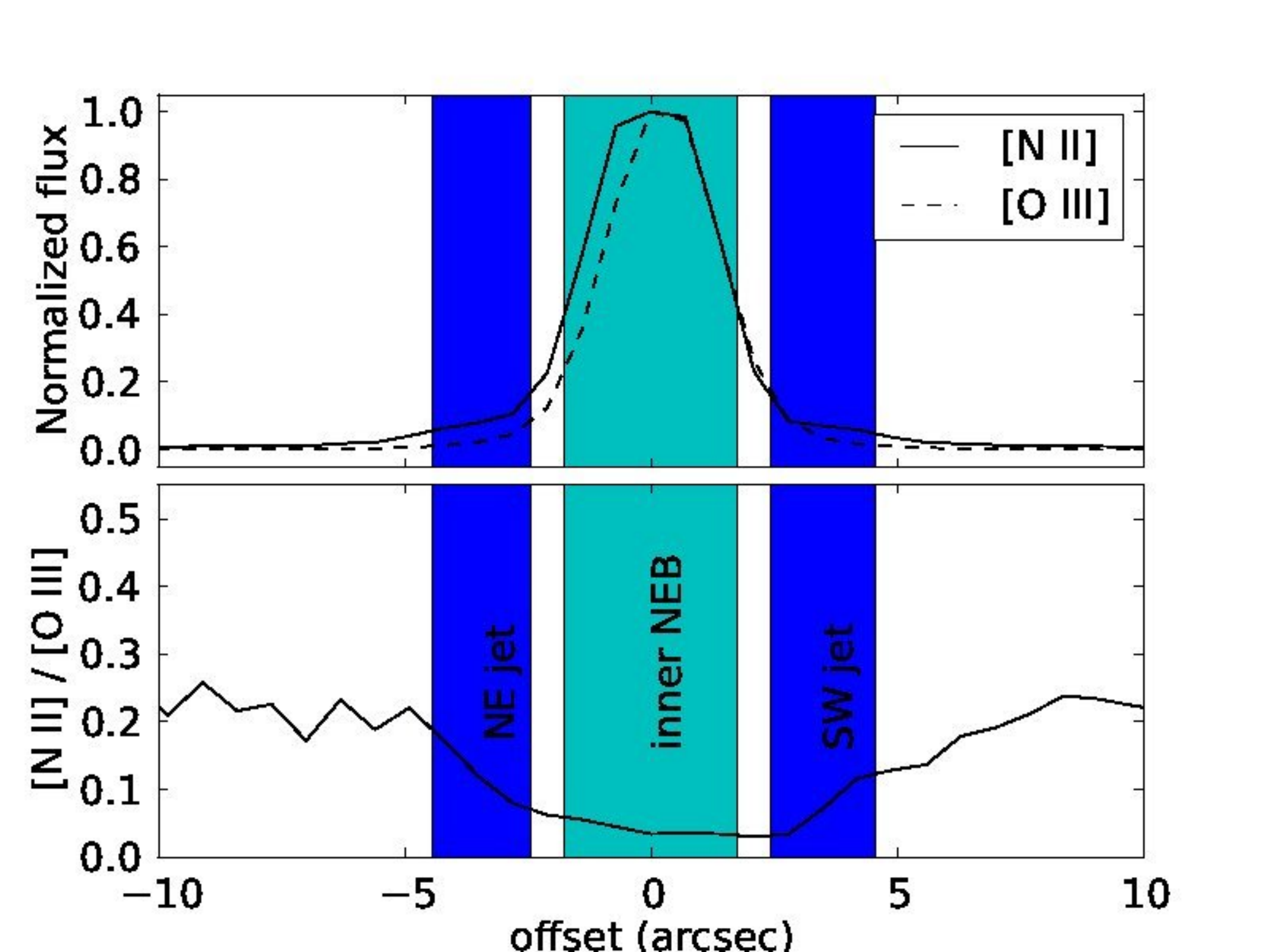}
\caption[]{Upper panel: The HST \ha\ image of IC~4846. The size of the field shown is 10$\times$10~arcsec$^2$.
The nebular components under analysis are indicated in the image. The extraction windows for these 
components are indicated by the corresponding boxes: the entire nebula (entire NEB: 10~arcsec, in red), the inner nebula 
(inner NEB: 3.5~arcsec, in cyan) and the NE and SW jets (2.1~arcsec, in blue). North is up, and East to the left.
Lower panels: The radial profile of the \nitrogen\ and \oxygeniii\ emission-line fluxes, normalized to 1.0 
and the \nitrogen/\oxygeniii\ line ratio, along the slit.}
\end{figure}

\subsection{IC~4846}

IC~4846 (PN G027.6-09.6) is a relatively compact nebula with a size of 10~arcsec. Low angular resolution 
images of this nebula (Miranda et al. 2001) show two ellipsoidal shells at PAs of 11\degree\ (inner shell) and 54\degree\ (outer shell). 
These authors also show the \nitrogen/\ha\ line ratio image which reveals the presence 
of features enhanced in N$^+$: i) a pair of knots, attached to the inner shell, at PA=11\degree; and ii) a pair of 
filaments, attached to the outer shell, at PA=54\degree.

A higher resolution \ha\ image of IC~4846 (Fig.~2), obtained with the HST in 2001 (program ID:8345; PI: Sahai), clarifies that 
the pair of filaments is made of highly collimated structures (jets) with some indications of precession. In this new image, the pair of 
knots along PA=11\degree\ appears to be the result of the interaction between the inner and outer shell. It is worth mentioning here 
the astonishing similarity between IC~4846 and Fleming~1 (Boffin et al. 2012). Both PNe exhibit a pair of knotty, precessing, bipolar 
jets or BRETS (L\'opez et al. 1995), likely associated with an accretion disk around a binary system. This binary system has already 
been confirmed in Fleming~1 (Boffin et al. 2012) but not yet in IC~4846.

\begin{figure}
\centering
\includegraphics[scale=0.27]{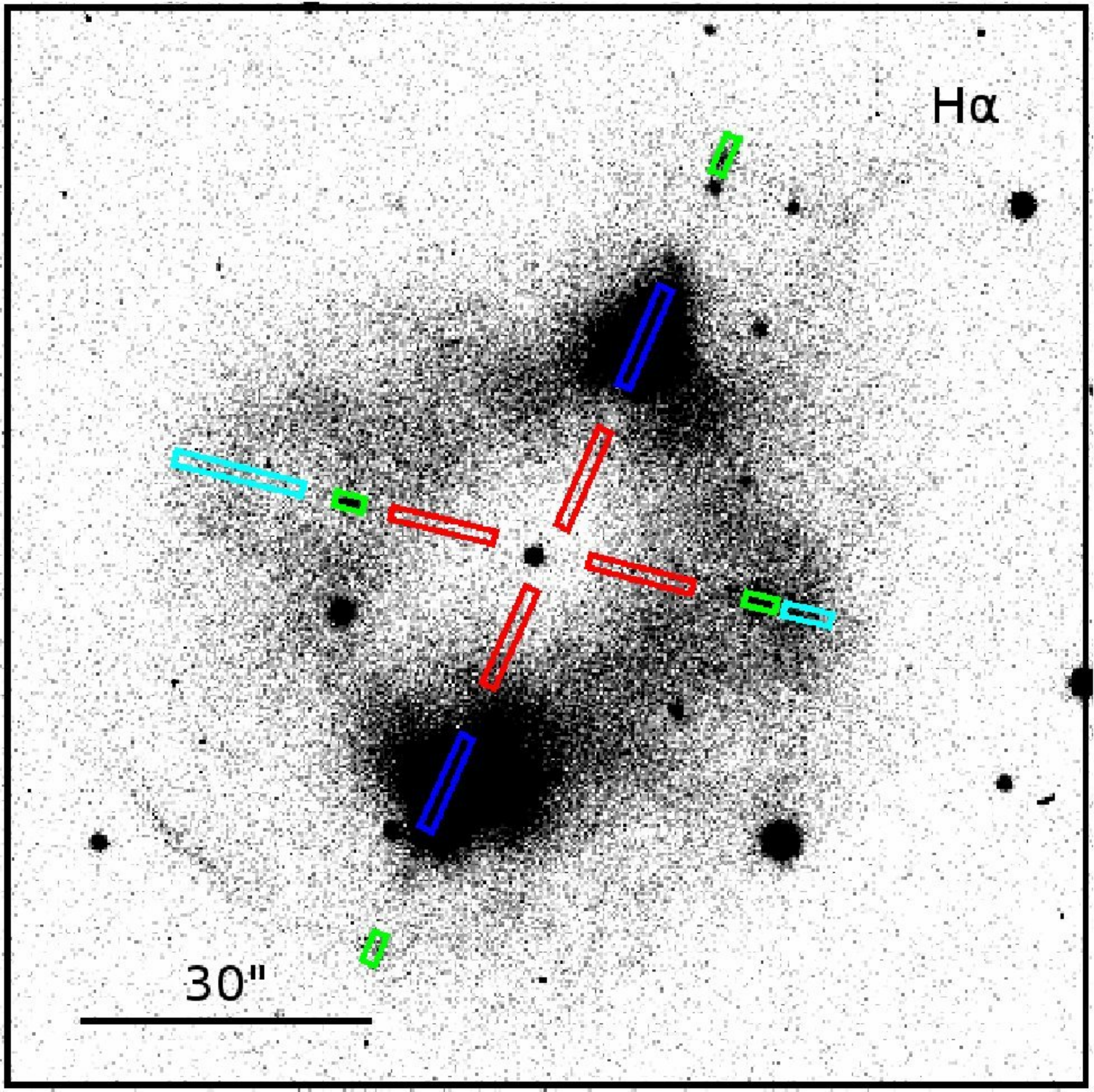}
\includegraphics[scale=0.41]{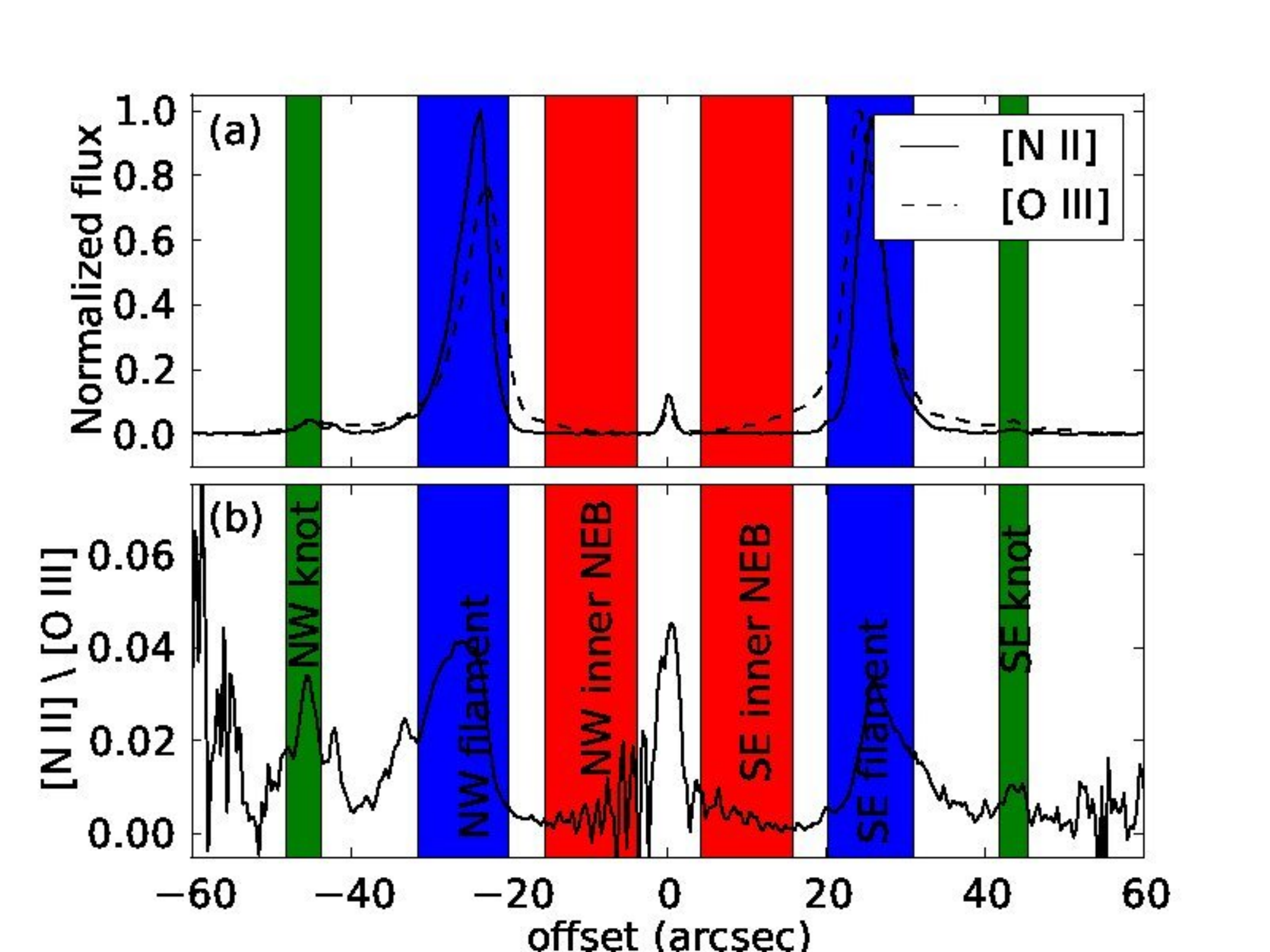}
\includegraphics[scale=0.41]{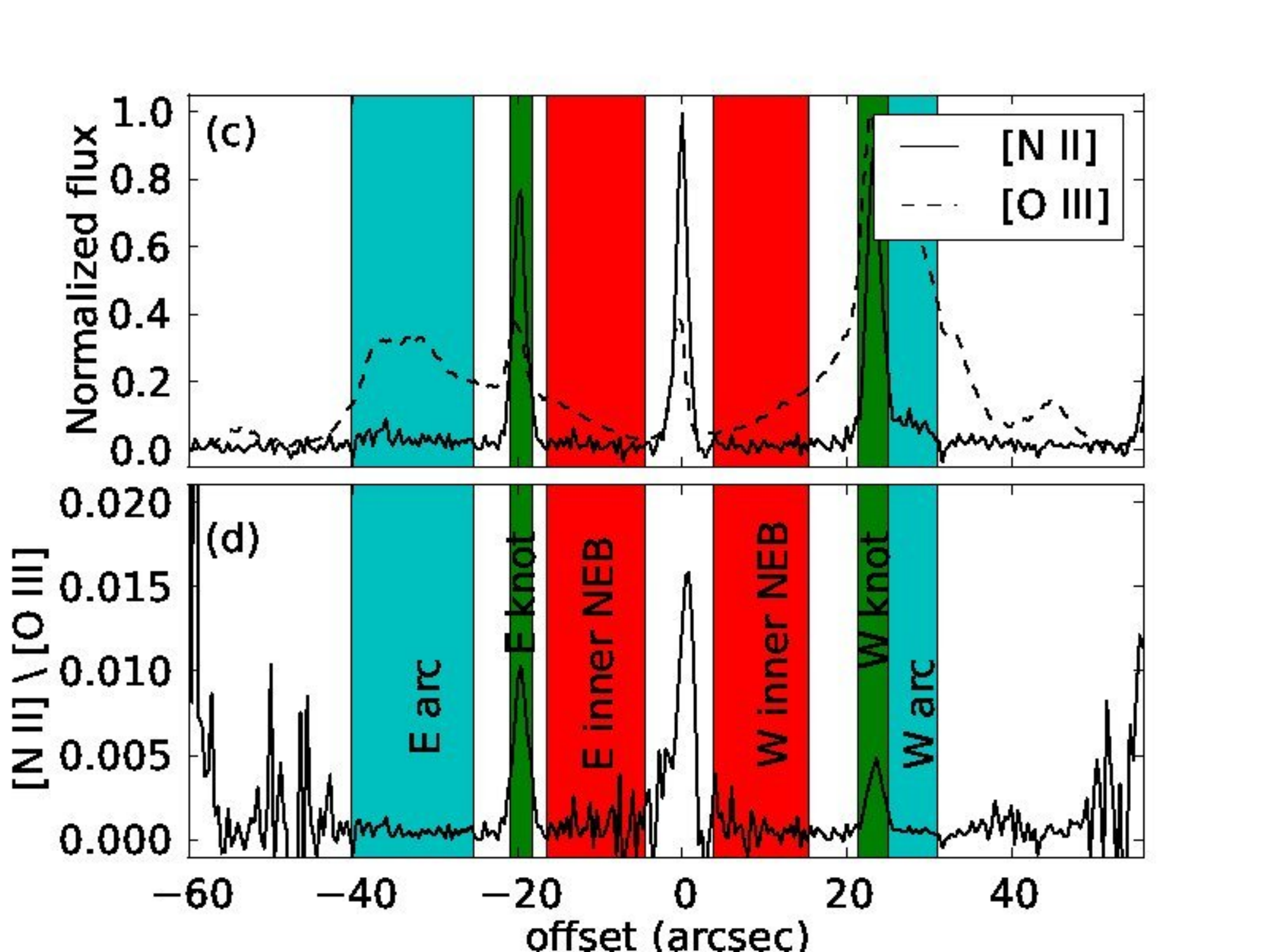}
\caption[]{Upper panel: \ha\ image of Wray~17-1 obtained from the ESO Archive. The field shown is 110$\times$110~arcsec$^2$. The nebular components under analysis are indicated in the image. 
The extraction windows for these components are indicated by the  
corresponding boxes: the inner nebular regions (inner NEB: 11.6~arcsec, in red); the NW and SE blobs (11.6~arcsec 
and 10.8~arcsec, in blue); the knots (NW knot of 4.4~arcsec, SE knot of 3.6~arcsec, NE knot of 2.8~arcsec, and 
SW knot of 3.6~arcsec, in green); and the arcs (E arc of 14.8~arcsec and W arc of 6~arcsec, in cyan). North is up, and 
East is to the left. Lower panels: The radial profile of the \nitrogen\ and 
\oxygeniii\ emission-line fluxes, normalized to 1.0 and the \nitrogen/ \oxygeniii\ line ratio along the slits. Panels (a) and (b) correspond to PA=155\degree, 
whereas (c) and (d) to PA=75\degree).} 
\end{figure}

Nevertheless, the central star of IC~4846 has been classified as a weak emission line or WEL-type star (Acker et al. 1992; G\'orny et al. 2009).
The typical line features of WELs and/or [WC] stars, such as the $\lambda 4650$ (C {\sc III}--C {\sc IV}) and the $\lambda 5805$ 
(C {\sc IV}) lines, were detected in our spectra with a full width at half maximum (FWHM) of 22$\pm$2 and 24$\pm$2~\AA, respectively. 
These detections, therefore, reinforce the possibility the central star of IC~4846 may belong to the rare group of binary systems with a WR or WEL-type companion.

In Table~2 we list the emission line fluxes of the inner shell, the pair of jets at PA=54\degree, 
and the integrated emission of the nebula along the slit (entire nebula). Unfortunately, no spectra 
along PA=15\degree\ were obtained. The exact location of the extracted 
portions of the spectra are shown, and labelled, in Fig.~2. 

IC~4846's interstellar extinction, $c$(\hbeta), is found to vary between 0.48$\pm$0.03 and 0.53$\pm$0.05, 
in excellent agreement with those derived from previous studies (0.47, Kaler 1986; 0.50, 
Cahn et al. 1992). Notice, however, that the most recently published value, by Wesson et al. (2005), 
is higher by a factor of 1.45. One possible explanation for this discrepancy may be the different 
slit positions (PA=0\degree\ in Wesson et al. 2005). The logarithmic \hbeta\ fluxes are calculated equal to 
-11.62, -13.49 and -13.45 and 11.54 for the inner shell, NE jet, SW jet, and the entire nebula, respectively. 
Acker et al. (1991) give two \hbeta\ fluxes for this nebula (-11.56 and -11.26), which were derived from the 
Haute--Provence Observatory spectroscopic data (OHP--CCD) with a slit width of 2.5~arcsec, and from the 
ESO--CCD, spectra with a slit width of 4~arcsec, respectively. Our value for the entire nebula better agrees 
with the value derived from the OHP--CCD data because of the same slit widths.

The lower panels in Fig.~2 show the fluxes, normalized to 1.0, of the \nitrogen\ (solid-line) and \oxygeniii\ 
(dashed-line) emission lines along the slit (panel a), as well as the \nitrogen/\oxygeniii\ line ratio (panel b). 
The (spatial) line profiles and the ratio of these two lines are important in order to better define the sizes of 
each structure along the slit. The size of each nebular component is indicated in the figures by the coloured regions.

$\it{N}_{\rm{e}}$ and $\it{T}_{\rm{e}}$ of IC~4846 are estimated using the diagnostic line ratios of sulfur, argon,
chlorine and oxygen. No significant difference is found among them (see Table 2). The 
comparison of $\it{N}_{\rm{e}}$ and $\it{T}_{\rm{e}}$ between the nebular components shows that the jets are significantly less 
dense than the inner shells, by a factor of ~4, whereas, within the errors, all structures have the same $\it{T}_{\rm{e}}$. 
Our $\it{N}_{\rm{e}}$ values are slightly higher than those found by Wang et al. (2004). We have to notice here that 
an extremely high $\it{N}_{\rm{e}}$\sulfurt\ of $\sim$20000~cm$^{-3}$ has been reported by Wesson et al. (2005), 
which is not confirmed by our analysis, neither for the jets nor for the inner shell. Our analysis reveals that 
the jets in IC~4846 have low $\it{N}_{\rm{e}}$, with respect to the main nebular component (inner NEB), 
and therefore effaces any argument that the LISs in this nebula could have Ne higher than the surrounding medium. 
The questionable line ratios given by Wesson et al. (2005) were also previously mentioned by Delgado-Inglada et al. (2015).

In Table 3 we present the ionic abundances (obtained using the {\sc ionic} task in \textsc{iraf}; Shaw \& Dufour 1995) 
as well as the total chemical abundances of IC~4846 for each nebular component. The total ones were computed using the ionization 
correction factor (ICF) from Kingsburgh \& Barlow (1994). These authors do not provide an ICF for Cl, thus we used the 
 equation by given Liu et al. (2000). 

The low He abundance and N/O ratio of IC~4846 indicate a non-type I nebula. We do not find any trend in chemical 
abundances among the nebular component; therefore, they are all constant within the uncertainties. 
A comparison between our total chemical abundances and those found in the literature 
show a good agreement within the errors, except for Ar (Perinotto et al. 2004; Wesson et al. 2005). 
The low N abundance in the SW jet is highly uncertain due to the ICF scheme for N (mentioned in the Introduction), 
as well as to the fact that only N$^{+}$ was measured in the different regions. However, both N/H of the SW and NE 
jet agree, within the errors.

\subsection{Wray~17-1}
Wray~17-1, or PN G258.0-15.7, displays a diffuse complex nebula of approximately 80~arcsec in diameter 
(Fig.~3). A pair of bright regions is seen in the \oxygeniii, \nitrogen\ and \sulfurt\ emission line images 
embedded in the diffuse nebula, along PA=155\degree. The \nitrogen /\oxygeniii\ line ratio image 
of the nebula unveils highly collimated jet-like structures (Corradi et al. 1999). 
At the end of these jet-like structures (hereafter filaments), a pair of knots can also be discerned, 
whereas a second pair of knots is also apparent along the PA=75\degree. The latter is found to 
be surrounded by two much fainter arc-like structures. These arcs show an enhanced \oxygeniii /\ha\ ratio, 
suggesting a shock interaction (Guerrero et al. 2013). 

The central star of Wray~17-1 has been found to be a H-deficient P Cygni star with a T$_{\rm{eff}}$=140000~K and 
$\rm{log}$(g)=6.3 (Rauch \& Werner 1997). The formation of jets and/or knots may be associated with the strong 
turbulent stellar winds from this star (Miszalski et al. 2009b) or a putative accretion disk around a close binary system 
(Soker \& Livio 1994, Miszalski et al. 2009b).

Several nebular regions/components are selected for the analysis of this nebula:  
the pair of filaments, the two pairs of knots, the two arcs and four inner nebular regions, for 
direct comparison with the properties of the entire nebula. In Fig.~3, we label each of these features with the spatial regions under analysis. The lower panels show the line profiles of the 
normalized fluxes of the \nitrogen\ and \oxygeniii\ emission lines, as well as their ratio. 
The position of the filaments and knots is noticeable in these diagrams. The emission line fluxes, the absolute flux of 
\hbeta, $c$(\hbeta), $\it{N}_{\rm{e}}$ and $\it{T}_{\rm{e}}$ for all these regions are presented in Tables~4 (PA=155\degree) 
and 6 (PA=75\degree). The $c$(\hbeta) of this nebula varies from 0.04 to 0.13 among the nebular component. 
The average value is 0.08$\pm$0.03, which is in a good agreement with the value derived by Tylenda et al. (1992). 
$\it{T}_{\rm{e}}$\oxygeniii\ and $\it{T}_{\rm{e}}$\nitrogen\ are found to vary from 11100~K to 14700~K and from 11625~K 
to 13625~K, respectively. Within the uncertainties, all these measurements agree with one another.

Unlike the $\it{T}_{\rm{e}}$, where no variation among the nebular components is found, 
$\it{N}_{\rm{e}}$ may show some variation. In particular, the LISs (knots and jet) have higher $\it{N}_{\rm{e}}$ 
compared to the inner nebular regions, but because of their high uncertainties, we can say whether the LISs are, 
in general, less or more dense than the inner nebular regions. Moreover, the $\it{N}_{\rm{e}}$ values of the inner 
nebular regions are close to the low density limit of 100~cm$^{-3}$, and they should be taken with caution.

We present here the first measurements of the ionic and total chemical abundances of Wray~17-1 (Tables 5 and 7). 
Surprisingly, the knots of Wray~17-1 do not present any He$^+$ (all He is doubly ionized) and hence the abundance of He  
cannot be calculated properly, similarly to K~1-2 (see below). The high effective temperature of the 
central star (T$_{eff}$=140000~K) is consistent with the high excitation of the nebula. Unlike the knots, the filaments in 
Wray~17-1 appear to have much lower excitation, where He$^+$ is more abundant than He$^{++}$.

\begin{figure}
\centering
\includegraphics[scale=0.27]{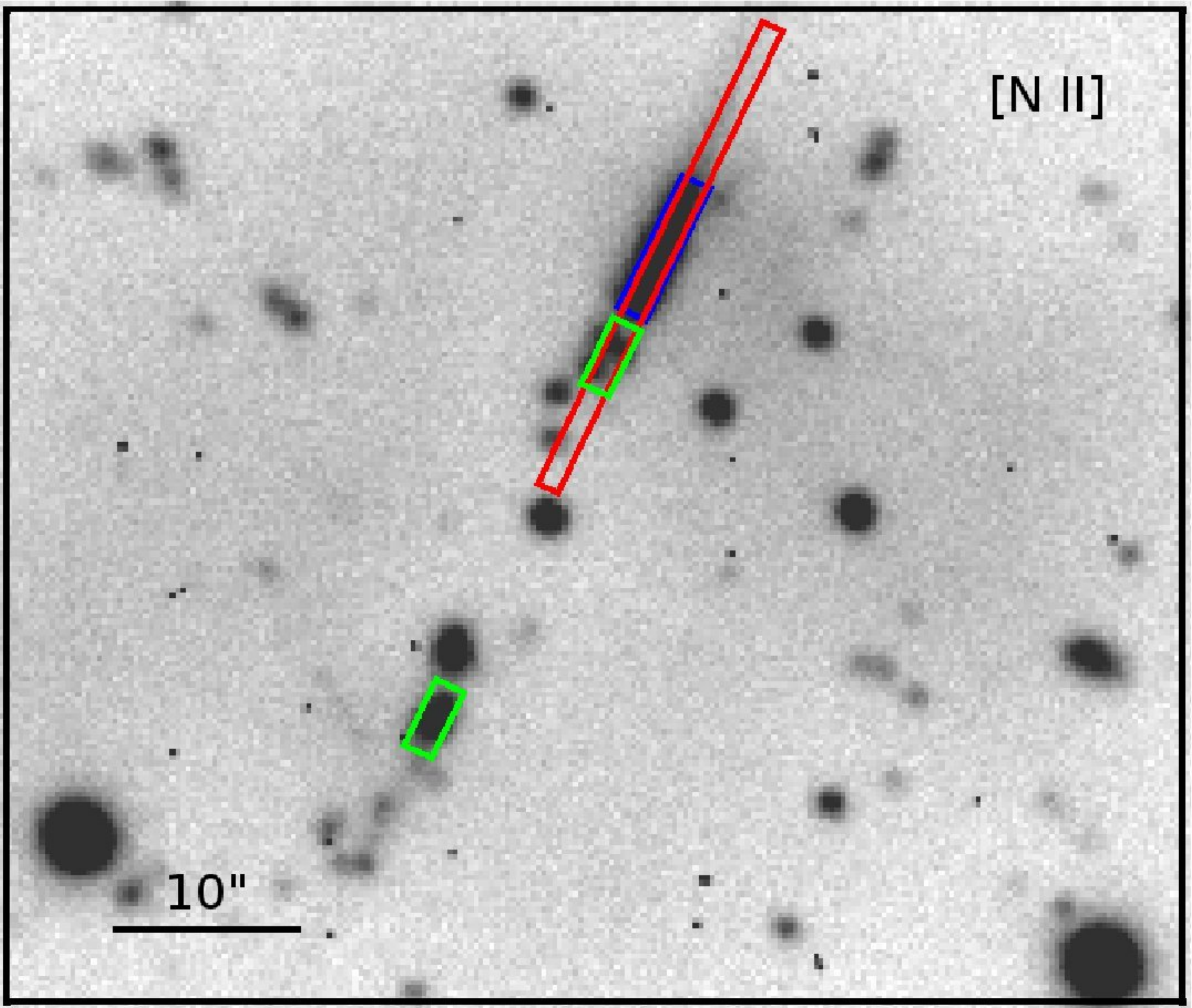}
\includegraphics[scale=0.43]{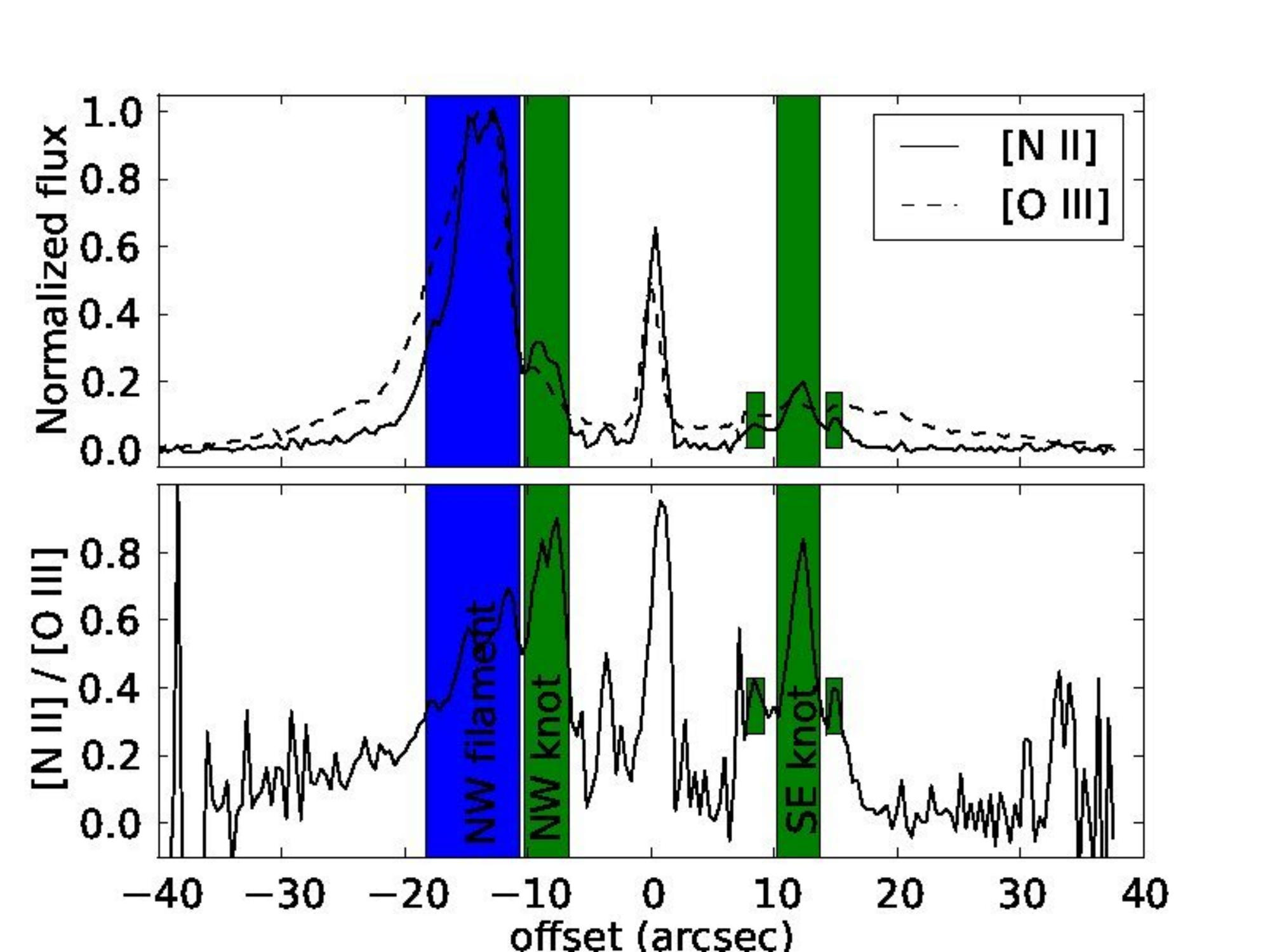}
\caption[]{Upper panel: \nitrogen\ image of K~1-2 obtained from the ESO archive. The size of 
the field shown is 62$\times$53~arcsec$^2$. The nebular components under 
analysis are indicated in the image. The extraction windows for these components are 
indicated by the corresponding boxes: the Northwest nebula (NW NEB: 35.6~arcsec, in red), 
the NW filament (7.6~arcsec, in blue) and the knots (NW and SE knots: 3.6~arcsec, in green). 
North is up, and East is to the left. Lower panels: The radial profile of 
the \nitrogen\ and \oxygeniii\ emission-line fluxes, normalized to 1.0,  and the \nitrogen/ \oxygeniii\ line ratio, along the slit.}
\end{figure}

\subsection{K~1-2}
K~1-2 is an elliptical nebula with very low surface brightness and a known binary system. 
Highly collimated structures with several knots on opposite directions of the central star and
a jet-like structure oriented at PA=153\degree\ can easily be discerned in the \nitrogen\ 
and \sulfurt, as well as the \oxygeniii\ and hydrogen recombination line images 
(Corradi et al. 1999). These authors also found several other knots almost 
perpendicular to the direction of the jet-like structure, similar to the case of Wray~17-1. 
The post-CE close binary system found at the centre of K~1-2 would be more 
easily consistent with the presence of the collimated jet-like structures and knots, as argued in 
the theoretical works from Soker and Livio (1994), a number of other later works, and the recent observational works by Tocknell et al. (2014) 
and Jones et al. (2014).

Our spectroscopic data of K~1-2 were obtained only along the PA=153\degree\ with the results presented 
in Table~8. The nebular regions under analysis in this nebula are shown in Fig.~4. We calculate $c$(\hbeta) between 0.22$\pm$0.06 and 
0.26$\pm$0.06 for the various nebular components (NW jet, NW knot and SE knot), which are close to the value  
derived by Exter el al. (2003) for both north and south jets, namely 0.25. It is worth mentioning here 
that the $c$(\hbeta) of the entire nebula has been estimated to be 0.11 (Exter et al. 2003), which suggests that the 
jet and knots are dusty features, at least much more than the surrounding nebula.

The absolute \hbeta\ fluxes for each nebular component are also given in Table~8. 
For the NW NEB component, the $F$(\hbeta) is 11.4 $\times10^{-15}$ ergs s$^{-1}$ cm$^{-2}$, which is in 
very good agreement with the value of the northern jet derived by Exter et al. (2003). 
We also confirm that $\it{T}_{\rm{e}}$\oxygeniii\ is $\sim$5000~K higher than $\it{T}_{\rm{e}}$\nitrogen, as previously found, 
but our values are 2000~K and 1000~K, respectively, lower than those in Exter et al. (2003). 
$\it{N}_{\rm{e}}$ is found to vary between 422$\pm$110~cm$^{-3}$ and 888$\pm$195~cm$^{-3}$ and, 
due to the large uncertainties, no unambiguous evidence of variation in $\it{N}_{\rm{e}}$ 
between the LISs and nebulae can be provided.

The lower panels of Fig.~4  displays the radial profiles of the \nitrogen\ and \oxygeniii\ 
emission-line normalized fluxes, as well as the \nitrogen/\oxygeniii\ line ratio, along the slit. 
The exact position of the knots can easily be discerned. Moreover, one can also see that the NW jet 
is a separate component from the NW knot. The \nitrogen/\oxygeniii\ is found to be very high at the 
position of the knot, close to 0.8, and decreases outwards. The positions of two knots (small 
green regions) with  offsets of 8 and 14~arcsec, respectively, are barely seen in the \nitrogen\ line, 
but they are further away with a sharp peak in the \nitrogen/\oxygeniii\ line ratio. These two knots 
have been noticed by Corradi et al. (1999).

Our NW NEB is the same region studied by Exter et al. (2003), and the chemical abundances are found to be in 
a good agreement, in the two works. NW NEB has high He$^{++}$/H$^{+}$ and low O$^+$/O$^{++}$ ionic abundance ratios 
indicating a high excitation nebula. Surprisingly, both knots, NW and SE, do not show He$^+$ lines. Moreover, the high 
excitation He$^{++}$ and S$^{+++}$ ions are found to be more abundant in these components. Why these 
knots are so highly excited is an open question. A possible explanation is that the contribution of the main nebula 
in these components is significant, especially in high-ionization lines (Exter et al. 2003).

\begin{figure*}
\centering
\includegraphics[scale=0.28]{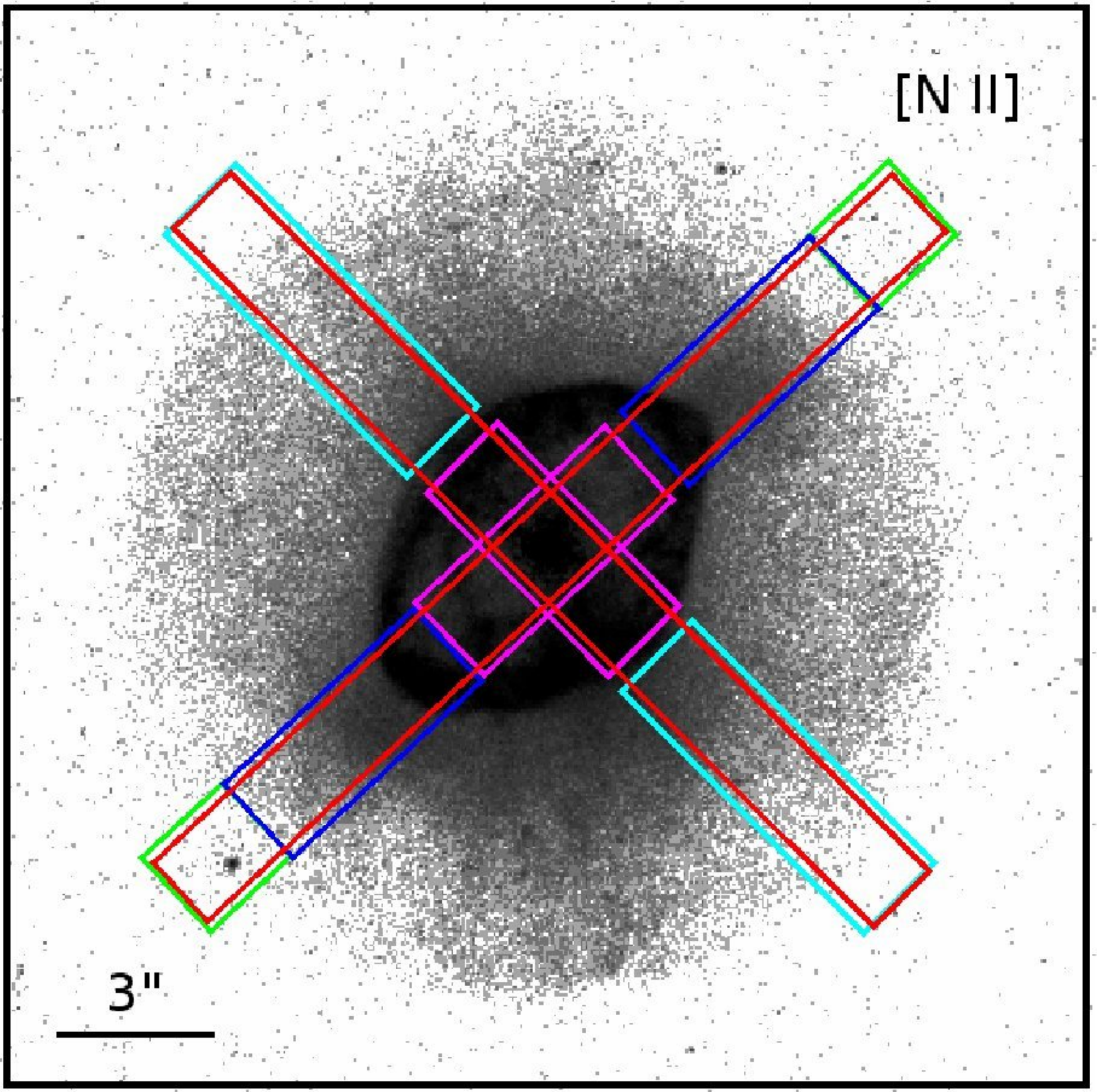}
\includegraphics[scale=0.265]{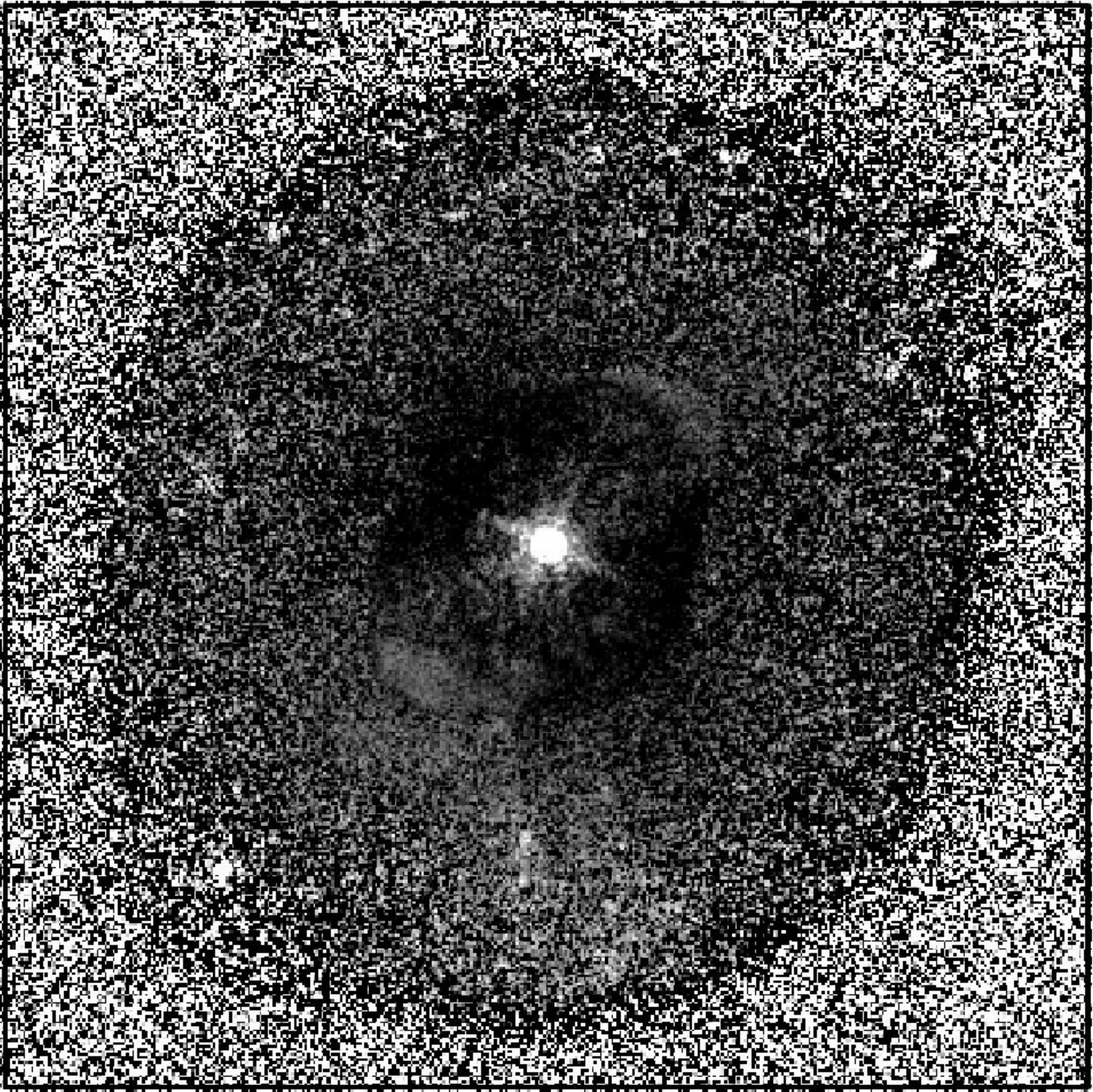}
\includegraphics[scale=0.42]{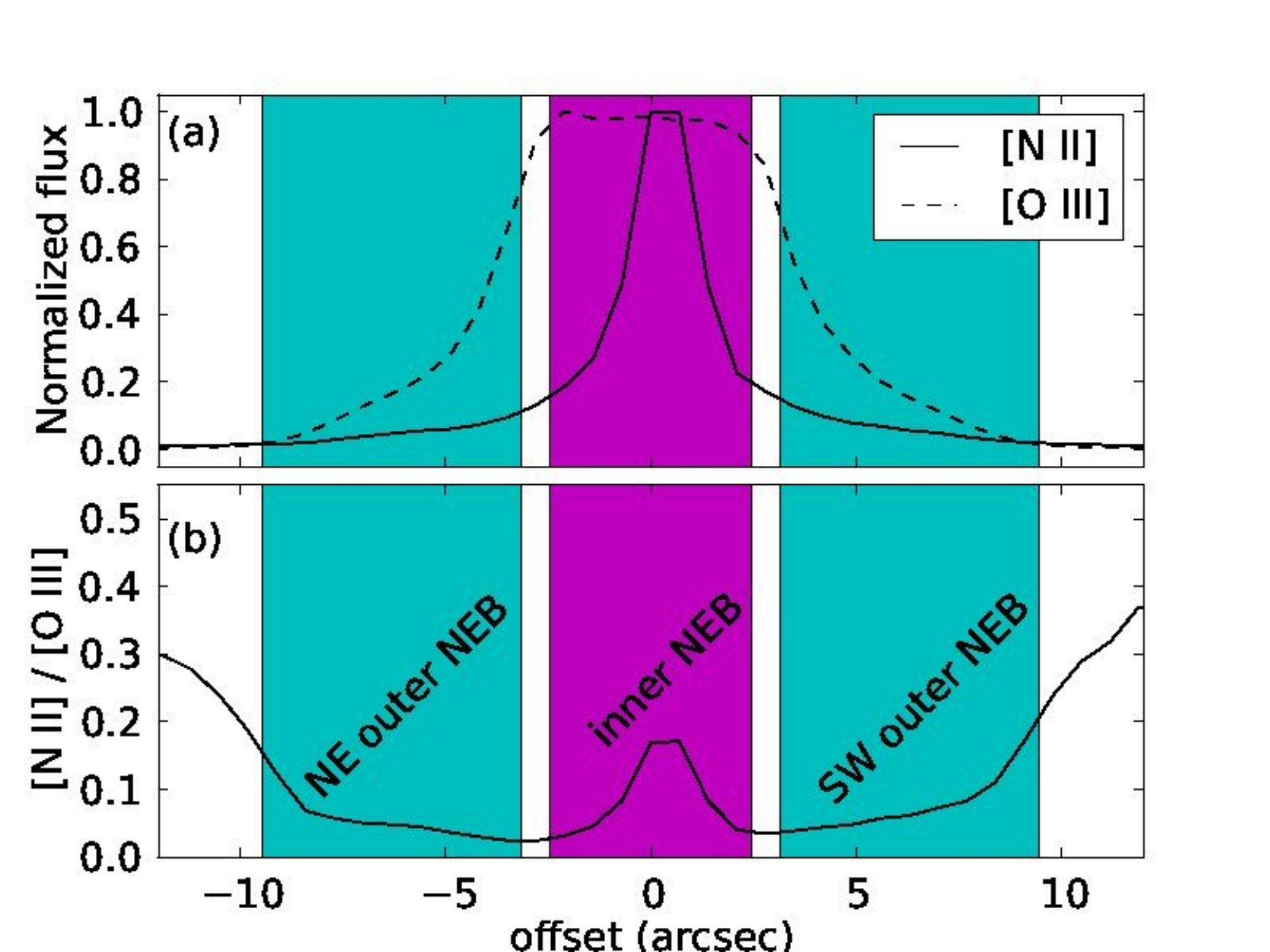}
\includegraphics[scale=0.42]{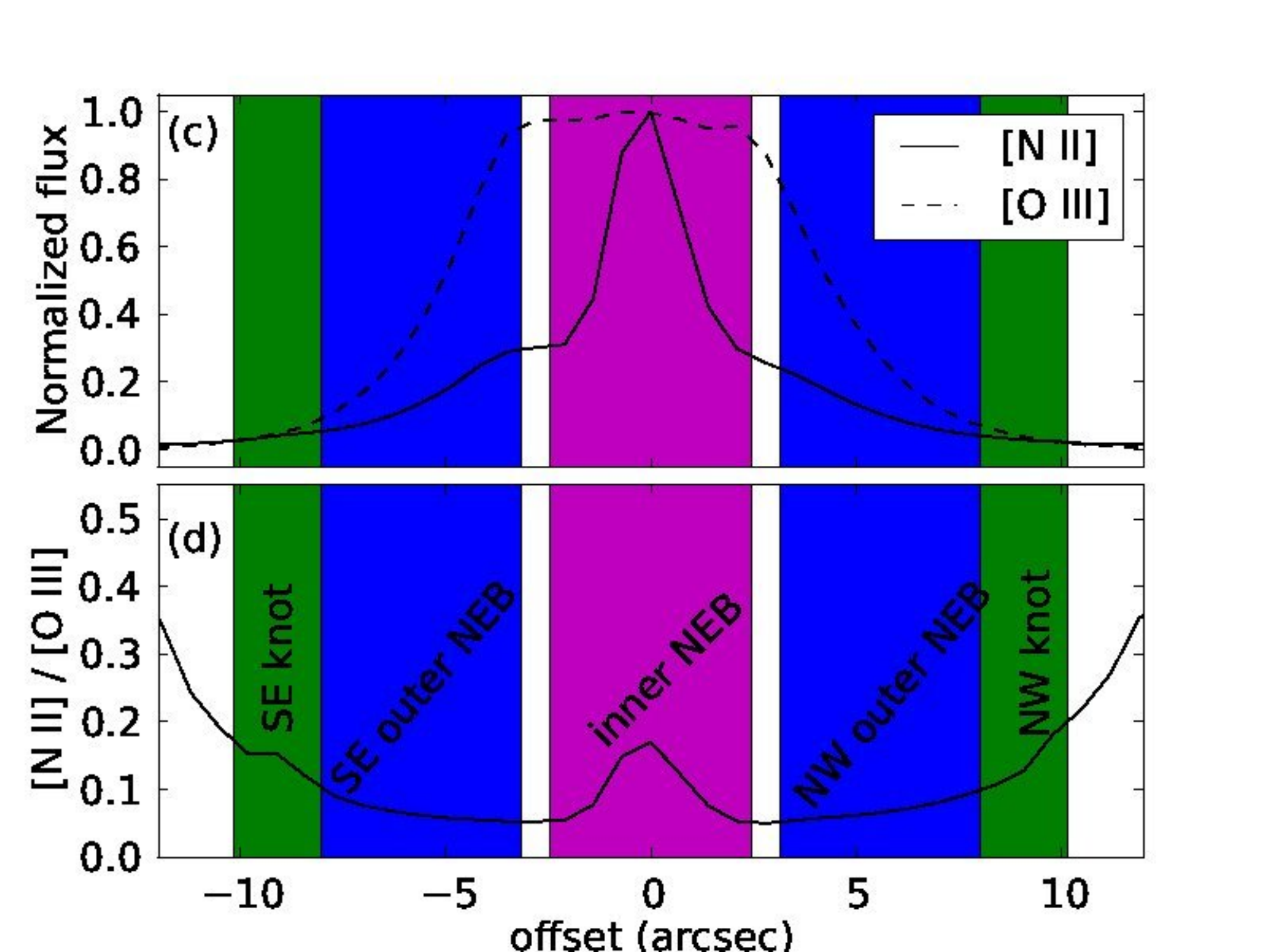}
\caption[]{Upper panels: HST \nitrogen\ image (left-hand panel) and \nitrogen/\oxygeniii\ line ratio image 
(right-hand panel) of NGC~6891. The size of the fields shown are 32$\times$32~arcsec$^2$ and 20$\times$20~arcsec$^2$, respectively. 
The nebular components under analysis are indicated. The extraction windows for these components 
are indicated by the corresponding boxes: the entire NW-SE and NE-SW nebular regions (entire NEB: 
20.3~arcsec, in red); the NW and SE outer nebular regions (4.9~arcsec, in blue); NE and SW outer nebular regions 
(6.3~arcsec, in cyan); the SE and NW knots (2.1~arcsec, in green); and the inner nebular regions (inner NEB: 4.9~arcsec, 
in magenta). North is up, and East is to the left.  Lower panels: The radial profile of
the \nitrogen\ and \oxygeniii\ emission lines fluxes, normalized to 1.0 and the \nitrogen/ \oxygeniii\ line ratio, along the slits: 
Panels (a) and (b) correspond to PA=45\degree\ whereas (c) and (d) to PA=135\degree).}
\end{figure*}

\subsection{NGC~6891}
NGC~6891 (PN G054.1-12.1) displays a bright inner prolate elliptical shell with 
the major axis oriented along PA=135\degree, and an inclination angle of 50--55\degree, surrounded by 
a diffuse spherically symmetric nebula (Guerrero et al. 2000; Palen et al. 2002). Guerrero et al. (2000) pointed out 
the presence of a (collimated) outflow between the inner and outer nebulae, as well as 
a pair of faint knots at the tip of these outflows, due to the interaction with the outer nebula (see Fig. 1 from Guerrero et al. 2000).
However, HST images of this nebula do not reveal the presence of any collimated outflow (Fig.~5). From our analysis, 
we found these regions as normal outer nebular regions (outer NEB), instead of LISs. 

The central star of NGC~6891 has been classified as an O-type star (McCarthy et al. 1990; 
Guerrero \& De Marco 2013) and as a WEL star (weak emission line, Tylenda et al. 1993). The strong P Cygni  profile 
found in its spectrum implies a terminal velocity of 1200--1400~\kms (Marcolino et al. 2007).
Neither the P Cygni nor any wind variability have been detected in the central star spectrum, by Guerrero \& De Marco (2013).
Nevertheless, the detection in its spectrum of the $\lambda 4650$ (C~III~4647/51 and 
C~IV~4658) and $\lambda 5805$ (C IV 5801--5812) line features, in conjunction with the absence of the emission line C~III~$\lambda 5696$ line, 
support the WEL-type classification. The \heliumb\ $\lambda\lambda 4541$, 5412 absorption lines, 
previously reported by Parthasarathy et al. (1998) and Marcolino \& de Ara\'ujo (2003), were also 
detected by us.

Two spectra were obtained for this nebula, with the slits positioned along PA 45\degree and 135\degree. 
The identification of all nebular components used for the analysis is given in Fig.~5. 
The logarithmic $F$(\hbeta) flux derived for the entire NEB are -11.29 and -11.31 for the PA=135\degree\ 
and PA=45\degree, respectively, in good agreement with those from Acker et al. (1991) work. $c$(\hbeta) 
is calculated between 0.03$\pm$0.02 and 0.06$\pm$0.04 (for the different nebular components). Be aware that, 
our values are substantially lower than those derived by Kaler et al. (1970) and Wesson et al. (2005), namely 0.41 and 0.29, 
respectively. The different slit positions between the aforementioned studies and ours may cause this discrepancy in $c$(\hbeta).

$\it{T}_{\rm{e}}$\oxygeniii\ of the entire nebula, for both PAs, are 9600$\pm$1630~K and 9500$\pm$1420~K, 
whereas $\it{T}_{\rm{e}}$\nitrogen\ is found to be higher by $\sim$3000~K (12720$\pm$6360 and 12190$\pm$5900~K). 
These value agree with the previously published ones: $\it{T}_{\rm{e}}$\oxygeniii=9200~K and $\it{T}_{\rm{e}}$\nitrogen=10600~K 
(Henry et al. 2004) and $\it{T}_{\rm{e}}$\oxygeniii=9330~K (Wesson et al. 2005). As for the $\it{N}_{\rm{e}}$, we derive 
$\it{N}_{\rm{e}}$\sulfurt=1290$\pm$390~cm$^{-3}$, $\it{N}_{\rm{e}}$\chloro= 2460$\pm$490~cm$^{-3}$, and 
$\it{N}_{\rm{e}}$\argon=2508$\pm$640~cm$^{-3}$, which also agree within the errors with those densities 
derived by Wesson et al. (2005). The high $\it{N}_{\rm{e}}$ value of 10000~cm$^{-3}$, reported by Henry 
et al. (2004), is not confirmed in this work, nor in any other so far available in the literature (Kaler 1970; Wesson et al. 2005).

A comparison of $\it{T}_{\rm{e}}$ and $\it{N}_{\rm{e}}$ among the inner NEB, outer NEB, and knots, do not reveal any 
important difference.  The surprisingly high $\it{N}_{\rm{e}}$\argon\ found in the inner NEB, may be indicating 
a density stratification in the inner core. 

Recent \nitrogen\ and \oxygeniii\ HST images of NGC~6891 (Project ID:8390; PI: Hajian) reveal
a prolate ellipsoidal nebula, as proposed by Guerrero et al. (2000), but neither the jet-like structures nor the other 
kinds of collimated outflows are apparent (Fig.~5, upper-left panels). Given that the \nitrogen/\oxygeniii\ line ratio 
image unveils the faint LISs within the much larger scale structures of PNe (Corradi et al. 1996), we constructed the 
same image using the recent high-resolution HST images (Fig.~5, upper-right panel), where no jet-like structures are 
discerned. This provides additional confirmation of our previous conclusion, based on our low-dispersion spectra, 
that there are no jet-like structures in NGC~6891. The bottom panels in Fig.~5 display the radial profile of 
the \nitrogen\ and \oxygeniii\ emission-line normalized fluxes (panels a and c) and the \nitrogen/\oxygeniii\ line ratio 
(panels b and d) along the slits for the two PAs of 45\degree\ and 135\degree, respectively.  

The ionic and chemical abundances of NGC~6891 are presented in Tables 11 and 13.  NGC~6891 is a non-type I nebula 
due to its low He and N/O abundances. The chemical abundances for all nebular components, along both slit positions, 
are found to be the same, within the errors. Our chemical abundances of NGC~6891 agree with those derived 
from previous work (Wesson et al. 2005).

\subsection{NGC~6572}
NGC~6572, or PN G034.6+11.8, is a widely studied PN. Narrow-band images show an elliptical 
shell with the major axis oriented along PA $\sim$ 0\degree\ and a seemingly toroidal structure 
(Miranda et al. 1999). The \nitrogen/\oxygeniii\ image ratio highlights the presence of two pairs of 
knots at PA=15\degree\ and PA=162\degree\ (see Fig. 1 from Miranda et al. 1999). Both pairs are located at 
the tips of two much fainter collimated outflows, thus suggesting shock interactions.
HST images of NGC 6572 do not confirm the presence of any knotty structure. Contrariwise, 
the HST H$\alpha$/\oxygeniii\ line ratio image reveals a filamentary structure around the collimated 
outflows indicating strong shock interactions (Guerrero et al. 2013), probably responsible for the narrow zones (shock fronts) 
and the enhancement of low-ionization lines.

\begin{figure}
\centering
\includegraphics[scale=0.305]{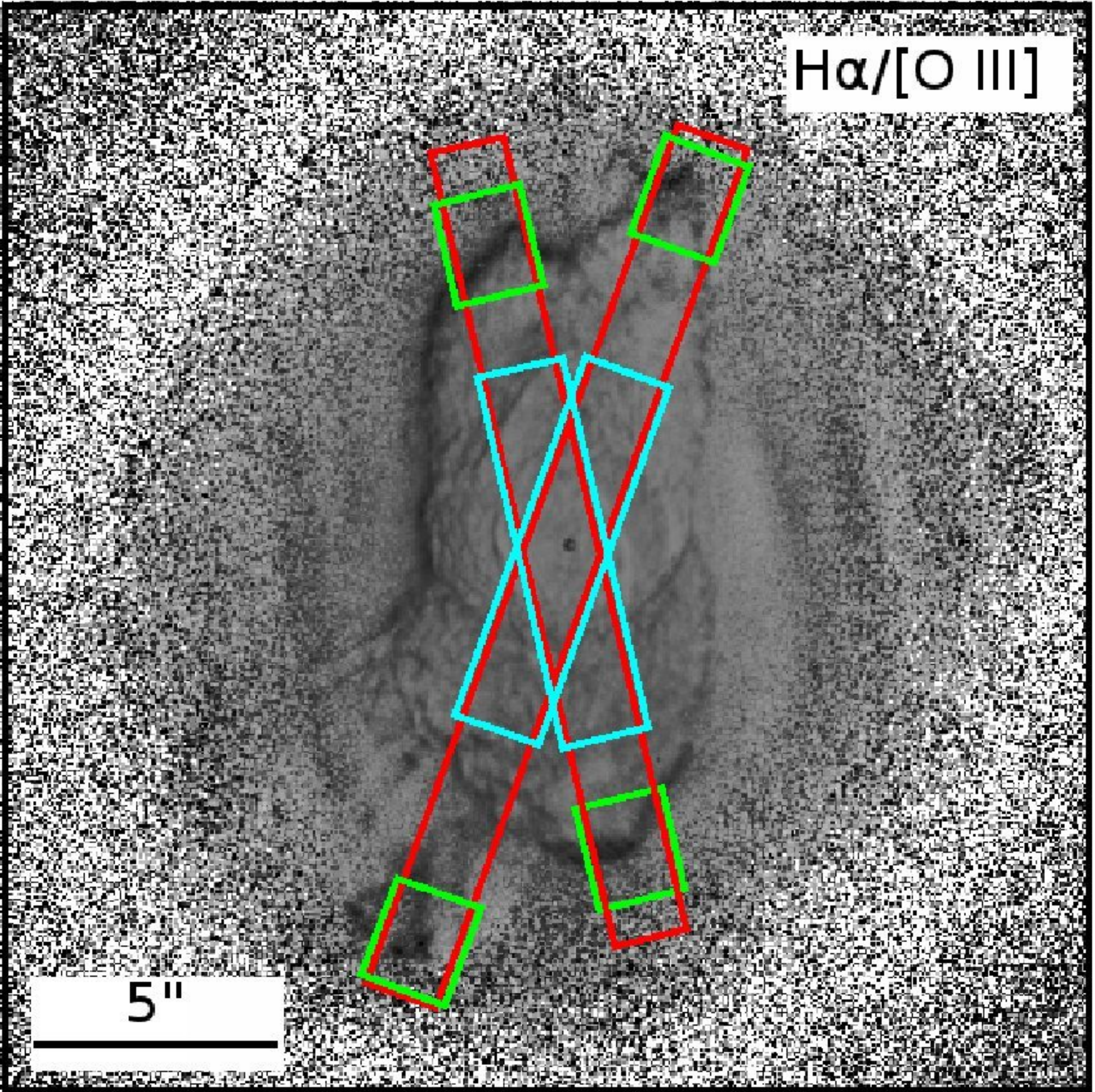}
\includegraphics[scale=0.41]{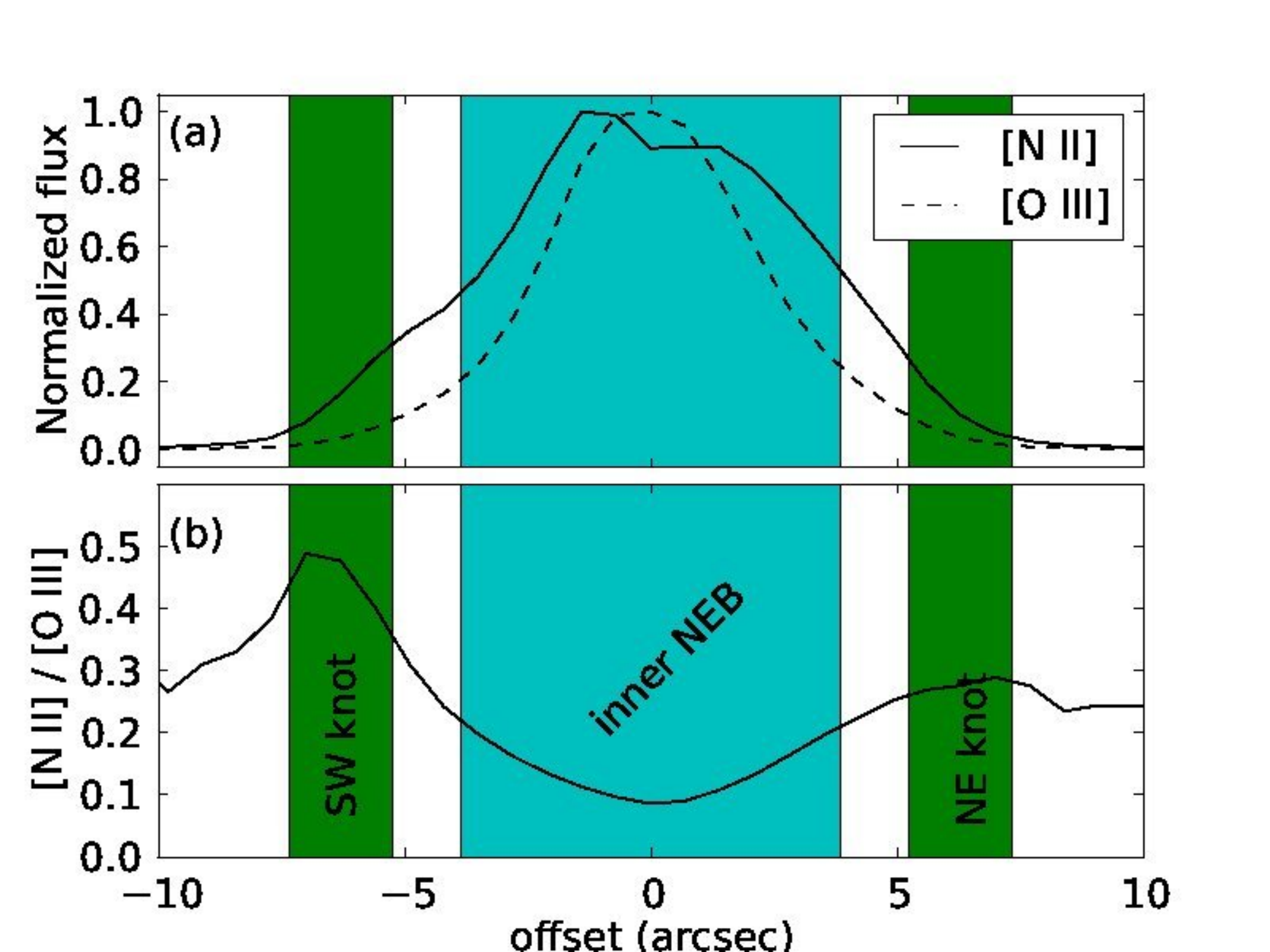}
\includegraphics[scale=0.41]{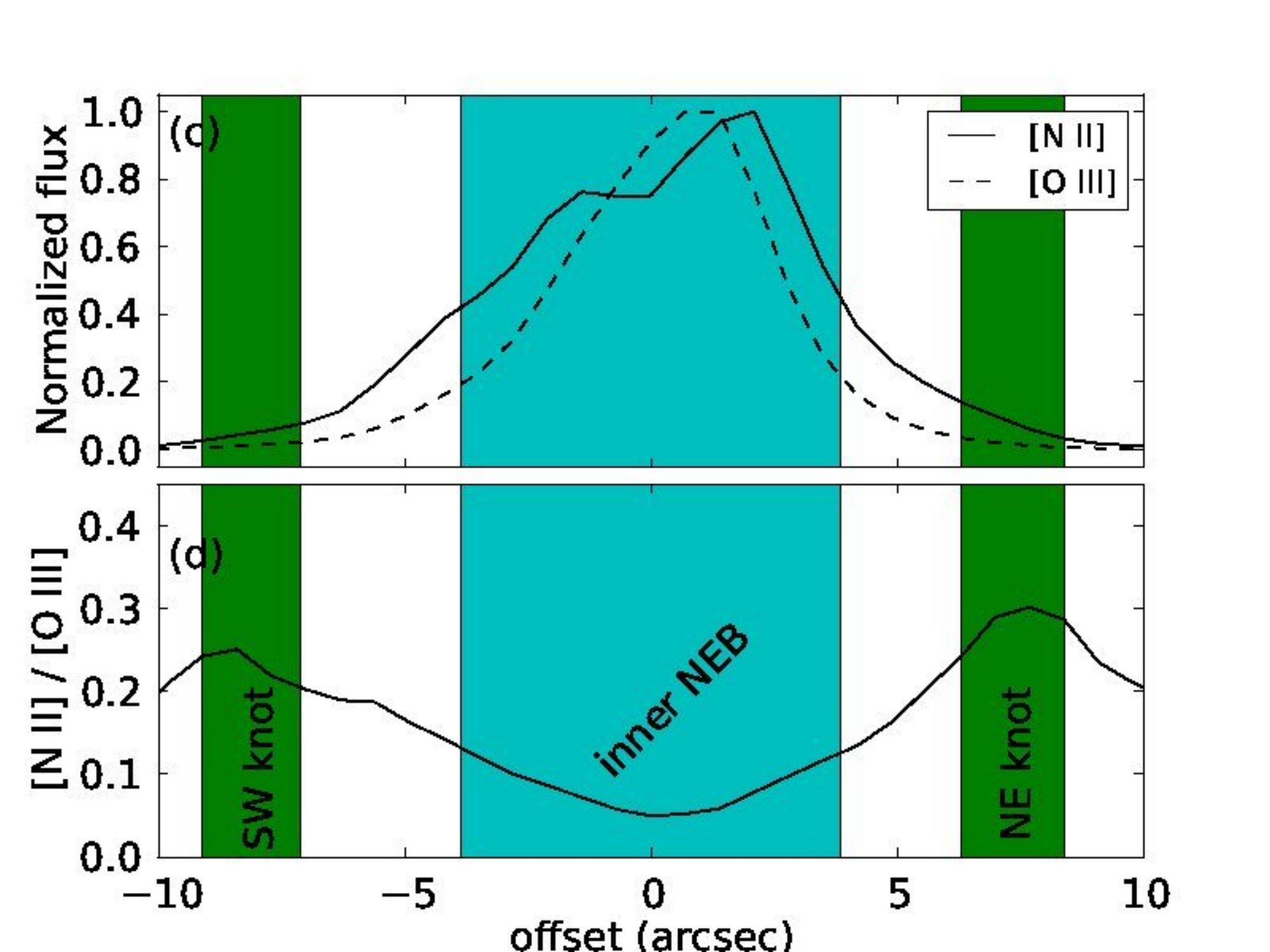}
\caption[]{Upper panel: HST \ha/\oxygeniii\ image of NGC~6572. The size of the field shown is 
22$\times$22~arcsec$^2$. The nebular components under analysis are indicated in the image. 
The extraction windows for these components are indicated by the corresponding boxes: the entire nebular regions 
(entire NEB: 17.5~arcsec, in red); the inner nebular regions (inner NEB: 7.7~arcsec, in cyan); 
and the four, NE, NW, SE and SW knots (2.1~arcsec, in green). North is up, and East is to the left. 
Lower panels: The radial profile of the \nitrogen\ and \oxygeniii\ emission-line fluxes, normalized to 1.0, and 
the \nitrogen/ \oxygeniii\ line ratio, along the slits. Panels (a) and (b) correspond to PA=11\degree, 
whereas (c) and (d) to PA=160\degree).} 
\end{figure}

The spectral type of NGC~6572 central star is still debatable. First, it was classified as a 
Of-WR type (Mendez et al. 1990), then as a WEL star (Tylenda et al. 1993; Parthasarathy 
et al. 1998), and more recently as a [WC]-PG 1159 star, by Marcolino et al. (2007). 
Narrow N~III $\lambda\lambda 4634-41$ and C~IV $\lambda 5808$ emission lines (with FWHM of 15$\pm$2 and 20$\pm$3~\AA, 
respectively) are detected in our spectra, which is in agreement with the previous WEL-type or [WR] classification. 

In order to study the physico-chemical properties of the nebula and the pairs of knots, 
two low-dispersion spectra were obtained along the PA=15\degree\ and 162\degree. The emission 
line fluxes as well as $c$(\hbeta), $\it{N}_{\rm{e}}$ and $\it{T}_{\rm{e}}$ are presented in the Tables 14 and 16. 
All the nebular structures used here are defined in Fig.~6. The \nitrogen/\oxygeniii\ 
ratio profile of NGC~6572, along the two slits, are shown in Fig.~6 (panels c and d).    
$c$(\hbeta) varies slightly among the nebular component from 0.24$\pm$0.05 to 0.29$\pm$0.06. These values are 
lower than those published by Liu et al. (2004; 0.48) and Frew et al. (2013; 0.41), but very close to the 
value derived by Kwitter et al. (2001; 0.32). Regarding the $\it{T}_{\rm{e}}$, no variations were found between the 
inner nebula and the pairs of knots (see Table 14 and 16). Unlike $\it{T}_{\rm{e}}$, $\it{N}_{\rm{e}}$ shows significant 
variations among the components. In particular, the knots are found to have lower density than the inner nebular regions, 
by a factor between 2 and 4. Moreover, the high densities derived from the Ar and Cl diagnostic line ratios imply a 
strong density stratification in its central region. $\it{N}_{\rm{e}}$\sulfurt\ is found to be lower than those previously 
published by Wang et al. (2004) and Kwitter et al. (2001), whilst $\it{N}_{\rm{e}}$\argon\ and $\it{N}_{\rm{e}}$\chloro\ are 
higher. The very high $\it{N}_{\rm{e}}$ of this nebula, close to the upper limit, indicate that any small change in the line 
ratio will results in substantially differences $\it{N}_{\rm{e}}$ values. This can explain the differences found in $\it{N}_{\rm{e}}$ 
among the aforementioned works and ours.

Tables 15 and 17 list the ionic and total abundances for all the nebular components of NGC~6572 and for both slit positions. 
We do not find any significant enrichment of N or S, with respect to the rest of the nebula that could be associated with 
the brighter \nitrogen\ and \sulfurt\ lines found in the knots. Chemical abundances for a given element are 
the same for all the components. Comparing our chemical abundances with those derived by Perinotto et al. (2004) and Henry et al. (2004), 
we turn out that He and N abundances agree with those from Perinotto et al. (2004), whilst O, Ne, and S abundances better agree with those from Henry et al. (2004). 
There is a substantial difference between these two studies. Moreover, the S knot (PA=160\degree) and the N knot 
(PA=15\degree) show enriched Ne abundance compared to the other two knots. Regarding our analysis, 
NGC~6572 is a non-type I PN, in contrast to the previous classification by Henry et al. (2004).

\section{Results: morpho-kinematic properties}

Here we discuss the main morphological and kinematic characteristics of each nebula, and their different 
components, based on previously published works and echelle data taken from the SPM KCGPN. The wide range of 
expansion velocities of LISs, from a few tens to hundreds \kms (Gon\c calves et al. 2001), may suggest 
that shocks play a major role in the formation and excitation of these structures. Therefore, a parallel study of 
the morpho-kinematic and the physico-chemical properties (the emission line ratios, ionic and total chemical 
structure) is necessary to enlighten our knowledge of these structures.
The de-projected expansion velocities  of the nebular components, for each PN, are given in Table~1.

\begin{table*}
\begin{center}
\caption[]{De-projected expansion velocities of LISs and large-scales components. Errors  
are between 10\% and 20\% due to the uncertainties on the inclination angle.}
\label{table5}
\begin{tabular}{lccccccc}
\hline
PN name    & PA    & Jet/Jet-like  & Jet/Jet-like & Knot            & Knot        & Inner NEB & Outer NEB  \\                           
           &       & or Filament       & or Filament       &               &             &           &            \\
           &(\degree)&  (\kms)      &  (\kms)        & (\kms)        & (\kms)      & (\kms)    & (\kms)     \\ 
\hline
IC~4846    & 54    & $\geq$100 (NE)  &$\geq$100 (SW)   & $-$            & $-$           & 20--25    & $-$\\    
Wray~17-1  & 75    &    $-$          & $-$             & 40--50 (W)     & 40--50 (E)    & 40        & 40$^a$\\       
Wray~17-1  & 155   &  70--80 (NW)    & 60--70 (SE)     & $\geq$100 (NW) & $\geq$40 (SE) & 40        & $-$ \\       
K~1-2      & 153   &  30--35         & $-$             & 30--35 (NW)    & 30-35 (SE)       & $-$       & $-$\\       
NGC~6891   & 45    &    $-$          & $-$             & $-$            & $-$           & 10        & 28 \\       
NGC~6891   & 135   &    $-$          & $-$             & 80 (NW)        & 80 (SE)       & 17        & 45 \\       
NGC~6572   & 15    &    $-$          & $-$             & 46 (NE)        & 46 (SW)       & 18        & $-$\\      
NGC~6572   & 169   &    $-$          & $-$             & 65 (NW)        & 55--65 (SE)   & 18        & $-$\\       
\hline
\end{tabular}
\medskip{}
\begin{flushleft}
$^a$ This expansion velocity refers to the E and W arcs.
\end{flushleft}
\end{center}
\end{table*}

\subsection{IC~4846}
A morpho-kinematic analysis of IC~4846, performed by Miranda et al. (2001), reveals 
the presence of three pairs of elongated structures labelled as A1--A2 along  PA=54\degree, and B1--B2 / C1--C2 
along PA=11\degree. The authors claim that the components A1--A2 correspond to a high collimated precessing 
jet. The high-resolution HST \ha\ image of IC~4846, presented here for the first time, provides further 
evidence that supports this hypothesis (see Fig.~2).

The projected expansion velocities of NE and SW jets range from 36 to 52~\kms, with an average 
value of 44~\kms\ (Miranda et al. 2001). Considering that the inclination angle of the precession axis, 
with respect to the plane of sky, is 13-26\degree, the de-projected $V_{exp}$ of the NE and SW jets are estimated to be 
between 100 and 190~\kms. The inner NEB of IC~4846 (or the C1--C2 features in Miranda et al. 2001) expands with a velocity of 
$\sim$20--25~\kms.

\subsection{Wray~17-1}
The kinematic study of this peculiar nebula has been performed by Corradi et al. (1999). The NW and SE filaments 
(labelled as A and B in Corradi et al. 1999) show a linear increase of expansion velocities up to $\sim$30~\kms\ in the \nitrogen\ emission line. 
Incidentally, the NW and SE knots at the outer parts of the nebula (labelled as A$^\prime$ and B$^\prime$ respectively in Corradi et al. 1999), 
show peculiar expansion velocities compared to the filaments. In particular, the NW and SE knots are found to be both 
red-shifted, with $V_{\rm{exp}}$  of $\sim$60\kms\ and $\sim$10\kms, respectively, whilst the NW and SE filaments are 
blue and red-shifted, respectively. On the other hand, the E and W  knots (respectively C and D, in Corradi et al. 1999) 
show expansion velocities of ~15\kms\ with the E knot being blue-shifted and the W knot red-shifted. The arc structures that 
surround these knots show similar expansion velocities.  

High-dispersion \oxygeniii\ PV diagrams along PA=60\degree\ and 135\degree, were acquired from the SPM KCGPN to better constrain the 
morpho-kinematic structure of Wray~17-1 (Fig.~7). These PV diagrams clearly show the presence of an outer shell bright in 
the \oxygeniii\ emission, but absent in the \nitrogen\ one. This structure expands with a velocity of 60~\kms, with the NW part 
red-shifted and the SE blue-shifted. The LISs seem to interact with the outer shell and indicate shock excitation.

\begin{figure}
\centering
\includegraphics[scale=0.24]{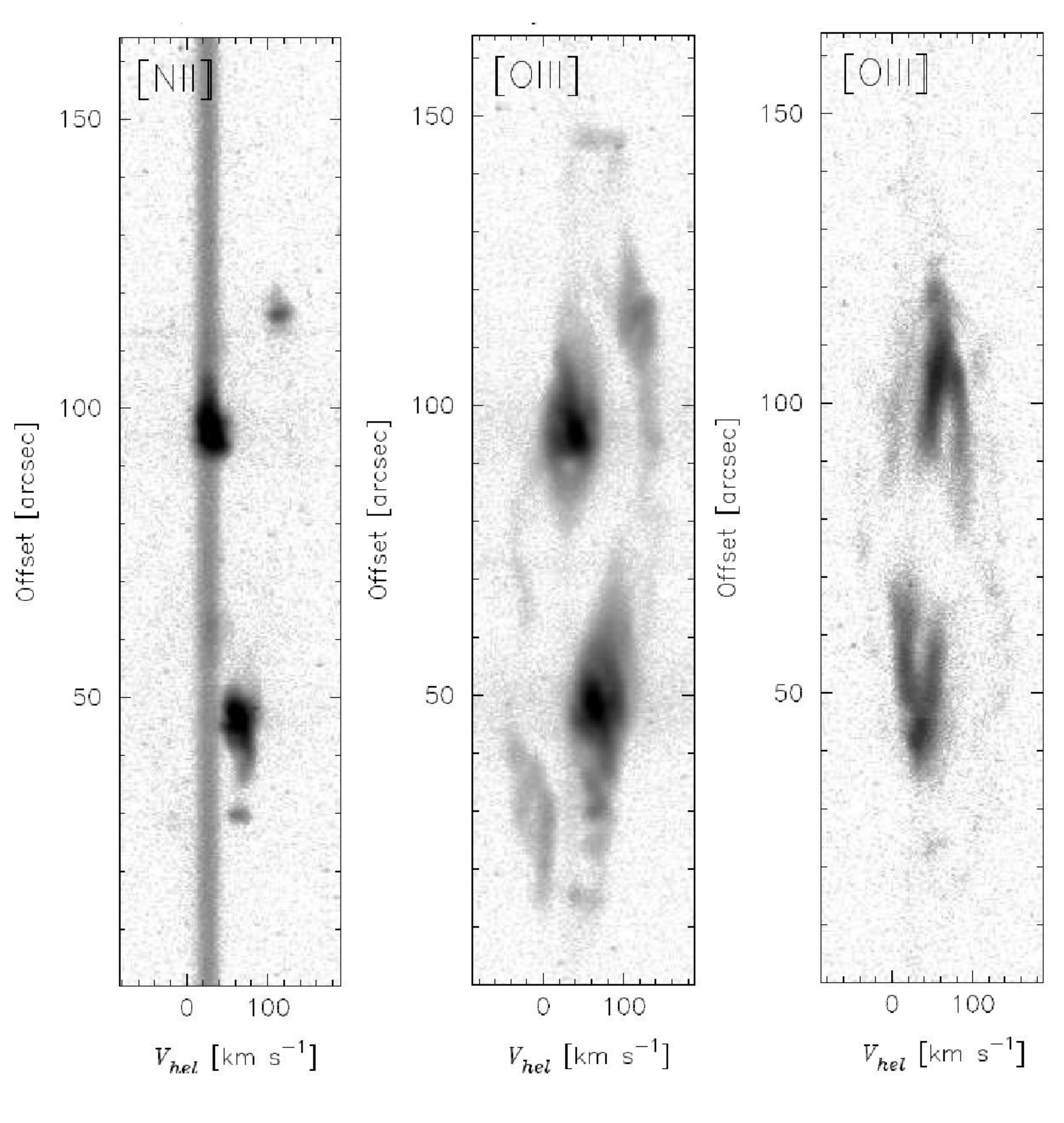}
\caption[]{Echelle spectra of Wray~17-1 drawn from KCGPN. Left panel: 
\nitrogen, along PA=155\degree. Middle panel: \oxygeniii, along PA=155\degree. Right panel: \oxygeniii, along PA=75\degree.}
\end{figure}

If the central star of Wray~17-1 ejected stellar material along a geometrical cone, this could give a good explanation for the peculiar 
velocity field of the filaments and knots. In particular, the filaments could be the result of the interaction between the cone 
and the outer \oxygeniii\ shell, whilst the NW knot, located exactly behind, and being well aligned with the filaments. 
Under this hypothesis, we can explain the blue-shifted NW filament and red-shifted NW knots. In the case of the SE filament and knot, 
this scenario implies that the two components are related. We calculate the inclination of this cone to 20--25\degree\ with 
respect to the line of sight. This implies that the de-projected velocities of the filaments and the knots are at least 2.5--3 times higher. 
On other hand, the E and W knots are found to share the same expansion velocities of the surrounding nebula, with a de-projected 
velocity $\sim$40--50~\kms. The inner regions, NEB (NE, SW, NW and SE) appear to expand with a velocity of $\sim$40~\kms.

\subsection{K~1-2}
K 1-2 shows the presence of a collimated filament and several knots oriented along PA=-27\degree\ (Corradi et al. 1999).
These authors pointed out that all component appear to move with a velocity similar to that of the main nebula, 25~\kms,
under the assumption that the NW filament (labelled A in Corradi et al. 1999) is located within the main 
spherically symmetric nebula. This assumption implies that the filaments and knots are oriented at a minimum 
inclination angle of 40\degree\ with respect to the line of sight. Given that the inclination angle of the spectroscopic binary 
has been determined 50\degree\ (Exter et al. 2003), the de-projected expansion velocity of the knots and jet-like structure 
would be $\geq$30--35~\kms.

\subsection{NGC~6891}
Guerrero et al. (2000) studied the kinematic properties of this nebula and reached the conclusion that a highly-collimated 
outflow (jet) is present along  PA=135\degree. The maximum de-projected velocity is 80~\kms, at the outer edges 
(A--A$^\prime$ component or the pair of knots), considering an inclination angle with respect to the plane of sky of 10\degree. 
Regarding the inner and outer NEB, the authors adopted an ellipsoidal shell model with a homologous expansion law 
to reproduce the observed data. They concluded that the inner NEB expands with a velocity of 10~\kms\ along the 
minor (PA=45\degree) and 17~\kms\ along the major axis (PA=135\degree), while the outer NEB expands faster 
with $V_{\rm{exp}}$=28$\pm$5~\kms\ along the minor (PA~70\degree) and 45$\pm$5\kms\ along the major axis (PA$\sim$160\degree).

\subsection{NGC~6572}
Following Miranda et al. (1999), NGC~6572  shows the presence of two pairs of low-ionization knots oriented along PA=15\degree\ 
and PA=-18\degree. Moreover, two, much fainter, ellipsoidal structures can also be discerned in these directions. Fig.~6 shows, in fact, 
that the two pairs of knots are not really low-ionization nebular condensations, but resemble those of knots due to the low angular 
resolution of the ground-based images. We think that these knots actually result from the interaction between the outflows and the outer shell 
(AGB envelope), consistent with theoretical shock models (Dopita 1997; Raga et al. 2008).  

The projected expansion velocity is 39~\kms\ for the NW and the SE knots, and 8~\kms\ for the NE and SW knots. 
It is worth mentioning here that the SE knot shows a peculiar decreasing velocity, from 38 to 33~\kms\ (Miranda et al. 1999). 
The NGC~6572 inner NEB shows a projected velocity of 14~\kms. Moreover, two clearly separated maxima are discerned in the PV diagrams, 
indicating the presence of a $\sim$14~\kms\ toroidal structure at the centre of the nebula. This toroidal component has 
an inclination angle of 38\degree\ with respect to the plane of sky (Hora et al. 1990; Miranda et al. 1999). 
In accordance with this inclination angle, the de-projected expansion velocities are calculated to 18~\kms\ for the torus and the 
inner NEB, $\sim$65~\kms\ for the NW knot, and $\sim$55--65~\kms\ for the SE knot. The ellipsoidal structure along PA=15\degree\ is 
less tilted, by $\sim$10~\degree, and we obtain a de-projected velocity of $\sim$46~\kms\ for both the NE and the SW knots.  

\section{Discussion} 

\subsection{Electron temperatures}

$\it{T}_{\rm{e}}$ was estimated for several individual regions e.g., knots, filaments, jets and jet-structures, inner and 
outer nebular regions and arcs, as well as the entire nebulae. We mainly used the \oxygeniii, \nitrogen\ and \sulfurt\ 
diagnostic lines. $\it{T}_{\rm{e}}$ is found to range between 9700~K and 14700~K, which is typical of photo-ionized nebula, 
with an average of $\it{T}_{\rm{e}}$\oxygenii\ of 11500~K and $\it{T}_{\rm{e}}$\nitrogen=11340~K. Moreover, among the various 
nebular components, no significant variation in $\it{T}_{\rm{e}}$ is found, which agrees with previous studies 
(e.g. Gon\c calves et al. 2009; and references therein). 

\begin{figure*}
\centering
\includegraphics[scale=0.145]{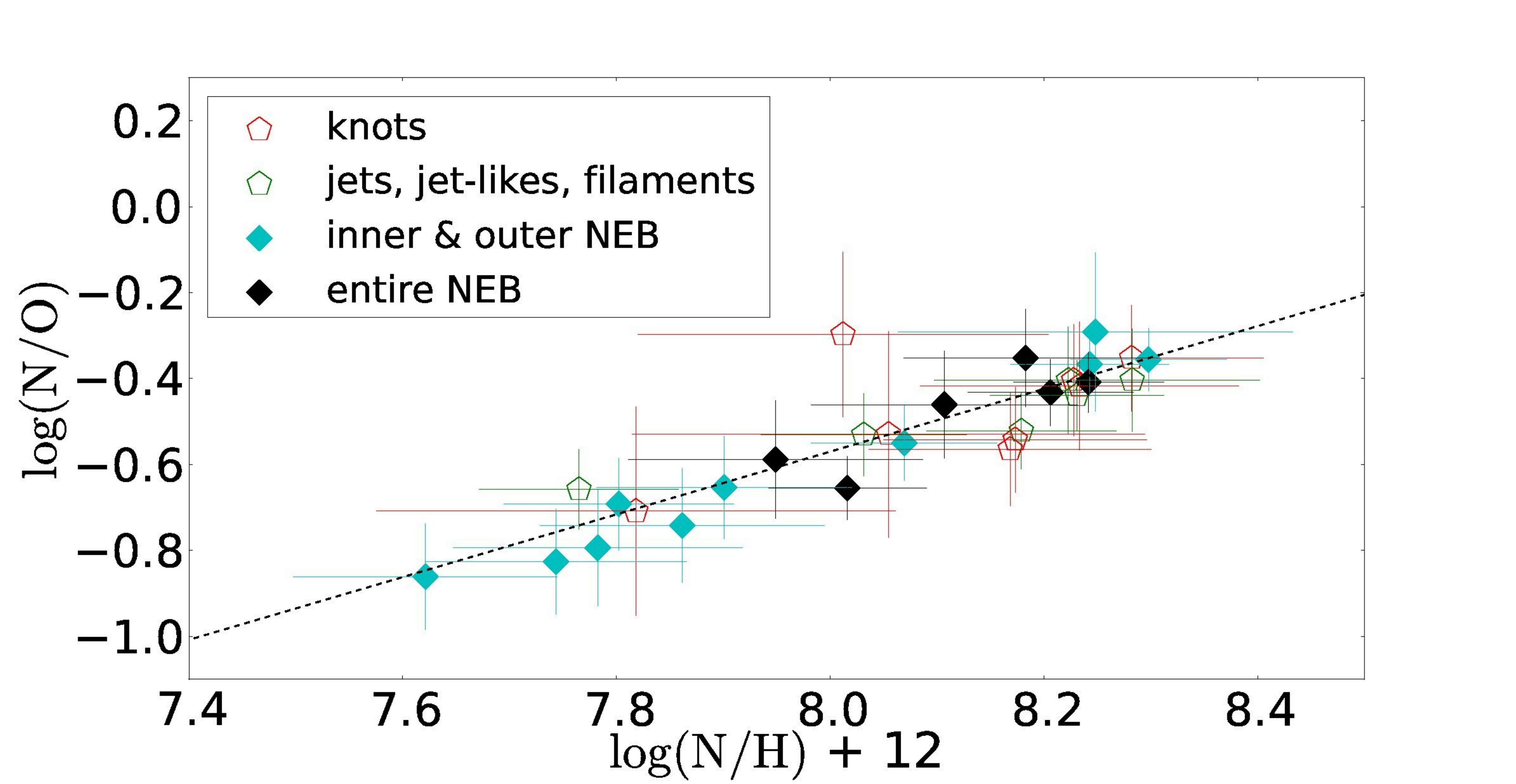}
\includegraphics[scale=0.145]{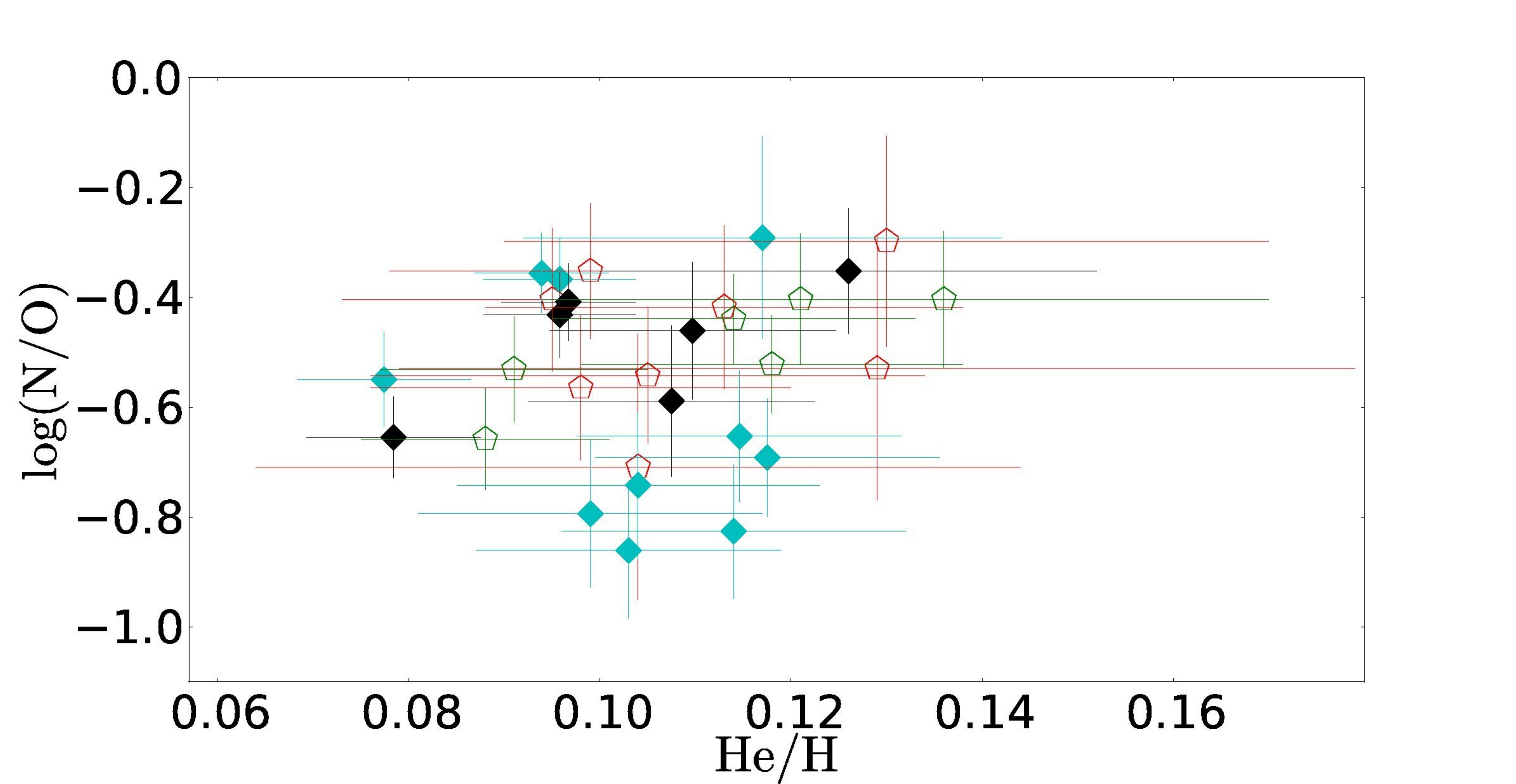}
\caption[]{Abundance patterns: $\rm{log}$(N/O) vs. He/H (left-hand panel) and $\rm{log}$(N/O) vs. $\rm{log}$(N/H)+12 
(right-hand panel). Red pentagons correspond to LISs, green pentagons to jets, filaments and jet-like structures. 
Nebular regions (inner and outer NEBs) are cyan filled diamonds whilst the entire NEBs are black filled diamonds. 
The dashed line represents the best linear fit to our data (see text).}
\end{figure*}

\subsection{Electron densities}

Regarding the electron densities, previous studies on the main nebular components (rims and shells) versus LISs have shown that the latter 
have systematically lower densities than the former, by a factor between 1 and 10 (see Gon\c calves et al. 2009; and references therein). 
This result comes into conflict with the theoretical predictions, on which the LISs are denser than the surrounding medium.

The most likely explanation for this discrepancy between models and observations may be that the models are referred to the total density of gas 
(dust, atomic and molecular as well) while $\it{N}_{\rm{e}}$ corresponds only to the ionized fraction of the gas (Gon\c calves et al. 2009). 
In this work, we confirm that LISs have the same or even lower densities than their surrounding medium. Indeed, in a paper in prep. 
we report our first successful detection of the molecular hydrogen emission, at 2.12~$\mu$m, associated to the LISs of two PNe (K~4-47 
and NGC~7662). Although, this detection was tried before by the group, the fact that LISs have in general smaller scales than the main 
bodies did not help. It seems that  only 8-m class telescopes can successfully detect the H$_2$ associated with LIS in Galactic PNe. 
See Manchado et al. (2015) for a similar work, though not associated with LISs directly, but with the molecular torus of NGC~2346. 
Moreover, it turns out that the molecular torus of this PN is composed of knots and filaments. On the other hand, in the Helix nebula 
(Meaburn et al. 1992; O’Dell \& Handron 1996; Huggins et al. 2002), which posses the best-known LISs of PNe, the association between 
LISs and the molecular gas is known for years.

Images of PNe in molecular hydrogen (H$_2$) and optical emission lines (e.g. \nitrogen, \sulfurt\ and \oxygeni) display similar morphologies 
(Guerrero et al. 2000; Bohigas 2001). In accordance with the theoretical work by Aleman \& Gruenwald (2011), the peak intensities of these 
optical lines and the H$_2$ v=1-0 co-rotational lines occurs in a narrow transition zone between the ionized and the neutral 
(photo-dissociation) regions. It is known that molecular hydrogen in PNe can be excited either by shocks or by the absorption 
of UV-photons (fluorescence) emitted by the central star. Therefore, the detection of strong \nitrogen\ and \oxygeni\ emission lines 
in LISs may be the result of shock excitation, due to the moderate expansion velocities (see \S~4).

Shock models can successfully reproduce the strong low-ionization lines from LISs (K 4-47; Gon\c calves et al. 2004, Raga et al. 2008). 
In particular, Raga et al. (2008) reached the conclusion that the lower the photo-ionization rate, the higher the $\it{N}_{\rm{e}}$ diagnostic line ratios, which then resemble the typical 
values of shock-excited regions. Besides the expansion velocities of LISs, the evolutionary age and photo-ionization rate of 
the central source are also key parameters for probing the excitation and formation mechanisms of LISs, thus  unveiling their true nature.

\subsection{Ionic and total abundances}

The chemical abundances were calculated for the several nebular components used here for the analysis of the 5 PNe. Comparison with 
those abundances derived in previous studies shows a reasonable agreement with previous results, not only for the entire nebulae but also for each component individually.

For a given PN, and element, all the components exhibit the same chemical abundances, within the uncertainties. 
This implies that the strong low--ionization lines, such as \nitrogen, \sulfurt\ or \oxygeni\ of LISs, may not be the 
result of an overabundance of these elements in the nebula. Fig.~8 displays $\rm{log}$(N/O) vs. $\rm{log}$(N/H)+12 
(left-hand panel) and $\rm{log}$(N/O) vs. He/H (right-hand panel) for all the PNe in our sample and their components. 
A linear relation between $\rm{log}$(N/O) and $\rm{log}$(N/H)+12 is found, which is consistent with the general picture 
for PNe (Perinotto et al. 2004). The best linear fit to our data yields the relation, 
$\rm{log}$(N/O)=0.74*($\rm{log}$(N/H)+12)-6.5, with R$^2$ (the goodness--of--fit) equals to 0.88, and it is in very good 
agreement with the relation derived from a sample of PNe with WR and WEL-type central stars (Garc\' ia--Rojas et al. 2013). 
Only E knot in Wray~17-1 appears as an outlier (easily identifiable in the figure), so not considered in the fitting process. 
The $\rm{log}$(N/O) vs. He/H plot does not show any correlation between the LISs and the other nebular components, 
whereas both show He/H$<$0.14 and $\rm{log}$(N/O)$<$-0.3. These figures, following Kingsburgh \& Barlow (1994) criteria, 
implies that all the PNe in our sample are non Type-I nebulae.
 
In Fig.~9 we plot $\rm{log}$(X/O) vs. $\rm{log}$(X/H)+12 for X=Ne, S and Ar, and clear linear trends are found.
We have to mention that the apparent distinct regions in these diagrams do not really reflect chemical abundance 
differences between LISs and other nebular components (NEBs), but potential differences between the PNe in our sample and their central stars. 
For instance the LISs with Ne/O$>$-0.55 and $\rm{log}$(Ne/H)+12$>$8.1 (as well as Ar/O$>$-2.5 and $\rm{log}$(Ar)+12$>$6.1) 
correspond to the LISs in K 1-2 and Wray 17-1, for which we were able to estimate their Ne and Ar abundances, whereas 
the NEBs are those from IC~4846, NGC~6572 and NGC~6891. Thus, K 1-2 and Wray 17-1 are likely richer, in Ne 
and Ar, than the rest of the PNe in our sample. Surprisingly, LISs in the former PNe are highly--ionized, with He being mainly 
doubly ionized (see Tables 5, 7 and 9). As for S, all the components (LISs and NEBs) are well mixed and cover the same range of values. 
Trends we found in Figure~9 are those usually seen in PNe between the $\alpha$-element to oxygen ratios.

In conclusion, the similarities in the chemical abundances among the components in each PN strongly suggest that the intense 
low-ionization emission lines are the result of their excitation mechanisms rather than the chemical overabundance of any element in the LISs.

Moreover, though a deep discussion about the abundance discrepancy factor ({\it adf}) is out of the scope of the present paper, 
it is relevant to note that {\it adf}s, close binaries as PN central stars (so, post CE stars) and the presence of LIS could somehow 
be related. As briefly discussed in the Introduction, the formation LIS, and particularly that of jets older than the main shells, 
is expected in the case of post common-envelope PN central stars (Gon\c calves et al. 2001; Miszalski et al. 2011).  Very recently, 
it was shown that the PNe having the highest {\it adf}s (10 to 300) all have post CE central stars (Corradi et al. 2015). 
In addition of confirming the mixing of the bi-chemistry in these PNe (hot normal component plus hydrogen-deficient cold 
inclusions; Liu et al. 2004), Corradi et al. (2015) point out the present difficult in depicting an evolutionary scenario 
consistent with all the observed properties of the highest {\it adf} PNe. Some of the scenarios these authors discuss could 
indeed be related with the presence of LISs via, for instance, the photo-ionization of the neutral material that was  in the 
orbital plane during the CE phase.

\subsection{Excitation mechanism}

\begin{figure}
\centering
\includegraphics[scale=0.145]{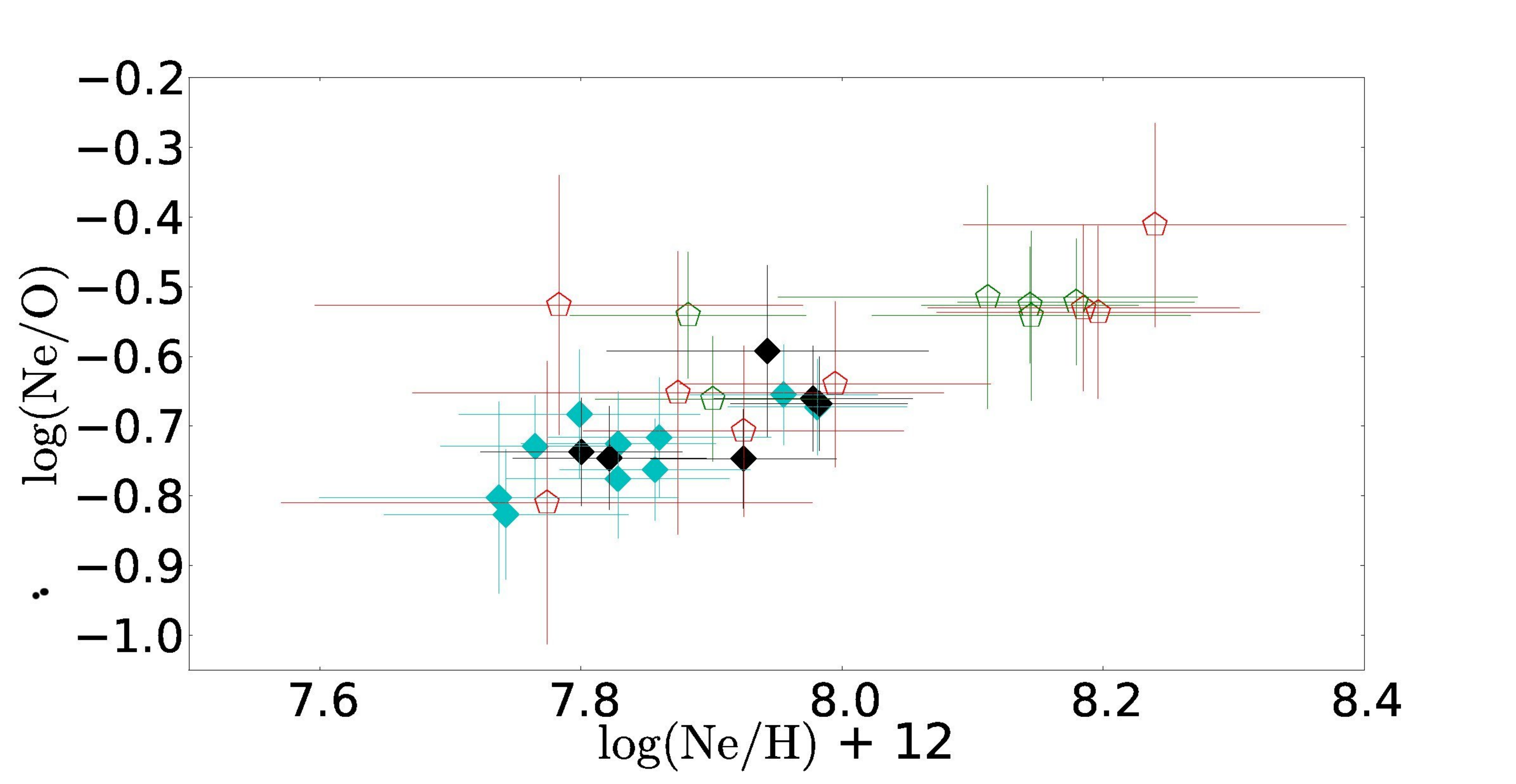}
\includegraphics[scale=0.145]{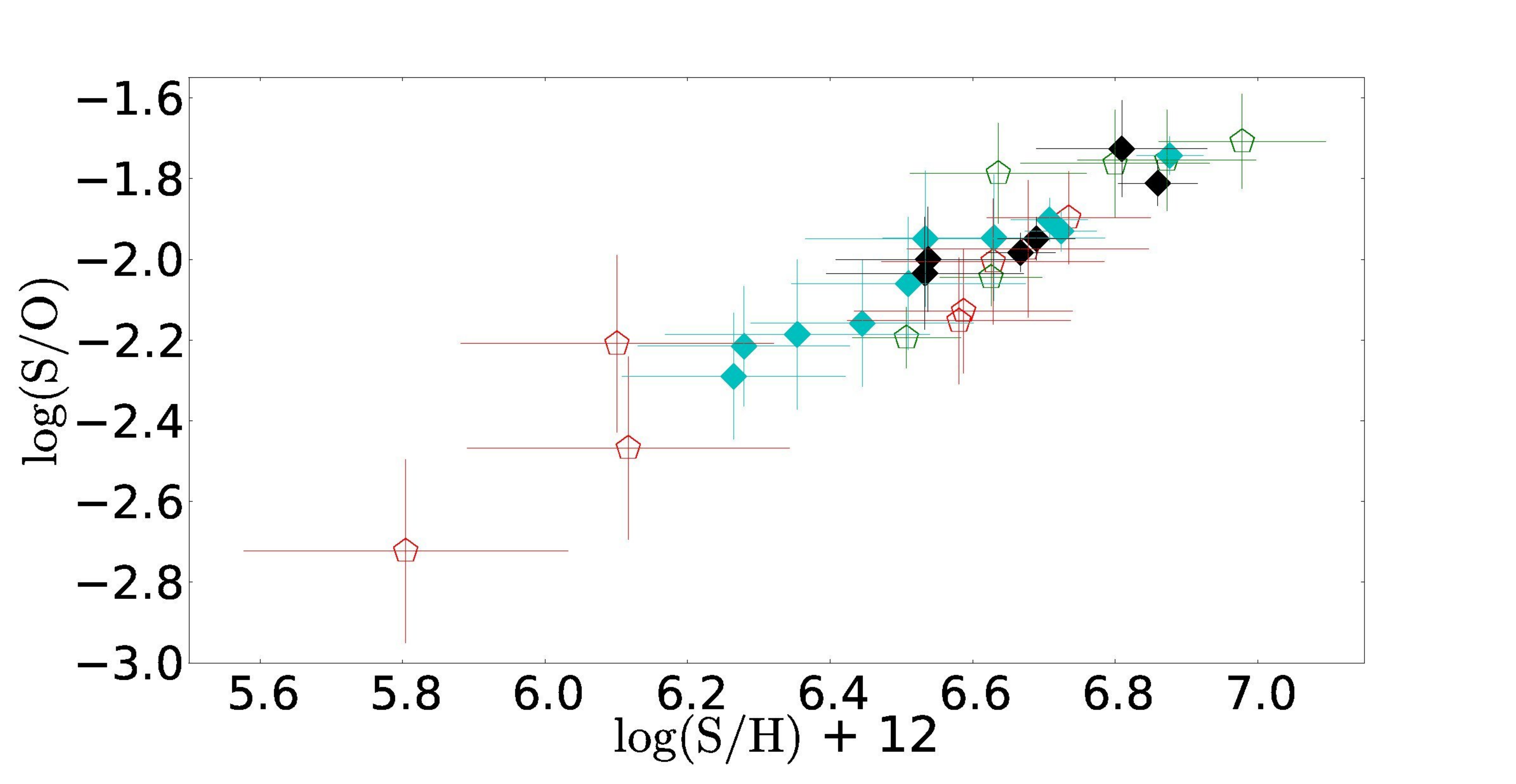}
\includegraphics[scale=0.145]{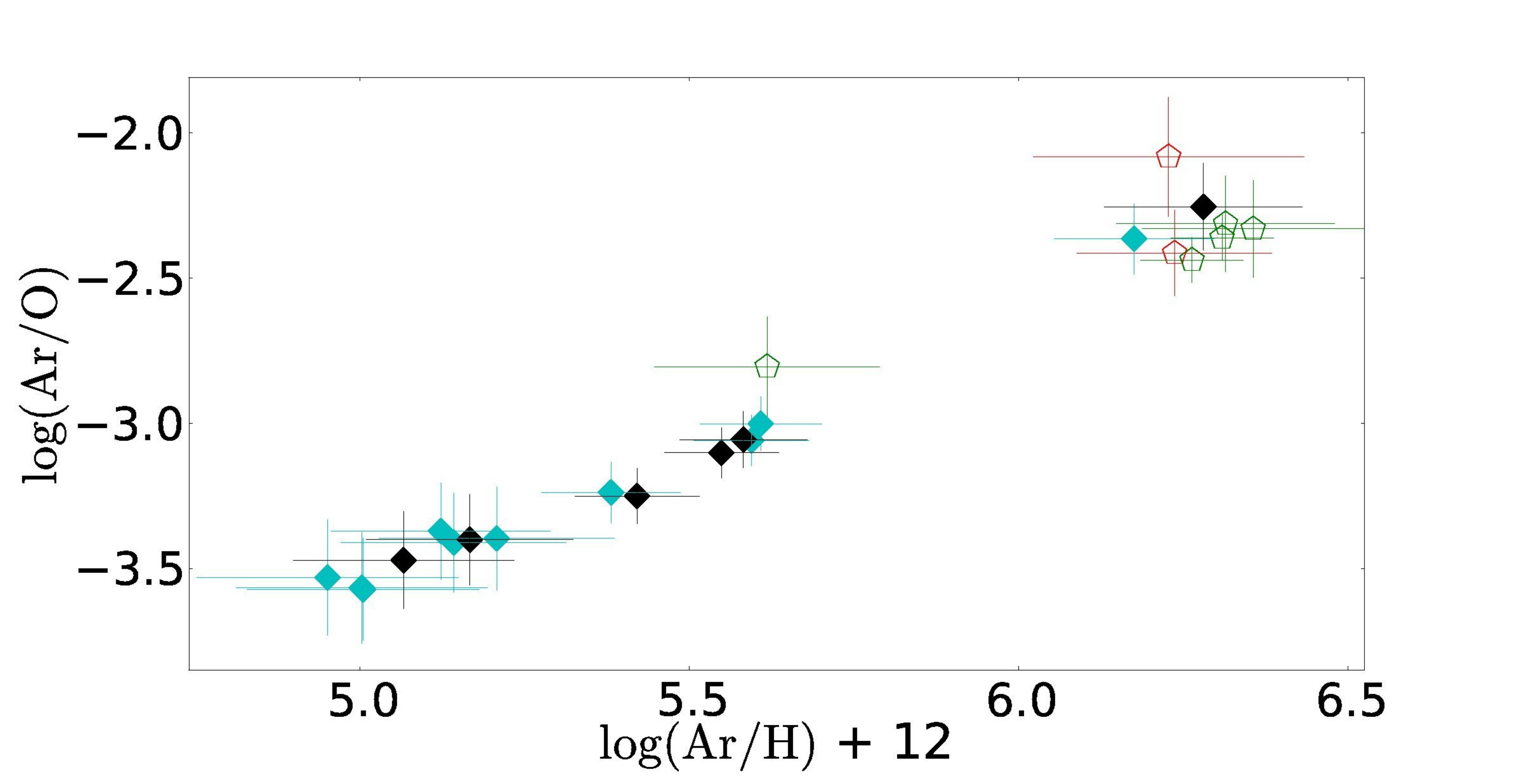}
\caption[]{Abundance patterns: $\rm{log}$(Ne/O) vs. $\rm{log}$(Ne/H)+12 (upper panel); $\rm{log}$(S/O) vs. $\rm{log}$(S/H)+12 (middle panel) and 
$\rm{log}$(Ar/O) vs. $\rm{log}$(Ar/H)+12 (lower panel). Symbols are the same as in Fig.~8. A trend of increasing Ne/O, S/O and 
Ar/O with increasing Ne, S and Ar, respectively, is found.}
\end{figure}

Besides the emission line ratios, kinematic information --in fact expansion velocities-- is essential to fathom the formation 
and excitation mechanisms of LISs in PNe. The diagnostic diagrams, based on the shock models from Raga et al. (2008), are used 
here in order to perform a more comprehensive study on the LISs and the other nebular components by combining their emission lines 
and expansion velocities.

In particular, Raga et al. (2008) modelled the evolution of a high-speed (100--150~\kms) and high-density 
($10^3$ cm$^{-3}$) knot moving away from a photo--ionizing source through a low-density medium, for a range of initial conditions. 
The simulated knot is a spherical cloudlet with temperature of 10000~K, and size of $10^6$~cm. The surrounding medium was 
assumed to be uniformly filled with a low-density gas of $10^2$~cm$^{-3}$ and the same temperature (10000~K) of the knot. 
A black-body approximation was assumed for the central photo-ionizing source, with a  $L$=5000~L$\odot$ 
and two $T_{\rm{eff}}$ of 50000~K and 70000~K, which correspond to an ionizing photon rate of 
$S_{\star}$=3.41$\times$10$^{47}$~s$^{-1}$ and 4.66$\times$10$^{47}$~s$^{-1}$, respectively. Moreover, the effect of 
the distance between the knot and the central source was also studied for three specific distances of 
3, 1, 0.3$\times$10$^{18}$~cm.

\begin{figure*}
\centering
\includegraphics[scale=0.145]{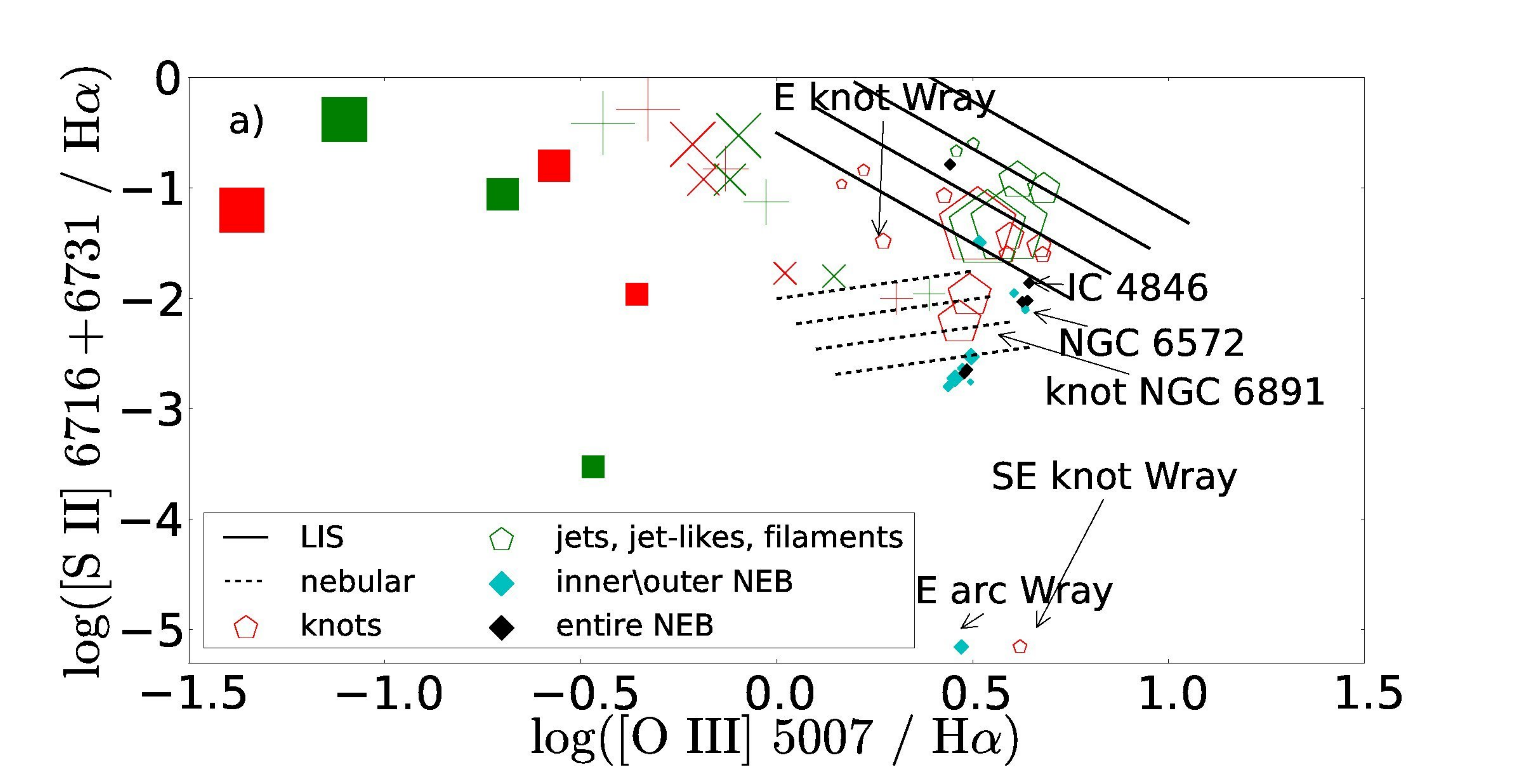}
\includegraphics[scale=0.145]{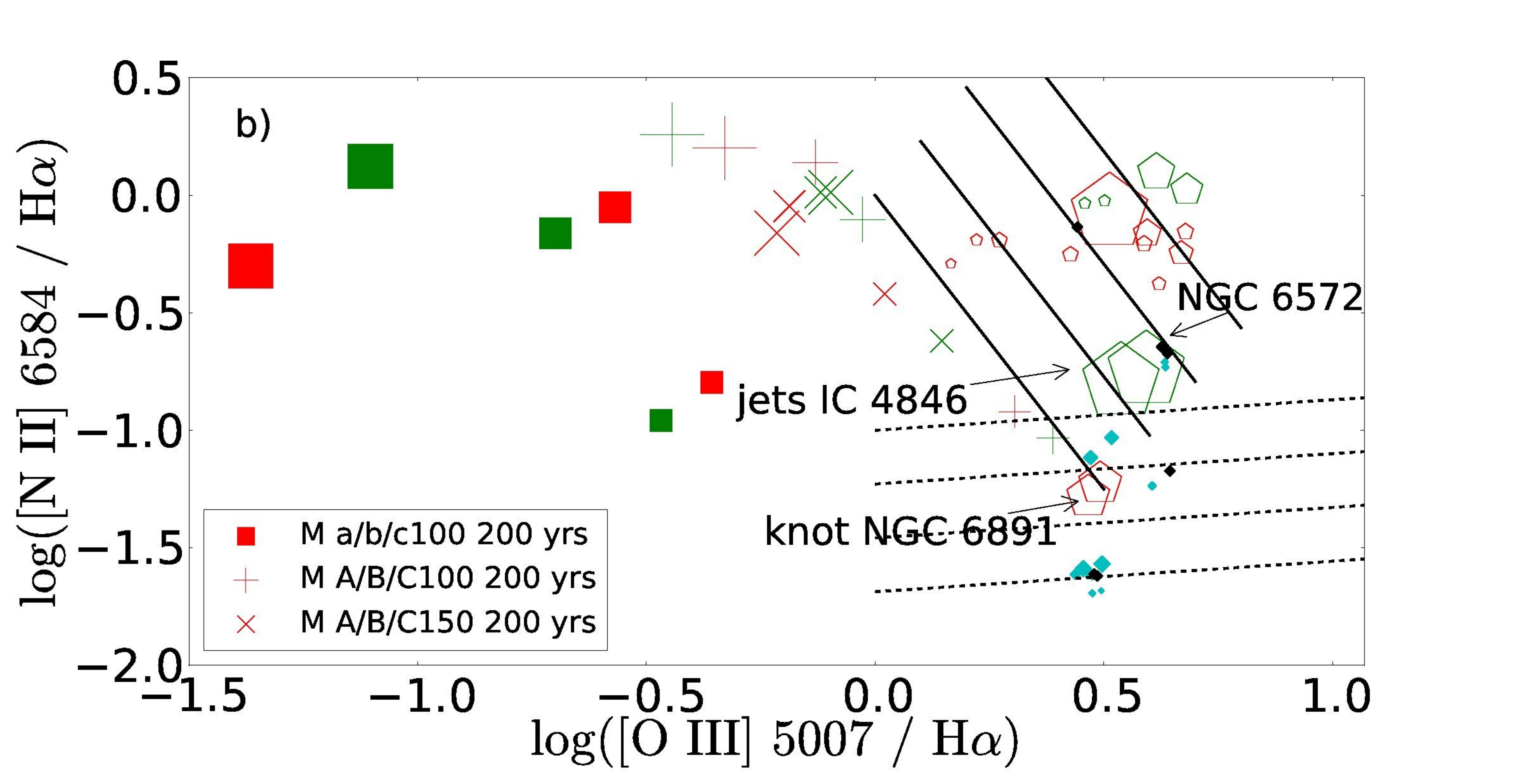}
\includegraphics[scale=0.145]{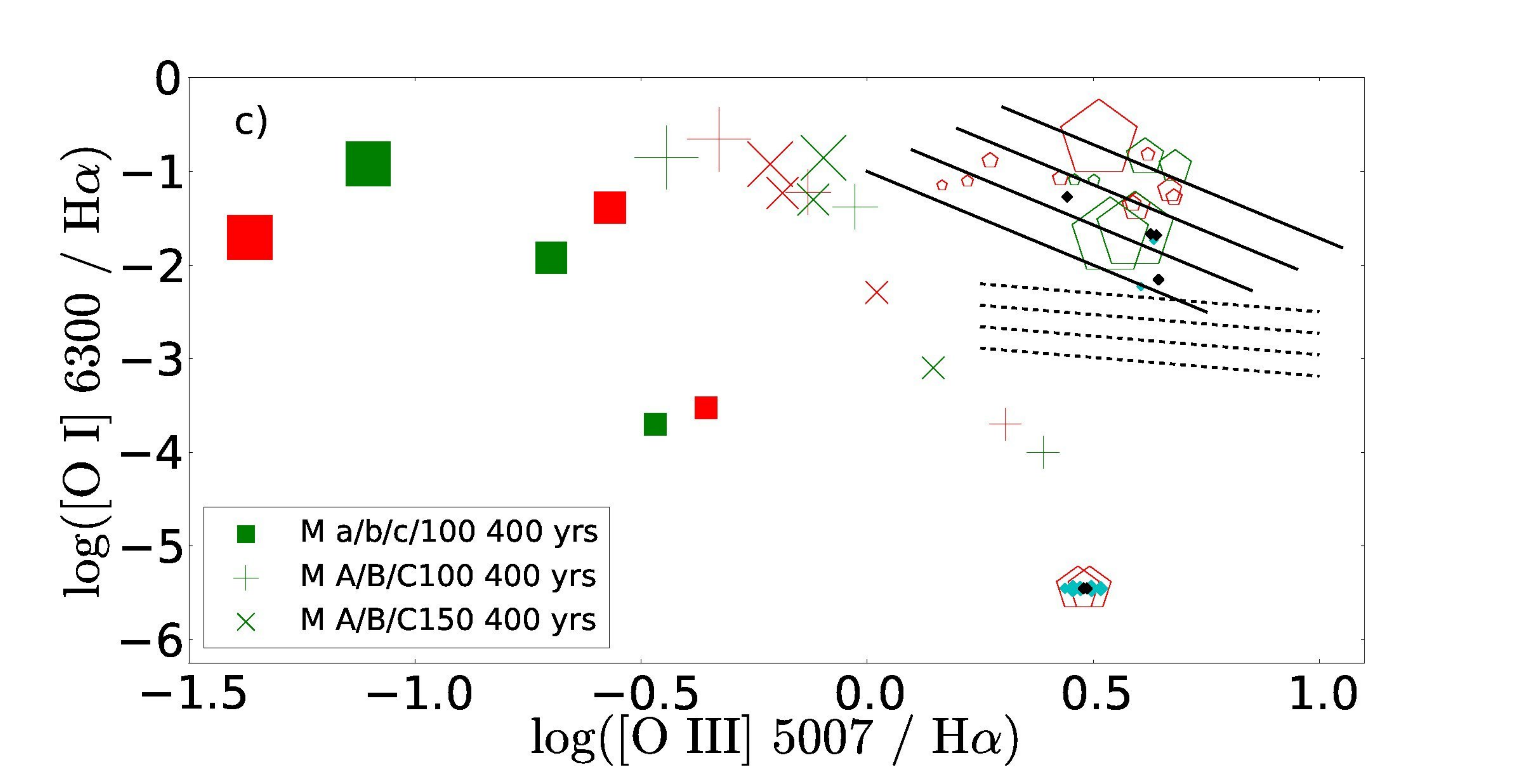}
\includegraphics[scale=0.145]{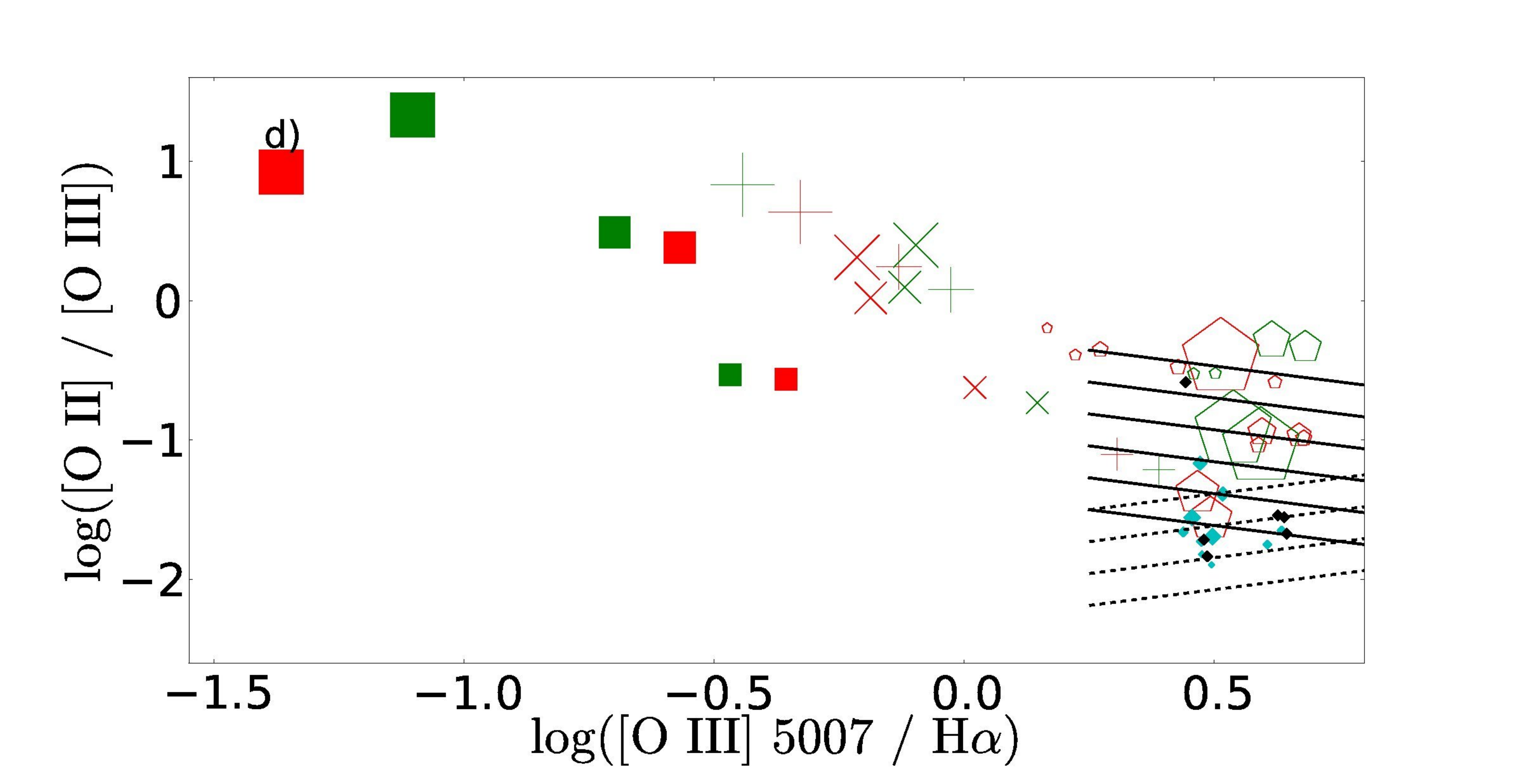}
\includegraphics[scale=0.145]{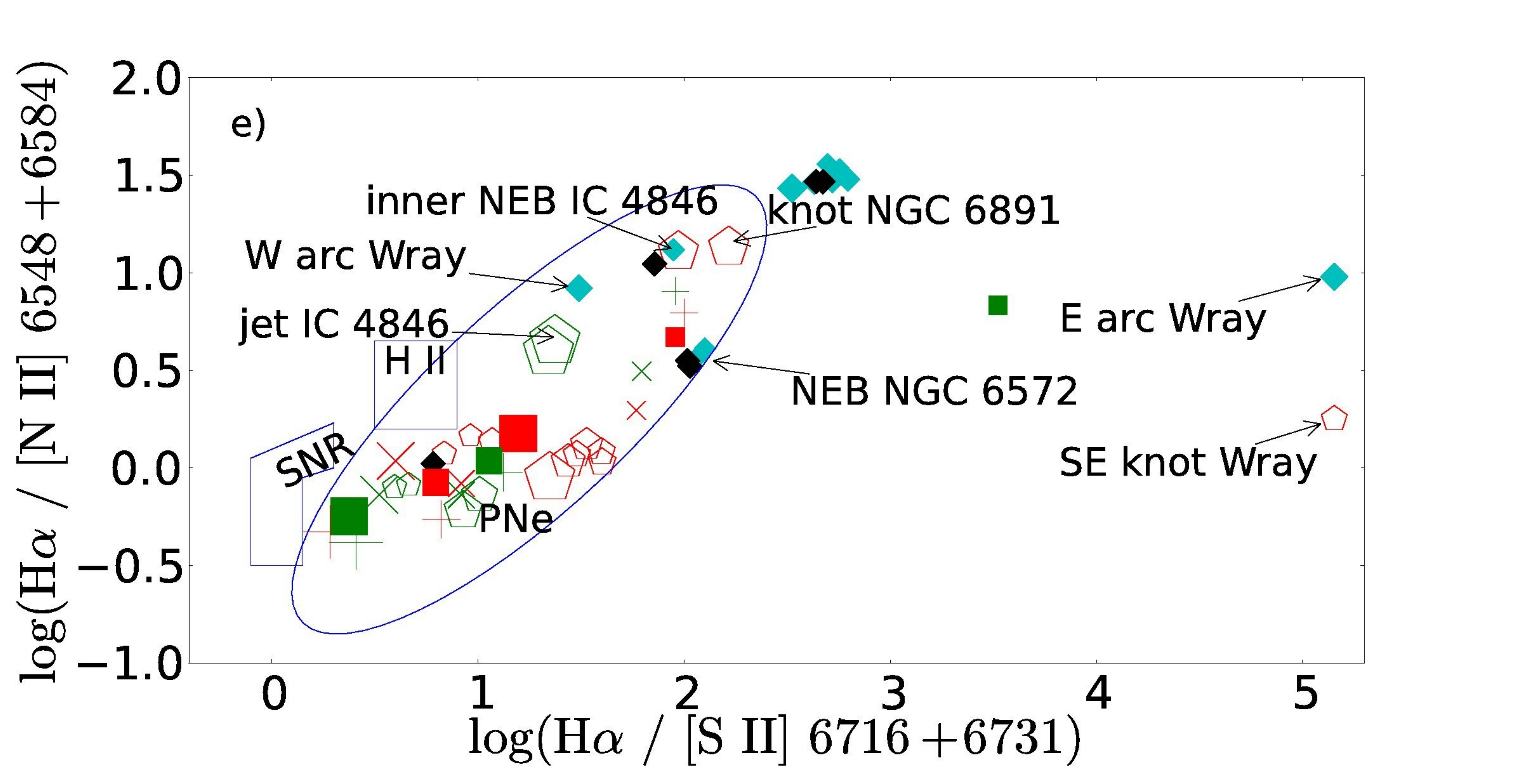}
\includegraphics[scale=0.145]{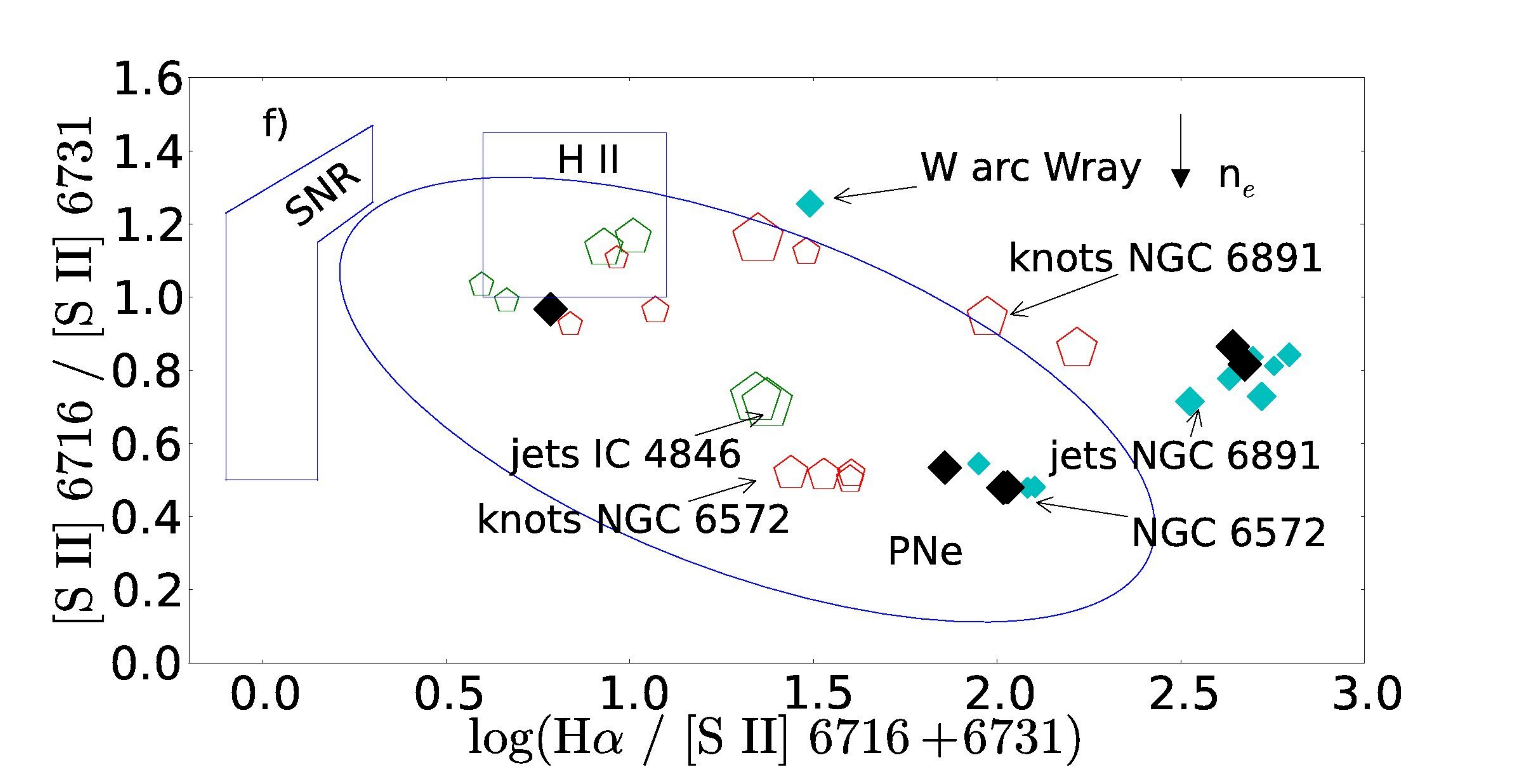}
\caption[]{Diagnostic diagrams distinguishing the shock-excited and the photo-ionized regions. Panels a, b, c, and d 
are the diagnostic diagrams from Raga et al.~(2008), on which we superimpose the results for our sample of PNe. 
The regions with diagonal solid lines represents the loci of fast LISs, whereas the regions with dashed lines represent 
the location of photo-ionized structures (e.g. inner and outer NEBs that are in fact shells and rims). Panels e and f 
display the diagnostic diagrams by Sabbadin et al. (1977), with the density ellipses introduced by Riesgo \& L\'opez (2006).}
\end{figure*}

\begin{figure*}
\centering
\includegraphics[scale=0.145]{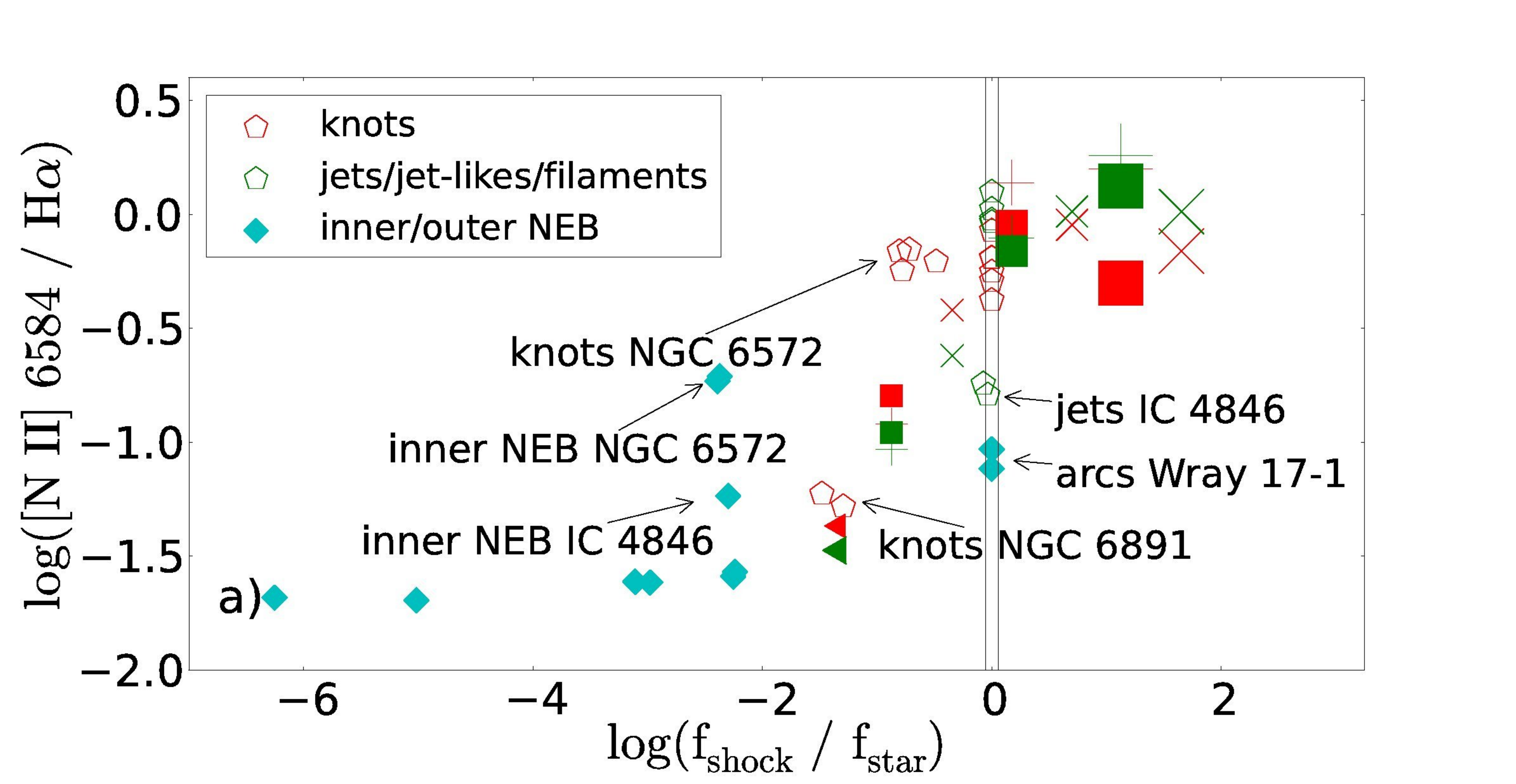}
\includegraphics[scale=0.145]{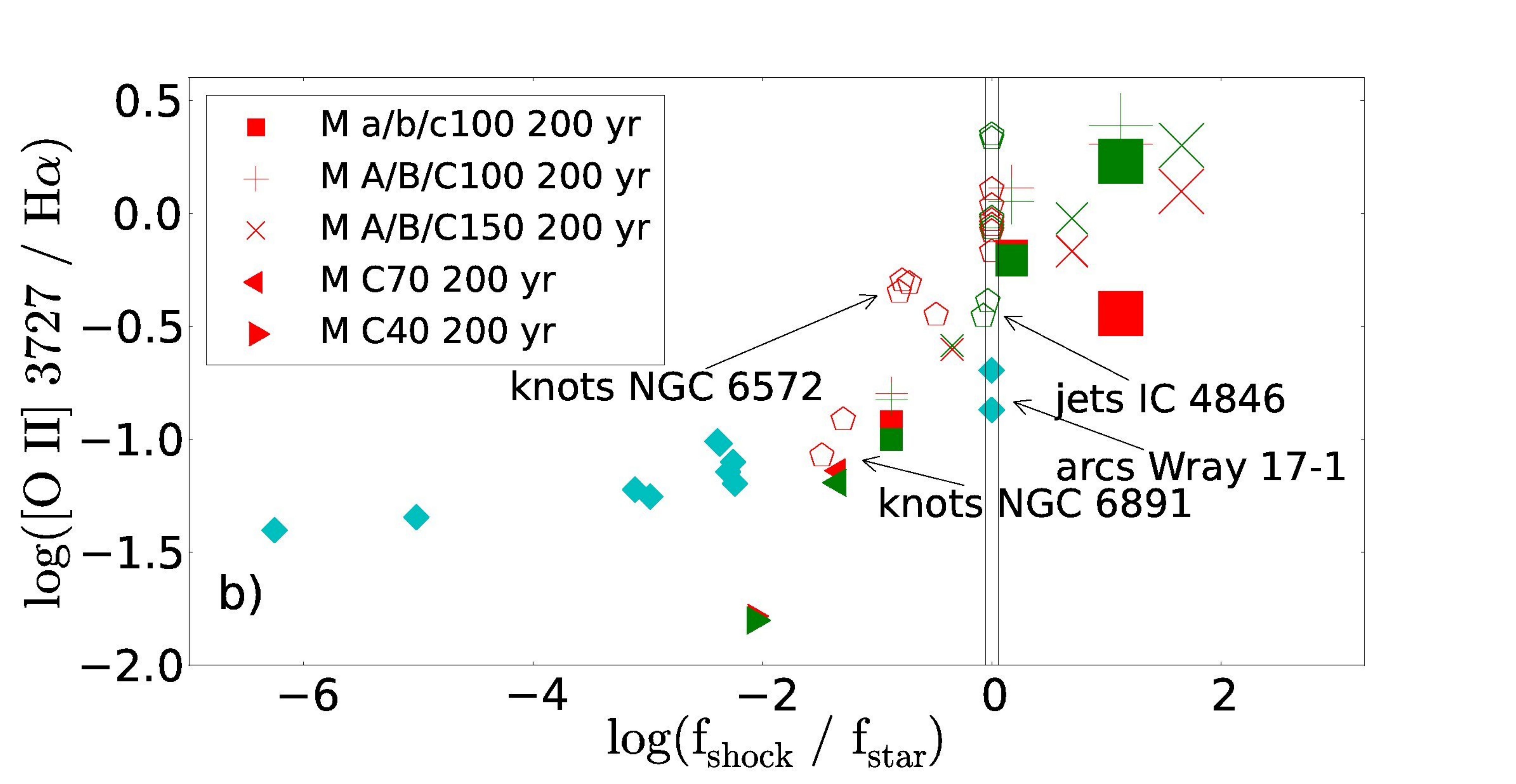}
\includegraphics[scale=0.145]{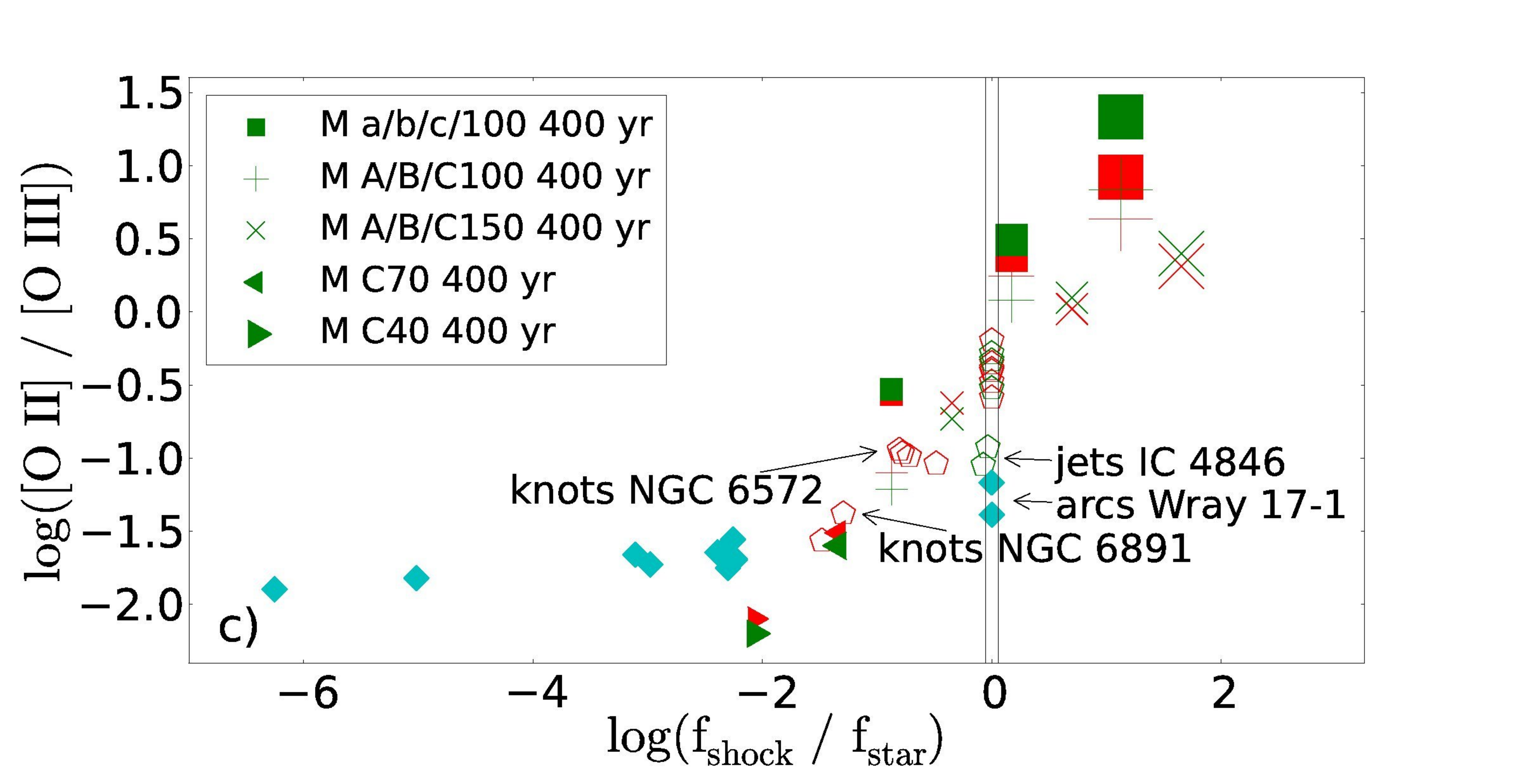}
\includegraphics[scale=0.145]{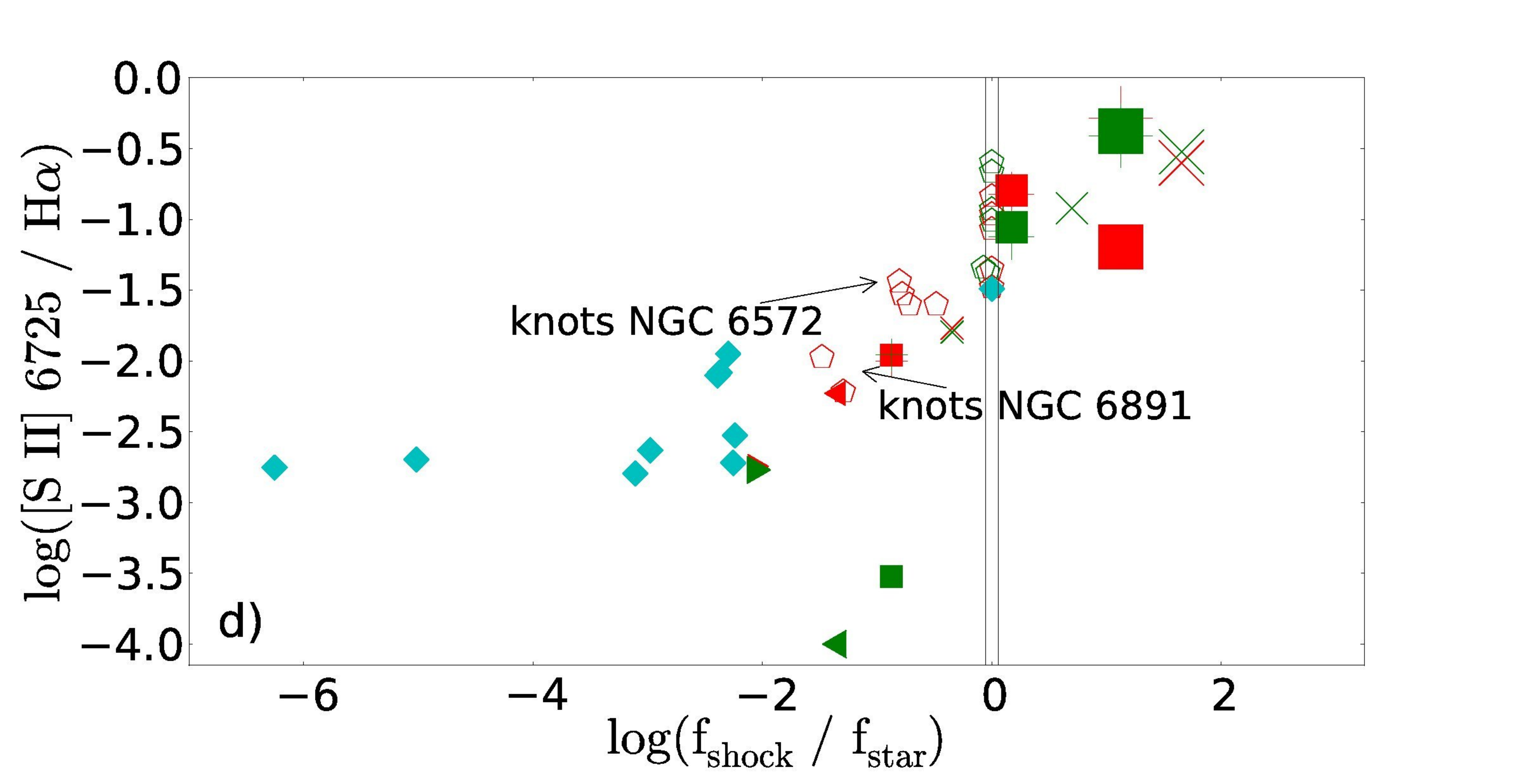}
\caption[]{Plots of \nitrogen/\ha\ (panel a), \oxygenii/\ha\ (panel b), \oxygenii/\oxygeniii\ (panel c) and \sulfurt/\ha\ (panel d) 
vs. $\rm{log}$(f$_{shocks}$/f$_{\star}$). The vertical strips represent $\rm{log}$(f$_{shocks}$/f$_{\star}$) = 0. K~1-2 and Wray~17-1 structures were 
arbitrarily placed into the strips, since there are no distance measurements for these PNe.}
\end{figure*}

The emission line ratios of the knots, for two different epochs of the simulations (200~yr in red and 400~yr, in green), are 
presented in Fig.~10, superposed to the observed emission line ratios from this work. Unfortunately, these models 
do not provide any predictions for the properties of the other nebular components like shells, 
rims and haloes. The symbols in Fig.~10 correspond to different expansion velocities, $T_{\rm{eff}}$, and distances to the central source.
--boxes:  $V_{exp}$=100~\kms\ and $T_{\rm{eff}}$=50000~K (models a/b/c100); crosses:  $V_{exp}$=100~\kms\ and $T_{\rm{eff}}$=70000~K (models A/B/C100); 
and x symbols: $V_{exp}$=150~\kms\ and $T_{\rm{eff}}$=70000~K (model A/B/C150). Letters from \lq\lq a" to \lq\lq c" (and \lq\lq A" to \lq\lq C") 
indicate the different distances from the central source and thus the different photo-ionization rates (small symbol $\equiv$ small distance $\equiv$ closer to the 
central source). The solid and dash-dotted lines enclose the regime of fast LISs (e.g. knots, jets) and rims and shells (or, more general, 
photo-ionized regions) taken from Raga et al. (2008). The open pentagons correspond to the LISs (knots: red and jets/jet-likes/filaments: 
green), the filled cyan diamonds correspond to the inner and outer NEB regions, and the filled black diamonds to the entire NEB. 
Moreover, the larger the points' size the higher the de-projected expansion velocity (see Table 1). The velocities are normalized to the 
average expansion velocity of PNe, 31\kms\ (Pereyra et al. 2013). For the entire NEB (black symbols), we considered an expansion 
velocity equal to the value just quoted, so emphasizing the different expansions among the nebular components in the diagnostic diagrams. 
No trend is found between the emission line ratios and $V_{exp}$ of the LISs and the nebular regions. This suggests that any attempt to 
disentangle the photo-ionized and shock-excited regions, or even investigate their excitation mechanisms, requires the evolutionary age 
and photo-ionization rates of the central star, apart from the expansion velocities themselves.
 
All LISs in this work are found to be located in the regime of fast LISs, except from the pair of knots in NGC~6891. The latter knots 
are displaced from the region of fast LISs, in each diagram, despite their high expansion velocities. Moreover, 
the NE and SW outer NEBs in NGC~6891 (the jet components in Guerrero et al. 2000) also lie in the regime of photo-ionized regions 
and close to the inner NEBs of the nebula (NE and SW outer NEBs). This is a further indication that there is no difference among 
the regions analyzed in NGC~6891. We already pointed out that there are no jets in NGC~6891 (see \S3.4). Knots, on the other hand, 
show low-ionization lines slightly enhanced. We believe that this may be associated with the low contribution of the photo-ionization 
in these knots (higher distance from the central star), resulting in more efficient shock excitation.

Adopting a distance of $~$2.5 kpc for NGC~6891 (Phillips 2005) and an angular distance from the central star of 9.1~arcsec, the linear 
distance is $\sim$3.4$\times$10$^{17}$~cm. Given that the $T_{\rm{eff}}$ and $L$ of the central star are 50000~K (Mendez et al. 1988) and 10000~L$\odot$ (Mendez et al. 
1992; re-scaled to the distance used here), respectively, the ionizing photon flux (f$_{\ast}$) reaching the knots is 
f$_{\ast}$\footnote{The ionizing photon flux, f$_{\star}$, is calculated by means the equation 
f$_{\star}$=$S_{\star}$4$\pi$ $d^2$, where $S_{\star}$ is the photon rate and d is the distance of the cloudlet from the central 
star.}=5.46$\times$10$^{11}$~cm$^{-2}$ s$^{-1}$. NGC~6891's knots lie closer to the Raga's C100 model ($V_{exp}$=100~\kms, 
$D$=3$\times$10$^{17}$~cm and f$_{\ast}$=3.73$\times$10$^11$~cm$^{-2}$ s$^{-1}$) than to the c100 model ($V_{exp}$=100~\kms, 
$D$=3$\times$10$^{17}$~cm and f$_{\ast}$=2.73$\times$10$^11$~cm$^{-2}$ s$^{-1}$). The main difference between 
these two models is the ionizing photon flux, which depends on the $T_{\rm{eff}}$. Models \lq\lq C" are those 
with the smallest distance from the central stars and therefore have  higher ionization rates. Hence, NGC~6891 appears to 
be a nebula with a higher ionization rate (so, mainly photo-ionized) whilst the contribution of shocks must be very low. 
Nevertheless, the high expansion velocities of the pair of knots ($\sim$80~\kms), results in the enhancement of 
the low-ionization lines.

As for NGC~6572, the inner NEBs are in a transition zone between the rims/shells and fast LISs 
regimes, whilst the two pairs of knots are well placed on the regime of the fast LISs. 
The de-projected expansion velocities of the two pair of knots are at least 2.5--3 times higher 
than the expansion velocities of the inner NEBs, then suggesting higher contribution from shocks. 
The angular distances of the knots from the central star are 6.67~arcsec (NE and SW), 
7.5~arcsec (NW) and 8.4~arcsec (SE). Considering a distance of 1.4~kpc (Mellema 2004), 
we obtain linear distances from the central star of 1.35, 1.5 and 1.71 $\times$ 10$^{17}$~cm. 
These distances give ionizing photon fluxes between 2.3 and 3.4$\times$10$^{10}$~cm$^{-2}$ s$^{-1}$, 
considering $T_{\rm{eff}}$=60000~K and $L$=2250~L$\odot$ (Shaw \& Kaler 1989; re-scaled to distance adopted in this work). 
This ionizing photon fluxes imply a low contribution in the excitation of the knots due to 
the absorption of UV photos. Therefore, the enhancement of the low-ionization lines in these structures should be 
a result of the shocks. The higher ionizing photon flux of the inner NEBs, due to their smaller distance to the central star, 
in conjunction with their much lower expansion velocities, explain their position on the diagnostic diagrams in the regime of photo-ionized regions.

Unlike the knots of NGC~6572, the jets of IC~4846 are located in the transition zone, despite their high expansion 
velocities ($>$100~\kms). For $T_{\rm{eff}}$=50000~K and $L$=2067~L$\odot$ (Shaw \& Kaler 1989; rescaled to the distance used here), 
the ionizing photon rate is 1.36 $\times$ 10$^{47}$~s$^{-1}$, and for an angular distance from the central of 3.5~arcsec and $D$=5.72 kpc 
(Phillips 2004), the linear distance is then 3$\times$10$^{17}$~cm. This implies an 
f$_{\ast}$=1.20$\times$10$^{11}$~cm$^{-2}$ s$^{-1}$, which is approximately three times lower than that of C100 model. 
Despite this low value, IC~4846 exhibits stronger \nitrogen, \sulfurt\ and \oxygeni\ emission lines than in model C100, 
which means that shock interactions are responsible for this enhancement.

All LISs of Wray~17-1  are well placed in the regime of fast LISs, with intense low-ionization lines. 
Unfortunately, there is no distance measurement for this object. Nevertheless, assuming a linear distance of 3$\times$10$^{17}$~cm  
between the knots and the central star (the smallest distance among the shock models), and an angular distance of 43~arcsec
from the central source, a $D=$0.46~kpc is obtained, which seems unrealistic. 
This may suggest that the knots are at a larger distance from the central star, e.g. 3$\times$10$^{18}$~cm (or $D\sim$5 kpc). 
We believe that the strong low-ionization lines found in this nebula are resulting from shocks, whilst the contribution of the 
stellar UV photons may be very low. The observed spectra from the inner NEBs are in agreement with this argument.

Regarding the \ha/\nitrogen\ vs. \ha/\sulfurt\ and \sulfurt\ $\lambda\lambda 6716/6731$  vs. \ha/\sulfurt\ diagnostic 
diagrams, introduced by Sabbadin et al. (1977), it is clear that all shock-excited features are located at the bottom-left 
region whilst the photo-ionized features at the upper-right region (lower panels in Fig.~10). The zone between 
0.2$<\rm{log}$(\ha/\nitrogen)$<$0.8 defines the transition zone where both 
excitation mechanisms are efficient. In general, hot and dim central stars are able to reproduce emission lines ratios 
that resemble those of shock-excited nebulae without necessarily being shocked-excited (Raga et al. 2008). 
This could explain the presence of some elliptical and round PNe in the region with $\rm{log}$(\ha/\nitrogen)$<$ 0 and 
$\rm{log}$(\ha/\sulfurt])$<$ 0.4 (Sabbadin et. 1977; Riesgo \& Lopez 2006). 

In accordance to the previous analysis, besides the emission line ratios, kinematic information is required in order 
to probe the excitation mechanisms of the LISs in PNe. Hence, a more general approach, which would take into account the 
contribution of shock- and photo-ionization is actually required here. We quantitatively investigate this by defining the 
dimensionless parameter $\rm{log}$(f$_{shocks}$/f$_{\star}$), where f$_{\rm{shocks}}$ is the ionizing photon flux in units 
of cm$^{-2}$ $s^{-1}$ for a shock with $V_{s}$, and f$_{\star}$ is the stellar ionizing photon flux at a given distance.
f$_{\rm{shocks}}$ is as follows: we calculate the total energy flux emitted due to the shock, by using the 
formula F$_{\rm{shocks}}$=2.28$\times$10$^{-3}$($V_{\rm{s}}$/100~\kms)$^3$ $\times$(n$_e$/cm$^{-3}$) in units of erg cm$^{-2}$ s$^{-1}$ 
from Dopita \& Sutherland (1996). Then, we convert this energy flux (F$_{\rm{shocks}}$) to photon flux (f$_{\rm{shocks}}$), 
by dividing by the average energy of a photon (L/S$_{\star}$) emitted from the central star.      
This ratio should give us of the contribution of each mechanism to the excitation of the gas.

In Fig.~11 we present the new diagnostic diagrams \nitrogen/\ha, \sulfurt/\ha, \oxygenii/\ha\ 
and \oxygenii/\oxygeniii\ vs. $\rm{log}$(f$_{shocks}$/f$_{\star}$) that allows us to distinguish nebular components with different excitation mechanisms. 
It is noticeable that there are two distinct regimes in these diagrams: i) the upper-right corner with the shock-excited 
structures and ii) the lower-left corner with the photo-ionized structures. A transition zone, where both excitation mechanisms 
are important, can also be considered. All the nebular regions (NEB) lie in the lower-left region, with $\rm{log}$(f$_{shocks}$/f$_{\star}$)$<$-2, 
whilst the knots and jets lie in the upper-right part of the diagrams, with $\rm{log}$(f$_{shocks}$/f$_{\star}$)$>$-1. 
Only the pair of knots in NGC~6891 show $\rm{log}$(f$_{shocks}$/f$_{\star}$)$<$-1, consistently with our previous conclusion 
that they must be mainly photo-ionized with a low contribution from shocks. The models a/b100, A/B100 and A/B150 also lie 
in the shock-excited regime (upper-right corner), whilst the models c100, C100 and C150 are found in the transition zone. 
The latter models have the higher ionizing photon flux among all the models, which means that the 
excitation due to the absorption of stellar UV photons is more efficient. Any decrease of $V_{\rm{exp}}$, for instance 
from C100 to C70 or C40, will result in further decrease of $\rm{log}$(f$_{shocks}$/f$_{\star}$) and will eventually place these models closer to 
the regime of photo-ionized regions.

The points representing K~1-2 and Wray~17-1 have been arbitrarily added to the diagram, at $\rm{log}$(f$_{shocks}$/f$_{\star}$)=0, 
since there are no distance measurements for these objects and the f$_{\star}$ cannot be calculated. Based on their emission line ratios, 
one can assume that the jets and knots of these two objects have, indeed, $\rm{log}$(f$_{shocks}$/f$_{\star}$) close to 0. 
Assuming $\rm{log}$(f$_{shocks}$/f$_{\star}$)=0 for the knots and jets, $\rm{log}$(f$_{shocks}$/f$_{\star}$)=-1 for the arcs, and $L$=5000~L$\odot$, 
we obtain a rough estimation of their distances. Wray~17-1 should be located between 3 and 7~kpc, where 
for K~1-2, we obtain a very high distance, $>$20~kpc. This suggests that the lumonosity of the central star in
K~1-2 is lower than 5000~L$\odot$ and/or the expansion velocities of the knots are higher than 35~\kms. For instance, 
a $L$=2000~L$\odot$ and $V_{exp}$=70~\kms\ gives us a more reasonable distance of $\sim$5~kpc. 

\subsection{Central Star}

Despite the various attempts to explain the formation of collimated (those that appear in pairs) LISs, their origin still remains 
debatable (see in Gon\c calves et al. 2001). The most prominent mechanisms is the interaction of a close binary system due to 
the mass-transfer exchange. The formation of an circumstellar/circumbinary disk via binary interaction provide those 
conditions to form bipolar outflows and/or collimated jets (Blackman et al. 2001, Nordhaus \& Blackman 2006). 
Some models suggest that the formation of jets occurs after the common envelope (CE) phase of a close binary star 
(Soker \& Livio 1994). Pairs of jets with kinematical ages much younger than the main nebular structures have indeed been 
found in some post-CE PNe, like NGC~6778 and NGC~6337 (Tocknell et al. 2014). Although, the majority of the post-CE PNe are 
found to exhibit jets that predate the main nebular structures, suggesting that the polar ejection occurs before the CE phase 
(see Tocknell et al. 2014; Jones et al. 2014). The discovery of several new close binary stars in PNe has provided strong 
evidence of a link between the close binary post-common envelope PNe and LISs (Miszalski et al. 2009b). 

Moreover, a possible link between the spectral type of the PN nucleus and LISs has also been pointed out by Miszalski et al. (2009a). 
According to these authors, more than half of PNe that are known to possess LISs ($\sim$65\%) have an emission line type central 
star (e.g. Of, [WR] or PG1159). Incidentally, all the PNe studied in this work have an emission-line star, except K 1-2. 
In particular, two of them have a WEL type nucleus (IC~4846, NGC~6891), one has a [WC]--PG1159 (NGC~6572), and one a PG--1159 
nucleus (Wray~17-1). The central star of K~1-2 has been classified as a K2 V but it is a known post-CE close binary system 
(Exter et al. 2003; De Marco 2009).

\section{Conclusions}

The properties like $\it{N}_{\rm{e}}$, $\it{T}_{\rm{e}}$, ionic and total chemical abundances were calculated for various nebular components, 
like nebular regions (inner and outer NEBs) and LISs (jets, jet-like, filaments and knots), as well as for 
the entire nebulae of a small sample of PNe (IC~4846, Wray~17-1, K~1-2, NGC~6891 and NGC~6572). 
We confirmed that LISs are less dense (in terms of $N_{\rm{e}}$) than the surrounding medium, whereas their $\it{T}_{\rm{e}}$ are the same 
as that of the other components of a given PNe.

There is no chemical abundance enrichment in LISs that could be attributed to the strong low-ionization lines. 
For a given PN, all the nebular components have the same chemical abundances within the errors, reflecting on 
the chemical composition of the entire nebula as well as the evolution of the central star. 

A comparison of between the emission lines of the nebular components and those predicted by the the shock models reveals 
that LISs are predominantly shock-excited regions. This shock-excitation of LISs results in the enhancement of the 
low-ionization lines, whereas the other nebular components are mainly photo-ionized, so less prominent in the low-ionization emission lines.

In order to probe the contribution of shocks and UV photons in the excitation of the LISs, we defined the parameter 
$\rm{log}$(f$_{shocks}$/f$_{\star}$), where f$_{shocks}$ and f$_{\star}$ are the ionizing photon fluxes due to the shock 
interactions and the central star continuum, respectively. We then proposed new diagnostic diagram that involves the 
emission-line ratios \nitrogen/\ha, \oxygenii/\ha, \sulfurt/\ha\ and \oxygenii/\oxygeniii\ versus the new excitation 
parameter, $\rm{log}$(f$_{shocks}$/f$_{\star}$). A robust relation differentiating the regimes of shock-exited ($\rm{log}$(f$_{shocks}$/f$_{\star}$)$>$-1) 
from that of the photo-ionized components ($\rm{log}$(f$_{shocks}$/f$_{\star}$)$<$-2), in addition to a transition zone (-2$<\rm{log}$(f$_{shocks}$/f$_{\star}$)$<$-1), 
where both mechanisms are equally important, is found. Most of the LISs analyzed in this work were found to be well 
located in the regime of shock-excited structures, apart from the pairs of jets and knots in NGC~6891. The former was found to be 
photo-ionized as much as its surrounding nebula, while the latter is placed in the transition zone, which indicates a contribution 
from shocks. All the inner and outer NEBs are well placed in the region of the photo-ionized components.

\section*{Acknowledgments} 
We thank, enormously, R. Corradi and H. E. Schwarz (in memoriam) who provide us the Danish telescope data we use in the 
present paper. We also would like to thank the anonymous referee for his/her detailed report on the manuscript, which indeed helped us to significantly improve the paper. 
The work of SA is supported by CAPES post-doctoral fellowship \lq Young Talents  Attraction' - Science Without Borders, A035/2013. 
This work was also partially supported by FAPERJ's grant E-26/111.817/2012 and CAPES's grant A035/2013. 
The analysis here presented is based on observations obtained at the 2.5-m INT (IDS), operated by the Isaac 
Newton Group on the island of La Palma in the Spanish Observatorio del Roque 
de los Muchachos of the Instituto de Astrof\'ı sica de Canarias. Our work is also based on the 1.54-m Danish 
telescope (DFOSC) data, operated by the European Southern Observatory (ESO) at La Silla, Chile, and on  
the 2.1-m SPM telescope, operated by the UNAM at the Observatorio Astron\'omico at San Pedro 
M\'artir Observatory in Baja California, Mexico. We also use NASA/ESA HST data, obtained at the 
Space Telescope Science Institute.

\bibliographystyle{mnras}

\begin{thebibliography}{99}

\bibitem[\protect\citeauthoryear{Acker}{1991}]{Ack91}
Acker A., Raytchev B., Stenholm B., Tylenda R., 1991, A\&AS, 90, 89

\bibitem[\protect\citeauthoryear{Acker}{1992}]{Ack92}
Acker A., Marcout J., Ochsenbein F., Stenholm B., Tylenda R., Schohn C., 1992, The
Strasbourg-ESO Catalogue of Galactic Planetary Nebulae. Parts I, II. 

\bibitem[\protect\citeauthoryear{Aleman}{2011}]{Ale11}
Aleman I., Gruenwald R., 2011, A\&A, 528A, 74	

\bibitem[\protect\citeauthoryear{Balick}{1987}]{Bal87}
Balick B., Preston H. L., Icke V., 1987, AJ, 94, 1641

\bibitem[\protect\citeauthoryear{Balick}{1993}]{Bal93}
Balick B., Rugers M., Terzian Y., Chengalur J. N. 1993, ApJ, 411, 778

\bibitem[\protect\citeauthoryear{Balick}{1994}]{Bal94}	
Balick B., Perinotto M., Maccioni A., Terzian Y.,  Hajian A. R., 1994, ApJ, 424, 800

\bibitem[\protect\citeauthoryear{Balick}{1998}]{Bal98}	
Balick B., Alexander J., Hajian A. R., Terzian Y., Perinotto M., Patriarchi P.,	
1998, AJ, 116, 360

\bibitem[\protect\citeauthoryear{Balick}{2002}]{Bal02}	
Balick B., Frank A., 2002, ARA\&A, 40, 439

\bibitem[\protect\citeauthoryear{Blackman}{2001}]{Bla01}	
Blackman E. G., Frank A., Welch C., 2001, ApJ, 546, 288

\bibitem[\protect\citeauthoryear{Boffin}{2012}]{Bof12}	
Boffin H. M. J., Miszalski B., Rauch T., Jones D., Corradi R. L. M., Napiwotski R., 
Day--Jones A. C.,  K\:oppen J., 2012, Science, 338, 773.

\bibitem[\protect\citeauthoryear{Bohigas}{2001}]{Boh01}	
Bohigas J., 2001, RMxAA, 37, 237

\bibitem[\protect\citeauthoryear{Boumis}{2003}]{Bou03}	
Boumis P., Paleologou E. V., Mavromatakis F., Papamastorakis J., 2003, MNRAS, 339, 735

\bibitem[\protect\citeauthoryear{Boumis}{2006}]{Bou06}	
Boumis P., Akras S., Xilouris E. M., Mavromatakis F., Kapakos E., 
Papamastorakis J., Goudis C. D., 2006, MNRAS, 367, 1551

\bibitem[\protect\citeauthoryear{Cahn}{1992}]{Cahn96}	
Cahn J. H., Kaler J. B., Stanghellini L., 1992, A\&AS, 94, 399

\bibitem[\protect\citeauthoryear{Corradi}{1996}]{Cor96}
Corradi R. L. M., Manso R., Mampaso A., Schwarz H. E., 1996, A\&A, 313, 913

\bibitem[\protect\citeauthoryear{Corradi}{1999}]{Cor99}
Corradi R. L. M., Perinotto M., Villaver E., Mampaso A., Gon\, calves D., 1999, AJ, 523, 721

\bibitem[\protect\citeauthoryear{Corradi}{2015}]{Cor15}
Corradi R. L. M. Garc\'\i a-Rojas J., Jones D., Rodr\'\i guez-Gil P., 
2015, ApJ, 807, 99

\bibitem[\protect\citeauthoryear{Delgado}{2015}]{Del15}	
Delgado-Inglada G., Rodr\'{i}guez M., Peimbert M., Stasi\'{n}ska, G., 
Morisset C., 2015, MNRAS, 449, 1797

\bibitem[\protect\citeauthoryear{Dopita}{1996}]{Dop96}	
Dopita M. A., Sutherland R. S., 1996, ApJS, 102, 161

\bibitem[\protect\citeauthoryear{Dopita}{1997}]{Dop97}
Dopita M. A., 1997, ApJ, 485, L41

\bibitem[\protect\citeauthoryear{DeMarco}{2009}]{DeM09}	
De Marco O., 2009, PASP, 121, 316

\bibitem[\protect\citeauthoryear{Exter}{2003}]{Ext03}	
Exter K. M., Pollacco D. L., Bell S. A., 2003, MNRAS, 341, 1349

\bibitem[\protect\citeauthoryear{Fitzpatrick}{1999}]{Fitz99}	
Fitzpatrick E.L., 1999, PASP, 111, 63

\bibitem[\protect\citeauthoryear{Frew}{2013}]{Frew13}		
Frew D. J., Boji\^ ci\'c I. S., Parker Q. A., 2013, MNRAS, 431, 2

\bibitem[\protect\citeauthoryear{Garcia}{2013}]{Gar13}		
Garc\'ia--Rojas J., Pe\~na M., Morisset C., Delgado--Inglada G., Mesa--Delgado A., Ruiz T. M., 2013, 2013, A\&A, 558A, 122

\bibitem[\protect\citeauthoryear{Garcia}{1999}]{Gar99}	
Garc\'ia--Segura, G., Langer, N., R\'o\.zyczka, M., Franco J., 1999, ApJ, 517, 767

\bibitem[\protect\citeauthoryear{Garcia}{2014}]{Gar14}	
Garc\'ia--Segura G., Villaver E., Langer N., Yoon S.-C., Manchado A., 2014, AJ, 783, 74

\bibitem[\protect\citeauthoryear{Goncalves}{2001}]{Gon01}
Gon\c calves D. R., Corradi R. L. M., Mampaso A., 2001, ApJ, 547, 302

\bibitem[\protect\citeauthoryear{Goncalves}{2003}]{Gon03}
Gon\c calves D. R., Corradi R. L. M., Mampaso A., Perinotto M., 2003, ApJ, 597, 975

\bibitem[\protect\citeauthoryear{Goncalves}{2004}]{Gon04r}
Gon\c calves D. R.,  2004, ASP Conference Proceedings, 313, 216

\bibitem[\protect\citeauthoryear{Goncalves}{2004}]{Gon04}
Gon\c calves D. R., Mampaso A.,  Corradi R. L. M., Perinotto M., Riera A., L\'opez-Mart\'in, 2004, MNRAS, 355, 37
\bibitem[\protect\citeauthoryear{Goncalves}{2006}]{Gon06}
Gon\c calves D. R., Ercolano B., Carnero A., Mampaso A., Corradi R. L. M., 2006, MNRAS, 365, 1039

\bibitem[\protect\citeauthoryear{Goncalves}{2009}]{Gon09}
Gon\c calves D. R., Mampaso A., Corradi R. L. M., Quireza C., 2009, MNRAS, 398, 2166-2176

\bibitem[\protect\citeauthoryear{Gorny}{1999}]{Gor99}
G\'orny S. K., Schwarz H. E., Corradi R. L. M., van Winckel H. 1999, A\&AS, 136, 145

\bibitem[\protect\citeauthoryear{Gorny}{2009}]{Gor09}	
G\'orny S. K., Chiappini C., Stasi\'nska G., Cuisinier F., 2009, A\&A, 500, 1089
	
\bibitem[\protect\citeauthoryear{Guerrero}{2000}]{Guer00}	
Guerrero M. A., Miranda L. F., Manchado A., V\'azquez R., 2000, MNRAS, 313, 1	

\bibitem[\protect\citeauthoryear{Guerrero}{2013}]{Guer13}	
Guerrero M. A., Toal\'a J. A., Medina J. J., Luridiana V., Miranda L. F., 
Riera A., Vel\'azquez P. F., 2013, A\&A, 557A, 121

\bibitem[\protect\citeauthoryear{Guerrero}{2014}]{Guer14}		
Guerrero M. A., De Marco O., 2013, A\&A, 553A, 126

\bibitem[\protect\citeauthoryear{Jones}{2014}]{Jon14}
Jones D., Santander-Garcia M., Boffin H. M. J., Miszalski B., Corradi R L. M., 
2014, in Asymmetrical Planetary Nebulae VI Conference, 43.		

\bibitem[\protect\citeauthoryear{Henry}{2004}]{Hen04}		
Henry R. B. C., Kwitter K. B., Balick B., 2004, AJ, 127, 2284

\bibitem[\protect\citeauthoryear{Hora}{1990}]{Hor90}		
Hora J. L., Hoffmann W. F., Deutsch L. K., Fazio G. G., 1990, ApJ, 353, 549

\bibitem[\protect\citeauthoryear{Kingsburgh}{1994}]{King94}	
Kingsburgh R. L., Barlow M. J., 1994, MNRAS, 271, 257

\bibitem[\protect\citeauthoryear{Kaler}{1970}]{Kal70}	
Kaler J. B., 1970, ApJ, 160, 887 

\bibitem[\protect\citeauthoryear{Kwitter}{2001}]{Kwit01}	
Kwitter K. B., Henry R. B. C.,	2001, ApJ, 562, 804

\bibitem[\protect\citeauthoryear{Kwok}{1978}]{Kwok78}
Kwok, S., Purton, C. R., Fitzgerald, P. M. 1978, ApJ, 219, L125

\bibitem[\protect\citeauthoryear{Leal-Ferreira}{2011}]{}
Leal-Ferreira M. L., Gon\c calves D. R., Monteiro H., Richards J. W.,
2011, MNRAS, 411, 1395

\bibitem[\protect\citeauthoryear{Liu}{2004}]{Liu04}
Liu Y., Liu X.-W., Luo S.-G., Barlow M. J., 2004, MNRAS, 353, 1231

\bibitem[\protect\citeauthoryear{Liu}{2004}]{Liu04}
Liu X.-W., Storey P. J., Barlow M. J., Danziger I. J., Cohen M., Bryce M., 
2000, MNRAS, 312, 585.

\bibitem[\protect\citeauthoryear{Lopez}{1995}]{Lop95}	
L\'opez J. A., V\'azquez R., Rodr\'iguez L. F., 1995, ApJ, 455, L63

\bibitem[\protect\citeauthoryear{Lopez}{2012}]{Lop12}	
L\'opez J. A., Richer M. G., Garc\'ia--D\'iaz Ma. T., Clark D. M., Meaburn J., 
Riesgo H., Steffen W., Lloyd, M., 2012, RevMexAA, 48, 3

\bibitem[\protect\citeauthoryear{Manchado}{2015}]{Man15}
Manchado A., Stanghellini L., Villaver E., Garc\'ia-Segura G., Shaw R. A., 
Garc\'ia-Hern\'andez D. A., 2015, ApJ, 808, 115

\bibitem[\protect\citeauthoryear{Manchado}{1996}]{Man96}
Manchado, A., Guerrero, M. A., Stanghellini, L., Serra-Ricart, M. 1996,
The IAC Morphological Catalog of Northern Galactic PNs (La Laguna : IAC)

\bibitem[\protect\citeauthoryear{Marcolino}{2003}]{Mar03}
Marcolino W. L. F., de Ara\'ujo F. X., 2003, AJ, 126, 887

\bibitem[\protect\citeauthoryear{Marcolino}{2007}]{Mar07}
Marcolino W. L. F., de Ara\'ujo F. X., Junior H. B. M., Duarte E. S., 2007, AJ, 134, 1380

\bibitem[\protect\citeauthoryear{McCarthy}{1990}]{McC90}
McCarthy J. K.m Mould J. R., Mendez R. H., Kudritzki R. P., Husfeld D., Herrero A., Groth H. G., 1990, ApJ, 351, 230

\bibitem[\protect\citeauthoryear{Meaburn}{1992}]{Mea92}
Meaburn J., Walsh J. R., Clegg R. E. S., Walton N. A., Taylor D., Berry
D. S., 1992, MNRAS, 255, 177

\bibitem[\protect\citeauthoryear{Mellema}{1991}]{Mell91}
Mellema G., Eulderink F., Icke V. 1991, A\&A, 252, 718

\bibitem[\protect\citeauthoryear{Mellema}{1995}]{Mell95}
Mellema G., 1995, MNRAS, 277, 173

\bibitem[\protect\citeauthoryear{Mellema}{1998}]{Mell98}
Mellema G., Raga A. C., Canto J., Lundqvist P., Balick B., 
Steffen W., Noriega-Crespo, A., 1998, A\&A, 331, 335

\bibitem[\protect\citeauthoryear{Mellema}{2004}]{Mell04}
Mellema G., 2004, A\&A, 416, 623

\bibitem[\protect\citeauthoryear{Mendez}{1990}]{Men90}
Mendez R. H., Herrero A., Manchado A., 1990, A\&A, 229, 152

\bibitem[\protect\citeauthoryear{Mendez}{1992}]{Men92}
Mendez R. H., Kudritzki R. P., Herrero A., 1992, A\&A, 260, 329

\bibitem[\protect\citeauthoryear{Mendez}{1998}]{Men98}	
Mendez R. H., Kudritzki R. P., Herrero, A., Husfeld D., Groth H. G., 1988, A\&A, 190, 113

\bibitem[\protect\citeauthoryear{Miranda}{1999}]{Mir99}
Miranda L. F., V\'azquez R., Corradi R L. M., Guerrero M., A.; L\'opez J. A., Torrelles J. M., 1999, ApJ, 520, 714

\bibitem[\protect\citeauthoryear{Miranda}{2001}]{Mir01}	
Miranda L. F., Guerrero M. A., Torrelles J. M., 2001, MNRAS, 322, 195

\bibitem[\protect\citeauthoryear{Miszalski}{2009}]{Mis09a}	
Miszalski B., Acker A., Parker Q. A., Moffat A. F. J, 2009a, A\&A, 505, 249

\bibitem[\protect\citeauthoryear{Miszalski}{2009}]{Mis09b}
Miszalski B., Acker A., Moffat A. F. J., Parker Q. A., Udalski A., 2009b, A\&A, 496, 813

\bibitem[\protect\citeauthoryear{Miszalski}{2011}]{Mis11}
Miszalski B., Jone, D., Rodr\'\i guez-Gil P., Boffin H. M. J., Corradi R. L. M.,  Santander-Garc\'\i a M. 2011, A\&A, 531, 158 

\bibitem[\protect\citeauthoryear{Monteiro}{2013}]{Mon}
Monteiro H., Gon\c calves D. R., Leal-Ferreira M. L., Corradi R. L. M., 2013, A\&A, 560, 102

\bibitem[\protect\citeauthoryear{Nordhaus}{2006}]{Nor06}
Nordhaus J., Blackman, E. G., 2006, MNRAS, 370, 2004

\bibitem[\protect\citeauthoryear{ODell}{1996}]{ODell96}
O’Dell C. R., Handron K. D., 1996, AJ, 111, 1630

\bibitem[\protect\citeauthoryear{Palen}{2002}]{Pal02}	
Palen S., Balick B., Hajian A. R., Terzian Y., Bond H., E., Panagia N., 2002, AJ, 123, 2666

\bibitem[\protect\citeauthoryear{Parker}{2006}]{Par06}	
Parker  Q. A., Acker A., Frew, D. J., et al., 2006, MNRAS, 373, 79

\bibitem[\protect\citeauthoryear{Parthasarathy}{1998}]{Par88}	
Parthasarathy M., Acker A., Stenholm B, 1998, A\&A, 329L, 9

\bibitem[\protect\citeauthoryear{Per}{2013}]{Per13}		
Pereyra M., Richer M. G., L\'opez J. A., 2013, ApJ, 771, 114

\bibitem[\protect\citeauthoryear{Per}{2000}]{Per00}		
Perinotto M., 2000, Ap\&SS, 274, 205

\bibitem[\protect\citeauthoryear{Per}{2004}]{Per04}	
Perinotto M., Morbidelli L., Scatarzi A., 2004, MNRAS, 349, 793

\bibitem[\protect\citeauthoryear{Phillips}{2004}]{Phip4}	
Phillips, J. P., 2004, MNRAS, 353, 589

\bibitem[\protect\citeauthoryear{Phillips}{2005}]{Phi05}	
Phillips, J. P., 2005, MNRAS, 362, 847

\bibitem[\protect\citeauthoryear{Raga}{2008}]{Rag08}	
Raga A. C., Riera A., Mellema G., Esquivel A., Vel\'azquez P. F., 2008, A\&A, 489, 1141

\bibitem[\protect\citeauthoryear{Rauch}{1997}]{Rau97}	
Rauch T., Werner K. 1997, in The Third Conference on Faint Blue
Stars, ed. A. G. D. Philip, J. Liebert, R. A. Saffer (Schenectady, NY: L.Davis Press), 217

\bibitem[\protect\citeauthoryear{Riesgo}{2006}]{Rie06}	
Riesgo H., L\'opez J. A., 2006, RMxAA, 42, 47

\bibitem[\protect\citeauthoryear{Sabbadin}{1977}]{Sab77}	
Sabbadin F., Minello S., Bianchini A. 1977, A\&A, 60, 147

\bibitem[\protect\citeauthoryear{Sabin}{2014}]{Sab14}	
Sabin L., Parker Q. A., Corradi R. L. M., et al. 2014, 2014, MNRAS, 443, 3388

\bibitem[\protect\citeauthoryear{Sahai}{2011}]{Sah11}	
Sahai R., Morris M. R., Villar G. G., 2011, AJ, 141, 134

\bibitem[\protect\citeauthoryear{Schwarz}{1993}]{Sch93}
Schwarz H. E., Corradi R. L. M., Stanghellini  L. 1993, in IAU Symp. 
155, Planetary Nebulae, ed. R. Weinberger \& A. Acker (Dordrecht: Kluwer), 214

\bibitem[\protect\citeauthoryear{Shaw}{1989}]{Shaw89}		
Shaw R. A., Kaler J. B., 1989, ApJS, 69, 495 

\bibitem[\protect\citeauthoryear{Shaw}{1995}]{Shaw95}	
Shaw R. A., Dufour R. J., 1995, PASP, 107, 896

\bibitem[\protect\citeauthoryear{Shaw}{2012}]{Shaw12}	
Shaw R. A. 2012, in IAU Symp. 283, Planetary Nebulae: An Eye to the Future,
ed. Manchado A., Stanghellini L., Sch\:onberner D., (Cambridge:Cambridge
Univ. Press), 156

\bibitem[\protect\citeauthoryear{Soker}{1994}]{Sok94}	
Soker N., Livio M., 1994, ApJ, 421, 219

\bibitem[\protect\citeauthoryear{Steffen}{2001}]{Stef01}
Steffen W., L\'opez J. A., Lim A., 2001, ApJ, 556, 823

\bibitem[\protect\citeauthoryear{Tcknell}{2014}]{Toc14}
Tocknell J., De Marco O., Wardle M., 2014, MNRAS, 439, 2014

\bibitem[\protect\citeauthoryear{Tylenda}{1992}]{Tyl92}
Tylenda R., Acker A., Stenholm B., Koeppen J., 1992, A\&AS, 95, 337

\bibitem[\protect\citeauthoryear{Tylenda}{1993}]{Tyl93}
Tylenda R., Acker A., Stenholm B., 1993, A\&AS, 102, 595

\bibitem[\protect\citeauthoryear{Wang}{2004}]{Wang04}
Wang W., Liu X.-W., Zhang Y., Barlow M. J., 2004, A\&A, 427, 873

\bibitem[\protect\citeauthoryear{Wesson}{2005}]{Wes05}
Wesson R., Liu X.-W., Barlow M. J., 2005, MNRAS, 362, 424

\end{thebibliography}

\newpage

\begin{table*}
\label{table2}
\begin{center}
\caption{IC~4846} 
{\tiny
\begin{tabular}{lcccc}
\hline
\hline
\noalign{\smallskip}
Line ID & \multicolumn{1}{c}{Inner NEB} & \multicolumn{1}{c}{NE jet} & 
\multicolumn{1}{c}{SW jet} & \multicolumn{1}{c}{Entire NEB}\\
        & \multicolumn{1}{c}{3.5\arcsec (5)$^a$} & \multicolumn{1}{c}{2.1\arcsec (3)} & 
\multicolumn{1}{c}{2.1\arcsec (3)} & \multicolumn{1}{c}{10\arcsec (15)}\\
\noalign{\smallskip}
\hline
\oxygenii  \ 3727  & 20.45 & 100.2 & 116.5 & 26.78\\ 
H12        \ 3750  & 2.421 &  $-$  & $-$   & 2.682 \\
H11        \ 3770  & 2.994 &  $-$  & $-$   & 3.364 \\
H10        \ 3798  & 4.415 &  $-$  & $-$   & 4.723 \\
\helium    \ 3819  & 0.933 &  $-$  & $-$   & 1.216\\
H9         \ 3835  & 5.792 &  $-$  & $-$   & 6.461 \\
\neon      \ 3869  & 66.91 & 84.11 & 102.9 & 76.43 \\
\neonHI    \ 3968  & 36.53 & 40.65 & 56.82 & 40.52 \\
\helium    \ 4026  & 2.086 & 2.986 & 3.812 & 2.403 \\
\sulfurt   \ 4070  & 1.553 & 3.014 & 2.191 & 1.655 \\
\hdelta    \ 4101  & 24.41 & 29.33 & 32.56 & 27.51 \\
\hc        \ 4340  & 46.83 & 50.58 & 52.54 & 52.34 \\
\oxygeniii \ 4363  & 7.274 & 7.952 & 7.768 & 7.866 \\
\helium    \ 4471  & 4.152 & 6.811 & 6.203 & 4.885 \\
\heliumb   \ 4686  & 0.461 &  $-$  &  $-$  & 0.526 \\
\helium    \ 4712$^{\alpha}$  & 0.606 & $-$ & 0.905 & 0.713\\ 
\argon     \ 4712  & 1.035 &  $-$  & 3.026 & 1.111 \\
\argon     \ 4740  & 1.349 &  $-$  & 2.587 & 1.453 \\
\hbeta     \ 4861  & 100.0 & 100.0 & 100.0 & 100.0 \\
\helium    \ 4924  & 1.198 &  $-$  &  $-$  & 1.267 \\
\oxygeniii \ 4958  & 378.2 & 349.5 & 337.5 & 421.1 \\
\oxygeniii \ 5007  & 1150  & 1116  & 982.9 & 1257  \\
\nitrogena \ 5200  & 0.173 &  $-$  &  $-$  & 0.201 \\
\ironiii   \ 5270  & 0.129 &  $-$  &  $-$  & 0.174 \\
\chloro    \ 5517  & 0.333 & 0.852 & 1.328 & 0.377 \\
\chloro    \ 5537  & 0.466 & 0.771 & 1.227 & 0.546 \\
\nitrogen  \ 5755  & 0.425 & 1.282 & 1.089 & 0.459 \\
\helium    \ 5876  & 14.19 & 15.59 & 14.14 & 14.18 \\
\kripto    \ 6101  & 0.093 &  $-$  &  $-$  & 0.108 \\
\oxygeni   \ 6300  & 1.676 & 6.549 & 5.691 & 1.993 \\
\sulfur    \ 6312  & 1.141 & 1.512 & 1.646 & 1.228 \\
\oxygeni   \ 6363  & 0.582 & 2.178 & 2.238 & 0.664 \\
\nitrogen  \ 6548  & 5.159 & 20.79 & 17.59 & 6.487 \\
\ha        \ 6564  & 285.0 & 285.0 & 285.0 & 285.0 \\
\nitrogen  \ 6584  & 16.54 & 51.45 & 45.97 & 19.13 \\
\helium    \ 6678  & 3.142 & 3.546 & 3.769 & 3.089 \\
\sulfurt   \ 6716  & 1.128 & 5.433 & 4.996 & 1.378 \\
\sulfurt   \ 6731  & 2.069 & 7.506 & 7.052 & 2.579\\ 
\\
log[$\it{F}$(\hbeta)      & --11.62& --13.49& --13.45& --11.54\\
$\it{c}$(\hbeta)          &  0.48$\pm$0.03 &  0.50$\pm$0.05 & 0.53$\pm$0.05 &  0.52$\pm$0.03 \\
\hline
Line Fluxes & \multicolumn{4}{c}{Percentage errors in line fluxes} \\
(0.001-0.01)\it{F}$_{{\rm H}\beta}$  & 13& 43 & 43& 12\\
(0.01-0.05)\it{F}$_{{\rm H}\beta}$   & 10& 27 & 26& 10\\
(0.05-0.15)\it{F}$_{{\rm H}\beta}$   &  9& 12 & 12&  8\\
(0.15-0.30)\it{F}$_{{\rm H}\beta}$   &  8& 10 & 10&  7\\
(0.30-2.0)\it{F}$_{{\rm H}\beta}$    &  7&  9 &  9&  7\\
(2.0-5.0)\it{F}$_{{\rm H}\beta}$     &  7&  8 &  7&  7\\
(5.0-10.0)\it{F}$_{{\rm H}\beta}$    &  6&  7 &  7&  6\\
$>$ 10 \it{F}$_{{\rm H}\beta}$       &  6 & 7 &  7&  6\\
\hline
& \multicolumn{4}{c}{Electron Densities (cm$^{-3}$) and Temperatures (K)} \\
$\it{N}_{\rm{e}}$\sulfurt   &  7200$\pm$1050 &  2150$\pm$400   &  2350$\pm$450 &  8050$\pm$1050  \\
$\it{N}_{\rm{e}}$\argon     &  8550$\pm$1400 &      $-$        &  2200$\pm$800 &  9100$\pm$1500 \\
$\it{N}_{\rm{e}}$\chloro    &  7400$\pm$1350 &  1650$\pm$950   &  1850$\pm$700 &  8050$\pm$1500 \\
$\it{T}_{\rm{e}}$\oxygeniii &  9900$\pm$1400 & 10350$\pm$1800  &11100$\pm$1950 &  9850$\pm$1400  \\
$\it{T}_{\rm{e}}$\nitrogen  & 11950$\pm$2250 & 12800$\pm$4150  &12450$\pm$4000 & 11400$\pm$2150 \\
$\it{T}_{\rm{e}}$\sulfurt   & 12700$\pm$2000 & 13700$\pm$4300  &13250$\pm$4150 &  9500$\pm$1550 \\ 
\noalign{\smallskip}
\hline
\end{tabular}
\begin{flushleft}
$^a$ The size of the component in arcsec. The number of pixels used to extract  
the spectra is indicated in the parentheses for each component.\\
\end{flushleft}
}
\end{center}
\end{table*}

\begin{table*}
\label{table3}
\begin{center}
\caption[]{Ionic and total abundances of IC~4846}
{\tiny
\label{table5}
\begin{tabular}{lcllllllllll}
\hline
Ionic abundances             & $\it{T}_{\rm{e}}$ \& $\it{N}_{\rm{e}}^a$ & Inner NEB              & NE jet           &  SW jet          & Entire NEB  \\
\hline
$\rm{He^{+}/H^{+}}$                  & 1        & 0.077$\pm$0.009            & 0.091$\pm$0.014             & 0.088$\pm$0.013          & 0.078$\pm$0.009\\
$\rm{He^{++}/H^{+}}$                 & 2        & 0.0004$\pm$0.0001          & $-$                         & $-$                      & 0.0004$\pm$0.0001\\
ICF(He)                            &           & 1.00                       & 1.00                       & 1.00                      & 1.00            \\
\textbf{$\rm{He/H}$ }              &           & 0.077$\pm$0.009           & 0.091$\pm$0.014              &0.088$\pm$0.013            & 0.078$\pm$0.009 \\ 
$10^5 \times \rm{O^{+}/H^{+}}$       & 1        & 0.78$\pm$0.08              & 2.07$\pm$0.28              & 2.55$\pm$0.33              & 1.31$\pm$0.13  \\
$10^5 \times \rm{O^{2+}/H^{+}}$      & 2        & 40.7$\pm$3.8               & 34.4$\pm$3.9               & 23.9$\pm$2.8               & 45.4$\pm$4.2 \\
ICF(O)                             &          & 1.01$\pm$0.01              & 1.00                        & 1.00                       & 1.01$\pm$0.01 \\
\hline
\end{tabular}
\begin{flushleft}
$^a$ 1: $\it{T}_{\rm{e}}$[N II] and $\it{N}_{\rm{e}}$[S II], 2: $\it{T}_{\rm{e}}$[O III] and $\it{N}_{\rm{e}}$[S II]
\end{flushleft}
}
\end{center}
\end{table*}

\begin{table*}
\begin{center}
\caption{Wray~17-1 (PA=155\degree)} 
{\tiny
\begin{tabular}{lcccccc}
\hline
\hline
\noalign{\smallskip}
Line ID & \multicolumn{1}{c}{NW filament} & \multicolumn{1}{c}{NW knot} &
\multicolumn{1}{c}{NW inner NEB} & \multicolumn{1}{c}{SE filament} & 
\multicolumn{1}{c}{SE knot} & \multicolumn{1}{c}{SE inner NEB}\\
        & \multicolumn{1}{c}{11.6\arcsec (29)} & \multicolumn{1}{c}{4.4\arcsec (11)} &
\multicolumn{1}{c}{11.6\arcsec (29)} & \multicolumn{1}{c}{10.8\arcsec (27)} & 
\multicolumn{1}{c}{3.6\arcsec (9)} & \multicolumn{1}{c}{11.6\arcsec (29)}\\
\noalign{\smallskip}
\hline
\oxygenii  \ 3727  & 611.3 & 364.2 &  $-$  & 636.1 & 309.3 &  $-$  \\
\neon      \ 3869  & 137.7 & 107.7 &  $-$  & 160.8 & 112.5 &  $-$  \\
\neonHI    \ 3968  & 54.22 &  $-$  &  $-$  & 65.39 & 43.71 &  $-$  \\
\hdelta    \ 4101  & 25.57 &  $-$  &  $-$  & 25.57 &  $-$  &  $-$  \\
\hc        \ 4340  & 46.09 & 48.56 & 45.31 & 49.85 &  $-$ & 44.21 \\
\oxygeniii \ 4363  & 14.18 & 10.48 & 6.044 & 16.37 &  $-$ & 8.292 \\
\hline
\end{tabular}
\begin{flushleft}
$^a$ $\it{N}_{\rm{e}}$[Ar IV] is considered as a lower limit since the contribution of \helium\ 4712 has not been subtracted.
\end{flushleft}
}
\end{center}
\end{table*}

\begin{table*}
\begin{center}
\caption[]{Ionic and total abundances of Wray~17-1 (PA=155\degree)}
\label{table5}
{\tiny
\begin{tabular}{lcllllllllll}
\hline
Ionic abundances& $\it{T}_{\rm{e}}$ \& $\it{N}_{\rm{e}}^a$& NW filament     & NW knot        & NW inner NEB     & SE filament      &SE knot$^b$       & SE inner NEB  \\
\hline
$\rm{He^{+}/H^{+}}$             & 1   &0.068$\pm$0.014  & $-$              &  $-$          &0.071$\pm$0.014    & $-$              & $-$          \\
$\rm{He^{++}/H^{+}}$            & 2   &0.046$\pm$0.005  &0.056$\pm$0.013   &0.098$\pm$0.02 &0.047$\pm$0.006    &0.053$\pm$0.015   &0.097$\pm$0.021\\
ICF(He)                       &     & 1.000           & 1.000             & 1.000         & 1.000             & 1.000            & 1.000          \\
$\rm{He/H}$                   &     & 0.114$\pm$0.019 & 0.056$\pm$0.013  & 0.098$\pm$0.02&0.118$\pm$0.020     & 0.053$\pm$0.015  & 0.097         \\ 
$10^5 \times \rm{O^{+}/H^{+}}$  & 1   & 12.5$\pm$1.3    & 6.61$\pm$1.61    & $-$             & 11.6$\pm$1.3    &5.98$\pm$1.57      & $-$        \\
$10^5 \times \rm{O^{2+}/H^{+}}$ & 2   & 20.7$\pm$2.1    & 17.7$\pm$4.1     & 4.21$\pm$0.82   & 24.2$\pm$2.6    &23.4$\pm$5.6       &5.35$\pm$0.99\\
ICF(O)                        &     & 1.41$\pm$0.09   & $-$              & $-$             & 1.40$\pm$0.09   & $-$                & 1.00       \\
\hline
\end{tabular}
\begin{flushleft}
$^a$ 1: $\it{T}_{\rm{e}}$[N II] and $\it{N}_{\rm{e}}$[S II], 2: $\it{T}_{\rm{e}}$[O III] and $\it{N}_{\rm{e}}$[N II]\\
$^b$ $\it{T}_{\rm{e}}$ and $\it{N}_{\rm{e}}$ are drawn from the NW knot. \\
$^c$ It is calculated by using the equation S$^{+2}$=S$^{+}$*(4.677+(O$^{+2}$/O)$^{0.433}$) from Kingsburgh \& Barlow (1994).
\end{flushleft}
}
\end{center}
\end{table*}

\begin{table*}
\begin{center}
 \caption{Wray~17-1 (PA=75\degree)} 
{\tiny
\begin{tabular}{lcccccccc}
\hline
\hline
\noalign{\smallskip}
Line ID & \multicolumn{1}{c}{E knot} & \multicolumn{1}{c}{E arc} & \multicolumn{1}{c}{E inner NEB} 
& \multicolumn{1}{c}{W knot} & \multicolumn{1}{c}{W arc} & \multicolumn{1}{c}{W inner NEB}\\
        & \multicolumn{1}{c}{2.8\arcsec (7)} & \multicolumn{1}{c}{14.8\arcsec (37)} & \multicolumn{1}{c}{11.6\arcsec(29)} 
& \multicolumn{1}{c}{3.6\arcsec (9)} & \multicolumn{1}{c}{6\arcsec (15)} & \multicolumn{1}{c}{11.6\arcsec (29)}\\
\noalign{\smallskip}
\hline
\oxygenii  \ 3727  &  237.4 & 57.55 &  $-$  & 253.3 & 38.46 & $-$\\
\neon      \ 3869  &  66.42 & 52.69 &  $-$  & 113.9 & 58.18 & $-$\\
\neonHI    \ 3968  &  28.35 & 31.61 &  $-$  & 47.23 & 21.86 & $-$\\
\hdelta    \ 4101  &  24.98 & 22.07 &  $-$  & 26.67 & 23.69 & $-$ \\
\hc        \ 4340  &  46.89 & 46.45 & 44.89 & 45.39 & 47.89 & 48.79\\
\oxygeniii \ 4363  &  9.577 & 9.244 &  $-$  & 9.039 & 12.83 & 4.863 \\
\hline
\end{tabular}
\begin{flushleft}
$^a$ $\it{N}_{\rm{e}}$[Ar IV] is considered as a lower limit since the contribution of \helium\ 4712 has not been subtracted.
\end{flushleft}
}
\end{center}
\end{table*}

\begin{table*}
\begin{center}
\caption[]{Ionic and total abundances of Wray~17-1 (PA=75\degree)}
{\tiny
\label{table5}
\begin{tabular}{lcllllllllll}
\hline
Ionic abundances& $\it{T}_{\rm{e}}$ \& $\it{N}_{\rm{e}}^a$ & E knot      & E arc        & E inner NEB$^b$     & W knot       & W arc       & W inner NEB  \\
\hline
$\rm{He^{+}/H^{+}}$             & 1   &0.047$\pm$0.020 &  $-$              &  $-$             &0.029$\pm$0.011 &0.032$\pm$0.012      &$-$          \\
$\rm{He^{++}/H^{+}}$            & 2   &0.083$\pm$0.020 &0.089$\pm$0.018    &0.090$\pm$0.017   &0.084$\pm$0.014 &0.085$\pm$0.013      &0.097$\pm$0.017\\
ICF(He)                       &     &1.000           & 1.000             & 1.000            & 1.000           & 1.000               & 1.000     \\
$\rm{He/H}$                   &     &0.130$\pm$0.040 & 0.089$\pm$0.018   & 0.090$\pm$0.017  &0.113$\pm$0.025  & 0.117$\pm$0.025      & 0.097$\pm$0.017      \\ 
$10^5 \times \rm{O^{+}/H^{+}}$  & 1   & 4.34$\pm$0.97  &0.99$\pm$0.20      & $-$               & 4.52$\pm$0.68   &0.45$\pm$0.08       & $-$        \\
$10^5 \times \rm{O^{2+}/H^{+}}$ & 2   & 6.01$\pm$1.19  &16.7$\pm$2.8       &2.81$\pm$0.55      & 13.6$\pm$1.9    & 14.2$\pm$1.9       &2.86$\pm$0.51\\
ICF(O)                        &     & 1.97$\pm$0.41  & $-$               & $-$               & 2.47$\pm$0.51   & 2.37$\pm$0.47      & $-$       \\
\hline
\end{tabular}
\begin{flushleft}
$^a$ 1:$\it{T}_{\rm{e}}$[N II] and $\it{N}_{\rm{e}}$[S II], 2: $\it{T}_{\rm{e}}$[O III] and $\it{N}_{\rm{e}}$[N II]\\
$^b$ $\it{T}_{\rm{e}}$ is drawn from the W nebula\\
$^c$ It is calculated by using the equation S$^{+2}$=S$^{+}$*(4.677+(O$^{+2}$/O)$^{0.433}$) from Kingsburgh \& Barlow (1994).
\end{flushleft}
}
\end{center}
\end{table*}

\begin{table*}
\begin{center}
\caption{K~1-2} 
{\tiny
\begin{tabular}{lcccccc}
\hline
\hline
\noalign{\smallskip}
Line ID & \multicolumn{1}{c}{NW filament} & \multicolumn{1}{c}{NW knot} & \multicolumn{1}{c}{NW filament+Knot} &  
\multicolumn{1}{c}{SE knot} & \multicolumn{1}{c}{NW NEB}\\
        & \multicolumn{1}{c}{7.6\arcsec (19)} & \multicolumn{1}{c}{3.6\arcsec (9)} & \multicolumn{1}{c}{11.6\arcsec (29)} &  
\multicolumn{1}{c}{3.6\arcsec (9)} & \multicolumn{1}{c}{35.6\arcsec (89)}\\
\noalign{\smallskip}
\hline
\oxygenii  \ 3727  & 272.1 & 193.1 & 245.2 & 265.6 & 205.3 \\
\neon      \ 3869  & 117.1 & 41.47 & 94.17 &  $-$  & 83.57 \\
\neonHI    \ 3968  & 57.85 & 20.94 & 48.83 &  $-$  & 38.51 \\
\hdelta    \ 4101  & 23.63 & 26.61 & 24.31 &  $-$  & 23.93 \\
\hc        \ 4340  & 45.86 & 50.61 & 45.82 & 49.21 & 48.55 \\
\oxygeniii \ 4363  & 16.87 & 7.828 & 12.95 & 7.118 & 13.94 \\
\hline
\hline
\end{tabular}
}
\end{center}
\end{table*}

\begin{table*}
\begin{center}
\caption[]{Ionic and total abundances of K~1-2}
{\tiny
\label{table5}
\begin{tabular}{lcllllllllll}
\hline
Ionic abundances       & $\it{T}_{\rm{e}}$ \& $\it{N}_{\rm{e}}^a$ & NW filament            & NW knot          & NW filament+Knot &  SE knot        & NW NEB  \\
\hline
$\rm{He^{+}/H^{+}}$                & 1          & 0.068$\pm$0.023  &  $-$             & 0.054$\pm$0.017  &   $-$            & 0.049$\pm$0.015\\
$\rm{He^{++}/H^{+}}$               & 2          & 0.068$\pm$0.011  &  0.082$\pm$0.014 & 0.067$\pm$0.010  & 0.092$\pm$0.020  & 0.077$\pm$0.011\\
ICF(He)                          &            & 1.000             &  1.00            & 1.000            & 1.000            & 1.000   \\
$\rm{He/H}$                      &            & 0.136$\pm$0.034   & 0.082$\pm$0.014  & 0.121$\pm$0.027  & 0.092$\pm$0.020  &  0.126$\pm$0.026\\ 
$10^5 \times \rm{O^{+}/H^{+}}$    & 1          & 16.8$\pm$2.4     & 2.35$\pm$0.46      & 17.9$\pm$2.5     & 2.81 $\pm$0.43     & 9.25$\pm$1.19 \\
$10^5 \times \rm{O^{2+}/H^{+}}$   & 2          & 9.85$\pm$1.34    & 5.83$\pm$1.02      & 10.5$\pm$1.4     & 5.07 $\pm$0.96     & 9.03$\pm$1.09 \\
ICF(O)                          &            & 1.59$\pm$0.20    &$-$                 & 1.71$\pm$0.22    &  $-$                & 1.88$\pm$0.26   \\
\hline
\end{tabular}
\medskip{}
\begin{flushleft}
$^a$ 1: $\it{T}_{\rm{e}}$[N II] and $\it{N}_{\rm{e}}$[S II], 2: $\it{T}_{\rm{e}}$[O III] and $\it{N}_{\rm{e}}$[S II]\\
\end{flushleft}
}
\end{center}
\end{table*}

\begin{table*}
\begin{center}
\caption{NGC~6891 (PA=135\degree)} 
{\tiny
\begin{tabular}{lcccccc}
\hline
\hline
\noalign{\smallskip}
Line ID & \multicolumn{1}{c}{Inner NEB} & \multicolumn{1}{c}{NW knot} &
\multicolumn{1}{c}{SE knot} & \multicolumn{1}{c}{NW outer NEB} & \multicolumn{1}{c}{SE outer NEB} 
& \multicolumn{1}{c}{Entire NEB}\\ 
        & \multicolumn{1}{c}{4.9\arcsec (7)} & \multicolumn{1}{c}{2.1\arcsec (3)} &
\multicolumn{1}{c}{2.1\arcsec (3)} & \multicolumn{1}{c}{4.9\arcsec (7)} & \multicolumn{1}{c}{4.9\arcsec (7)} 
& \multicolumn{1}{c}{20.3\arcsec (29)}\\
\noalign{\smallskip}
\hline
\oxygenii  \ 3727  & 12.89 & 34.87 & 24.18 &  22.59  & 18.15 & 16.55\\
H12        \ 3750  & $-$   & $-$   & $-$   &  3.345  & 2.239 & 2.266\\
H11        \ 3770  & $-$   & $-$   & $-$   &  4.037  & 3.069 & 2.978 \\
H10        \ 3798  & $-$   & $-$   & $-$   &  5.341  & 4.001 & 3.381\\
\helium    \ 3819  & $-$   & $-$   & $-$   &  1.191  & 1.161 & 1.489\\
H9         \ 3835  & $-$   & $-$   & $-$   &  7.707  & 5.777 & 4.999\\
\hline
\hline
\end{tabular}
\medskip{}
\begin{flushleft}
\end{flushleft}
}
\end{center}
\end{table*}

\begin{table*}
\begin{center}
\caption[]{Ionic and total abundances of NGC~6891 (PA=135\degree)}
{\tiny
\label{table5}
\begin{tabular}{lcllllll}
\hline
Ionic abundances&$\it{T}_{\rm{e}}$ \& $\it{N}_{\rm{e}}^a$ & Inner NEB           & NW knot       & SE knot       & NW outer NEB        & SE outer NEB        & Entire NEB \\    
\hline
$\rm{He^{+}/H^{+}}$           & 1  &0.116$\pm$0.018      &0.104$\pm$0.040  &0.129$\pm$0.050  &0.103$\pm$0.016  &0.114$\pm$0.018  &0.107$\pm$0.015\\  
$\rm{He^{++}/H^{+}}$          & 2  &0.0015$\pm$0.0003    &  $-$            & $-$             & $-$             &  $-$            &0.0005$\pm$0.0002\\ 
ICF(He)                     &    & 1.000               & 1.000            & 1.000           & 1.000           &  1.000          & 1.000\\
$\rm{He/H}$                 &    & 0.118$\pm$0.018    & 0.104$\pm$0.040  & 0.129$\pm$0.050 & 0.103$\pm$0.016 & 0.114$\pm$0.018 & 0.108$\pm$0.015\\
$10^5 \times \rm{O^{+}/H^{+}}$ & 1   & 0.59$\pm$0.09     & 1.79$\pm$0.54   & 1.32$\pm$0.41   &1.09$\pm$0.18   &1.34$\pm$0.21       & 0.31$\pm$0.05\\
$10^5 \times \rm{O^{2+}/H^{+}}$& 2   & 30.3$\pm$3.1      & 31.9$\pm$7.8    & 37.1$\pm$9.1    & 29.3$\pm$3.5   &35.8$\pm$4.3       &34.1$\pm$3.7\\
ICF(O)                        &     & 1.01$\pm$0.00    & 1.00            & 1.00            & 1.00            & 1.00              & 1.00$\pm$0.00 \\
\hline
\end{tabular}
\medskip{}
\begin{flushleft}
$^a$ 1: $\it{T}_{\rm{e}}$[N II] and $\it{N}_{\rm{e}}$[S II], 2: $\it{T}_{\rm{e}}$[O III] and $\it{N}_{\rm{e}}$[S II].\\
$^c$ It is calculated by using the equation S$^{+2}$=S$^{+}$*(4.677+(O$^{+2}$/O)$^{0.433}$) from Kingsburgh \& Barlow (1994).
\end{flushleft}
}
\end{center}
\end{table*}

\begin{table*}
\begin{center}
\caption{NGC~6891 (PA=45\degree)} 
{\tiny
\begin{tabular}{lcccc}
\hline
\hline
\noalign{\smallskip}
Line ID & \multicolumn{1}{c}{Inner NEB} & \multicolumn{1}{c}{NE NEB} &
\multicolumn{1}{c}{SW NEB} & \multicolumn{1}{c}{Entire NEB}\\
        & \multicolumn{1}{c}{4.9\arcsec (7)} & \multicolumn{1}{c}{6.3\arcsec (9)} &
\multicolumn{1}{c}{6.3\arcsec (9)} & \multicolumn{1}{c}{20.3\arcsec (29)}\\
\noalign{\smallskip}
\hline
\oxygenii  \ 3727  & 11.31 & 15.92 & 17.08 & 12.77\\
H12        \ 3750  & $-$   & 2.752 & 2.714 & 2.326\\
H11        \ 3770  & $-$   & 3.204 & 3.144 & 2.633\\
H10        \ 3798  & $-$   & 4.656 & 4.713 & 3.099\\
\helium    \ 3819  & $-$   & 1.978 & 1.573 & 1.259\\
H9         \ 3835  & $-$   & 6.902 & 7.545 & 4.285\\
\hline
\hline
\end{tabular}
\medskip{}
\begin{flushleft}
\end{flushleft}
}
\end{center}
\end{table*}

\begin{table*}
\begin{center}
\caption[]{Ionic and total abundances of NGC~6891 (PA=45\degree)}
{\tiny
\label{table5}
\begin{tabular}{lcllllll}
\hline
Ionic abundances&$\it{T}_{\rm{e}}$ \& $\it{N}_{\rm{e}}^a$& Inner NEB            & NE NEB           & SW NEB          & Entire NEB \\    
\hline
$\rm{He^{+}/H^{+}}$           & 1  &0.113$\pm$0.017    &0.104$\pm$0.019   &0.099$\pm$0.018   & 0.109$\pm$0.015\\ 
$\rm{He^{++}/H^{+}}$          & 2  &0.0016$\pm$0.0004  &  $-$             & $-$              &0.0007$\pm$0.0002\\ 
ICF(He)                     &    & 1.000              &  1.000           & 1.000            &  1.000         \\
$\rm{He/H}$                 &    & 0.115$\pm$0.017   & 0.104$\pm$0.019  & 0.099$\pm$0.018  & 0.110$\pm$0.015 \\
$10^5 \times \rm{O^{+}/H^{+}}$ & 1   & 0.63$\pm$0.09    & 1.05$\pm$0.20   & 1.12$\pm$0.21    &0.26$\pm$0.04 \\
$10^5 \times \rm{O^{2+}/H^{+}}$& 2   & 34.9$\pm$3.5     & 39.1$\pm$4.4    & 36.6$\pm$4.1     &36.5$\pm$3.6\\
ICF(O)                        &    & 1.01$\pm$0.00    & 1.00            & 1.00             & 1.00$\pm$0.00    \\
\hline
\end{tabular}
\medskip{}
\begin{flushleft}
$^a$ 1: $\it{T}_{\rm{e}}$[N II] and $\it{N}_{\rm{e}}$[S II], 2: $\it{T}_{\rm{e}}$[O III] and $\it{N}_{\rm{e}}$[S II]
\end{flushleft}
}
\end{center}
\end{table*}

\begin{table*}
\begin{center}
\caption{NGC~6572 (PA=160\degree)} 
{\tiny
\begin{tabular}{lcccc}
\hline
\hline
\noalign{\smallskip}
Line ID & \multicolumn{1}{c}{Inner NEB} & \multicolumn{1}{c}{N knot} &
\multicolumn{1}{c}{S knot} & \multicolumn{1}{c}{Entire NEB}\\
        & \multicolumn{1}{c}{7.7\arcsec (11)} & \multicolumn{1}{c}{2.1\arcsec (3)} &
\multicolumn{1}{c}{2.1\arcsec (3)} & \multicolumn{1}{c}{17.5\arcsec (25)}\\
\noalign{\smallskip}
\hline
\oxygenii  \ 3727  & 27.25 & 127.8 & 143.6 & 34.75\\
H12        \ 3750  & 3.901 & $-$   & $-$   & 3.325\\
H11        \ 3770  & 4.193 & $-$   & $-$   & 3.815\\
H10        \ 3798  & 5.295 & $-$   & $-$   & 5.671\\
\helium    \ 3819  & 1.735 & $-$   & $-$   & 1.771\\
H9         \ 3835  & 7.183 & $-$   & $-$   & 7.059\\
\hline
\hline
\end{tabular}
}
\end{center}
\end{table*}

\begin{table*}
\begin{center}
\caption[]{Ionic and total abundances of NGC~6572 (PA=160\degree)}
{\tiny
\label{table5}
\begin{tabular}{lcllll}
\hline
Ionic abundances& $\it{T}_{\rm{e}}$ \& $\it{N}_{\rm{e}}^a$& Inner NEB      & N knot         & S knot          & Entire NEB \\    
\hline
$\rm{He^{+}/H^{+}}$           & 1  &0.095$\pm$0.008    &0.099$\pm$0.021   &0.105$\pm$0.023  &0.095$\pm$0.008  \\ 
$\rm{He^{++}/H^{+}}$          & 2  &0.0008$\pm$0.0001   &  $-$            & $-$              &0.0008$\pm$0.0001 \\
ICF(He)                      &    & 1.00              &  1.00            & 1.0             & 1.00         \\
$\rm{He/H}$                  &    & 0.096$\pm$0.008  & 0.099$\pm$0.021  & 0.105$\pm$0.029 & 0.096$\pm$0.008 \\
$10^5 \times \rm{O^{+}/H^{+}}$  & 1   & 1.77$\pm$0.13    & 7.02$\pm$1.19   & 10.21$\pm$1.73  & 2.38$\pm$0.20\\
$10^5 \times \rm{O^{2+}/H^{+}}$ & 2   & 38.7$\pm$2.7     & 36.1$\pm$5.4    & 41.7$\pm$6.3    & 40.9$\pm$3.2\\
ICF(O)                        &     & 1.01$\pm$0.11    & 1.00            & 1.00            & 1.01$\pm$0.11 \\
\hline
\end{tabular}
\medskip{}
\begin{flushleft}
$^a$ 1: $\it{T}_{\rm{e}}$[N II] and $\it{N}_{\rm{e}}$[S II], 2: $\it{T}_{\rm{e}}$[O III] and $\it{N}_{\rm{e}}$[Ar IV] and 3: $\it{T}_{\rm{e}}$[O III] and $\it{N}_{\rm{e}}$[Cl III].\\
$^b$ It is calculated by using the equation S$^{+2}$=S$^{+}$*(4.677+(O$^{+2}$/O)$^{0.433}$) from Kingsburgh \& Barlow (1994).
\end{flushleft}
}
\end{center}
\end{table*}

\begin{table*}
\begin{center}
\caption{NGC~6572 (PA=15\degree)} 
{\tiny
\begin{tabular}{lcccc}
\hline
\hline
\noalign{\smallskip}
Line ID & \multicolumn{1}{c}{Inner NEB} & \multicolumn{1}{c}{N knot} &
\multicolumn{1}{c}{S knot} & \multicolumn{1}{c}{Entire NEB}\\
        & \multicolumn{1}{c}{7.7\arcsec (11)} & \multicolumn{1}{c}{2.1\arcsec (3)} &
\multicolumn{1}{c}{2.1\arcsec (3)} & \multicolumn{1}{c}{17.5\arcsec (25)}\\
\noalign{\smallskip}
\hline
\oxygenii  \ 3727  & 27.91 & 139.9 & 101.3 & 34.69\\
H12        \ 3750  & 3.337 & $-$   & $-$   & 2.773\\
H11        \ 3770  & 3.667 & $-$   & $-$   & 3.531\\
H10        \ 3798  & 4.948 & $-$   & $-$   & 4.986\\
\helium    \ 3819  & 1.243 & $-$   & $-$   & 1.208\\
H9         \ 3835  & 7.024 & $-$   & $-$   & 6.933\\
\end{tabular}
}
\end{center}
\end{table*}

\begin{table*}
\begin{center}
\caption[]{Ionic and total abundances of NGC~6572 (PA=15\degree)}
{\tiny
\label{table5}
\begin{tabular}{lcllll}
\hline
Ionic abundances& $\it{T}_{\rm{e}}$ \& $\it{N}_{\rm{e}}^a$& Inner NEB     & N knot         & S knot          & Entire NEB \\    
\hline
$\rm{He^{+}/H^{+}}$           & 1  &0.093$\pm$0.007     &0.098$\pm$0.022  &0.095$\pm$0.022   &0.096$\pm$0.007  \\ 
$\rm{He^{++}/H^{+}}$          & 2  &0.0009$\pm$0.0001   &  $-$            & $-$              &0.0007$\pm$0.0001 \\
ICF(He)                      &    & 1.00              &  1.00            & 1.00            & 1.00         \\
$\rm{He/H}$                  &    & 0.094$\pm$0.007  & 0.098$\pm$0.022  & 0.095$\pm$0.022 & 0.097$\pm$0.007      \\
$10^5 \times \rm{O^{+}/H^{+}}$  & 1   & 1.52$\pm$0.11   & 19.56$\pm$3.59   & 8.59$\pm$1.57    & 2.51$\pm$0.18\\
$10^5 \times \rm{O^{2+}/H^{+}}$ & 2   & 43.2$\pm$3.1    & 34.5$\pm$5.1      & 34.2$\pm$5.1    & 41.9$\pm$3.1\\
ICF(O)                        &     & 1.01$\pm$0.10   & 1.00             & 1.00             & 1.01$\pm$0.10       \\
\hline 
\end{tabular}
\medskip{}
\begin{flushleft}
$^a$ 1: $\it{T}_{\rm{e}}$[N II] and $\it{N}_{\rm{e}}$[S II], 2: $\it{T}_{\rm{e}}$[O III] and $\it{N}_{\rm{e}}$[Ar IV] and 3: $\it{T}_{\rm{e}}$[O III] and $\it{N}_{\rm{e}}$[Cl III].\\
$^b$ It is calculated by using the equation S$^{+2}$=S$^{+}$*(4.677+(O$^{+2}$/O)$^{0.433}$) from Kingsburgh \& Barlow (1994).
\end{flushleft}
}
\end{center}
\end{table*}

\end{document}